\documentclass[twocolumn, twocolappendix]{aastex631}

\usepackage{amsmath}
\usepackage{soul}

\accepted{May 3, 2025}

\submitjournal{ApJ}

\begin{document}

\title{Generation of Large-scale Magnetic Fields Upstream of Gamma-Ray Burst Afterglow Shocks}

\correspondingauthor{Ryan Golant}
\email{ryan.golant@columbia.edu}

\author[0000-0001-6603-1983]{Ryan Golant}
\affiliation{Department of Astronomy and Columbia Astrophysics Laboratory, Columbia University, New York, NY, 10027, USA}

\author[0000-0002-3643-9205]{Arno Vanthieghem}
\affiliation{Sorbonne Universit\'e, Observatoire de Paris, Universit\'e PSL, CNRS, LERMA, F-75005, Paris, France}
\affiliation{Department of Astrophysical Sciences, Princeton University, Princeton, NJ 08544, USA}
\affiliation{Department of Astro-fusion Plasma Physics (AFP), Headquarters for Co-Creation Strategy, NINS, Tokyo 105-0001, Japan}

\author[0000-0002-5408-3046]{Daniel Gro\v selj}
\affiliation{Centre for mathematical Plasma Astrophysics, Department of Mathematics, KU Leuven, B-3001 Leuven, Belgium}

\author[0000-0002-1227-2754]{Lorenzo Sironi}
\affiliation{Department of Astronomy and Columbia Astrophysics Laboratory, Columbia University, New York, NY, 10027, USA}
\affiliation{Center for Computational Astrophysics, Flatiron Institute, 162 5th Avenue, New York, NY 10010, USA}

\begin{abstract}

The origins of the magnetic fields that power gamma-ray burst (GRB) afterglow emission are not fully understood. One possible channel for generating these fields involves the pre-conditioning of the circumburst medium: in the early afterglow phase, prompt photons streaming ahead of the GRB external shock can pair produce, seeding the upstream with drifting electron-positron pairs and triggering electromagnetic microinstabilities. To study this process, we employ 2D periodic particle-in-cell simulations in which a cold electron-proton plasma is gradually enriched with warm electron-positron pairs injected at mildly relativistic speeds. We find that continuous pair injection drives the growth of large-scale magnetic fields via filamentation-like instabilities; the temporal evolution of the field is self-similar and depends on a single parameter, $\left[\alpha/(t_{\rm f} \omega_{\rm pi})\right]^{1/2} t\omega_{\rm pi}$, where $\alpha$ is the ratio of final pair beam density to background plasma density, $t_{\rm f}$ is the duration of pair injection, and $\omega_{\rm pi}$ is the plasma frequency of background protons. Extrapolating our results to parameter regimes realistic for long GRBs, we find that upstream pair enrichment generates weak magnetic fields on scales much larger than the proton skin depth; for bright bursts, the extrapolated coherence scale at a shock radius of $R \sim 10^{17} \, {\rm cm}$ is $\left<\lambda_y\right> \sim 100 \; c/\omega_{\rm pi}$ and the corresponding magnetization is $\sigma \sim 10^{-8}$ for typical circumburst parameters. These results may help explain the persistence of magnetic fields at large distances behind GRB shocks. 
    
\end{abstract}


\section{Introduction} \label{sec:introduction}

Despite decades of observational data from facilities like CGRO, Fermi, and Swift, our understanding of the physics of gamma-ray bursts (GRBs) -- the most luminous explosions in the Universe -- remains incomplete \citep{meszaros2006}. It is generally accepted that GRBs are powered by relativistic jets launched from the compact byproducts of cataclysmic stellar events, such as the merger of two neutron stars (for short bursts) or the collapse of massive Wolf-Rayet stars (for long bursts) \citep{meszaros2002, fireball, piran, kumar}. In the standard model of GRBs, the jet gives rise to two main phases of radiation: first, dissipative processes within the jet produce the so-called ``prompt'' emission, the initial burst of MeV photons; and second, the relativistic external shock at the head of the jet produces the broadband, long-lasting ``afterglow'' emission as it sweeps up the surrounding (``circumburst'') medium, powering synchrotron emission in the post-shock magnetic field. However, within this model, questions still remain concerning the progenitors and structure of the jet, the processes that yield the prompt emission, and the origin and nature of the magnetic fields and particles that generate the afterglow. 

Our uncertainty regarding the nature of GRB afterglows is intimately tied to our incomplete understanding of relativistic, weakly magnetized, collisionless shocks (e.g., \citealt{sironimaxenergy, sironishocks,shock_rev}). The low densities of the environments surrounding GRB progenitors yield Coulomb mean free paths that far exceed the system size, meaning that GRB external shocks are mediated not by binary particle collisions but by collective plasma interactions \citep{Moiseev_1963,medvedev, gruzinov1999, lyubarsky, brainerd, wiersma, kato}. The ambient magnetic fields of circumburst environments are typically very weak, with $\sigma_0 \equiv B_0^2/4\pi n_{\rm i} m_{\rm i} c^2 \sim 10^{-9} - 10^{-5}$, where $n_{\rm i}$ is the proton density, $m_{\rm i}$ is the proton mass and $c$ is the speed of  light. In this weakly magnetized, collisionless regime, we expect shocks to be mediated by magnetic turbulence produced by the filamentation (or Weibel) instability \citep{weibel, fried}, which channels the free energy of an anisotropic flow into magnetic energy, effectively generating magnetic fields from scratch \citep{achterberg1, achterberg2, Bret_2010a, lemoine1, lemoine2, rabinak, nakar, shaisultanov}. Unfortunately, the evolution of this magnetic turbulence is highly nonlinear -- with the magnetic fields backreacting on the same particles that produce them -- making Weibel-mediated shocks nearly intractable to study analytically \citep{kirk, keshet2005, keshet2009, Bret_2013, Bret_2014, Pelletier_2019, Lemoine_2019_II, Lemoine_2019_III, Lemoine_2019_PRL}; only fully kinetic numerical simulations can completely capture these nonlinear interactions \citep{spitkovsky, birdsall, buneman}, but even these simulations are strained by limited computational resources, leaving many questions still unanswered \citep{keshet2009}.

A crucial question in the modeling of GRB afterglows concerns the extent of the afterglow emission region. If the magnetic fields powering the synchrotron afterglow are indeed generated by the filamentation instability, then theory predicts that these fields should grow on scales comparable to the proton skin depth \citep{medvedev}, leading to fast decay downstream of the shock \citep{chang, gruzinovdecay, lemoinedecay}; instead, observations combined with analytical models suggest that these downstream fields may extend
up to $10^8$ proton skin depths behind the shock \citep{waxman}. In addition, if particles at the shock were to scatter in exclusively microscale fields,
their non-thermal acceleration would be limited \citep{sironimaxenergy, Reville2014, Huang2022}.
One possible solution to this problem involves the feedback of particles accelerated at the shock \citep{keshet2009, groseljlong}: as particles are accelerated to higher and higher energies via the Fermi process \citep{blandford}, they will penetrate 
farther into the upstream, seeding progressively 
larger-scale magnetic fluctuations that 
decay more slowly and scatter particles more efficiently.
An alternative solution -- particularly relevant to long bursts, in which the GRB's external shock propagates through the dense wind of the Wolf-Rayet progenitor star -- involves the loading of the circumburst medium with electron-positron pairs prior to the passage of the shock \citep{thompson, beloborodovradiation, kumarwind, ramirezruiz, derishev, groseljpair}.  
Early in the afterglow phase (i.e., within
a few minutes of the burst), 
prompt photons stream just ahead of the shock front, where a small fraction could scatter off of electrons in the upstream plasma; the scattered photons are then susceptible to collisions with other outgoing prompt photons, producing warm electron-positron pairs.\footnote{The photons required for upstream pair production may alternatively be sourced by synchrotron radiation downstream of the shock \citep{derishev}.} As each pair increases the optical depth of the upstream medium -- thus increasing the probability of photon scattering -- the rate of pair creation increases exponentially, ultimately resulting in a pair density profile that decreases away from the shock front \citep{beloborodovgev}. The streaming of these electron-positron pairs through the circumburst plasma should trigger the filamentation instability well before the shock arrives, potentially pre-seeding the upstream with magnetic fields; if these fields are coherent on sufficiently large scales, they could survive far into the downstream. 

\citet{garasev} investigated the pair loading of the circumburst medium by conducting 2D kinetic simulations in a periodic box representing a fluid element of the upstream plasma approaching the shock front. In these simulations, electron-positron pairs with anisotropic temperature were continuously and uniformly injected into the box over time, capturing the progressive pair enrichment of the upstream fluid. This study found that the continuous injection of pairs -- which acts to maintain the particle distribution's anisotropy over macroscopic timescales -- continually fueled the filamentation instability, allowing the filamentation-generated fields to cascade to scales much larger than typical plasma scales. As the duration of pair injection increased, the final magnetic energy decreased but the average spatial scale of the magnetic field increased, in accordance with the phase mixing model of  \citet{gruzinovdecay} and \citet{chang}; this implied that long-duration pair enrichment might be able to produce large-scale fields capable of surviving far downstream of the shock. However, this work made a few unrealistic assumptions: the background plasma was composed of electrons and positrons, and the injected pairs were non-relativistic and carried no net momentum. In reality, the circumburst medium consists of an electron-proton plasma, and the pairs are injected with mildly relativistic speeds and net mean momentum in the upstream frame. 

In this paper, we overcome the limitations of \citet{garasev} by carrying out periodic, 2D, fully kinetic simulations of the continuous enrichment of an electron-\emph{proton} plasma (with realistic mass ratio) by \emph{relativistic} electron-positron pairs \emph{carrying net momentum} in the fluid frame. We also explore the physics of this system in depth, identifying trends that allow us to extrapolate our simulation results to regimes realistic for long GRBs. We conduct an extensive parameter scan over two key quantities: $t_{\rm f}$ -- the pair injection duration (or, equivalently, the time for the upstream fluid element to encounter the shock front) -- and $\alpha$ -- the ratio of the final pair beam density to the background plasma density. In the range of $t_{\rm f}$ and $\alpha$ that we explore, we find that continuous pair injection drives the growth of large-scale magnetic fields via filamentation-like instabilities; the temporal evolution of the system is self-similar and depends on a single parameter, $\left[\alpha/(t_{\rm f} \omega_{\rm pi})\right]^{1/2} t\omega_{\rm pi}$, where $\omega_{\rm pi}$ is the plasma frequency of the background protons. 
At the end of the pair injection phase, the mean wavenumber of the magnetic field (in units of $\left(c/\omega_{\rm pi}\right)^{-1}$) scales approximately with $\alpha^{-1/12} \, \left(t_{\rm f}\omega_{\rm pi}\right)^{-2/3}$ and the self-generated magnetic energy density (in units of the proton rest mass energy density) scales with $\alpha^{5/8} \, \left(t_{\rm f}\omega_{\rm pi}\right)^{-3/4}$. Using these trends to extrapolate to realistic regimes, we find that upstream pair enrichment generates weak magnetic fields on scales much larger than the proton skin depth;
for bright bursts, the extrapolated coherence scale at a shock radius of $R \sim 10^{17} \, {\rm cm}$ is $\left<\lambda_y\right> \sim 100 \; c/\omega_{\rm pi}$ and the corresponding magnetization is $\sigma \sim 10^{-8}$ for typical circumburst parameters.
Since larger-scale fields decay more slowly, our results may help explain the persistence of magnetic fields at large distances behind GRB shocks.     

This paper is structured as follows: in Section \ref{sec:setup}, we detail our simulation setup; in Section \ref{sec:case}, we analyze an example simulation, highlighting the salient features of the magnetic field evolution; in Section \ref{sec:comparison}, we compare how the magnetic field evolution varies with $\alpha$ and $t_{\rm f}$; in Section \ref{sec:extrapolation}, we extrapolate our simulation results to realistic regimes; and in Section \ref{sec:conclusions}, we summarize our main findings and conclude. We provide supplementary information in the appendices: in Appendix \ref{sec:3D}, we compare our 2D results with a 3D simulation; in Appendix \ref{sec:lowmass}, we show results from simulations with reduced mass ratios to confirm that our aforementioned extrapolations are fully justified; in Appendix \ref{sec:gammab}, we compare simulations with varying beam Lorentz factors; in Appendix \ref{sec:kint_appendix}, we demonstrate that changing our definition of the dominant magnetic wavenumber negligibly changes our conclusions; and in Appendix \ref{sec:convergence}, we show that the results presented in this paper are converged numerically.

\section{Simulation Setup} \label{sec:setup}

We employ the electromagnetic particle-in-cell (PIC) code TRISTAN-MP \citep{tristan, spitkovsky}, which is especially well-suited for handling relativistic flows in collisionless plasmas. The code solves the Maxwell-Vlasov system using a second-order Finite-Difference Time-Domain (FDTD) scheme on a standard Yee mesh \citep{yee} and pushes particles using the Boris algorithm \citep{boris} and first-order shape functions. The electromagnetic fields are extrapolated to particle positions using a bilinear interpolation function (or a trilinear interpolation function for our 3D run). At each time step, after depositing the electric current to the grid using the charge-conserving zig-zag scheme \citep{zigzag}, we apply twenty passes of a 3-point (1-2-1) digital current filter in each direction to smooth out non-physical short-wavelength oscillations. 

We run fully periodic 2D simulations (in the $xy$-plane) of a cold, initially unmagnetized electron-proton plasma (the ``background'') that is gradually enriched with warm, mildly relativistic electron-positron pairs (the ``beam''). The background particles and the beam particles are both distributed homogeneously throughout the simulation box. The background is initially at rest and follows a Maxwellian with $\frac{k_{\rm B} T_{\rm i}}{m_{\rm i} c^2} = \frac{m_{\rm e}}{m_{\rm i}} \frac{k_{\rm B} T_{\rm e, \, bg}}{m_{\rm e} c^2} = 10^{-7}$ (where $k_{\rm B}$ is Boltzmann's constant, $c$ is the speed of light, $T_{\rm i}$ is the proton temperature, $T_{\rm e, \, bg}$ is the background electron temperature, $m_{\rm i}$ is the proton mass,  $m_{\rm e}$ is the electron mass, and $m_{\rm i}/m_{\rm e} = 1836$). The beam pairs are drawn from a Maxwell-J\"uttner distribution with $\frac{k_{\rm B} T_{\rm b}}{m_{\rm e} c^2} = 1$ (for $T_{\rm b}$ the initial temperature of both beam species), drifting along the positive $x$ direction with bulk Lorentz factor $\gamma_{\rm b} = 1.5$. 

The beam particles are injected in pairs at a constant rate throughout the duration of the simulation. The injection rate is determined by two key parameters: $t_{\rm f}$ -- the injection duration -- and $\alpha$ -- the ratio of the total beam density to the total background density at the end of injection. In terms of these quantities, $n_{\rm bg} \times \frac{\alpha}{t_{\rm f}\omega_{\rm pb}}$ electron-positron pairs are injected per plasma time ($\omega_{\rm pb}^{-1}$). Here, $n_{\rm bg}$ is the apparent number density of the background plasma (i.e., the total number density of the background electrons and the protons) and $\omega_{\rm pb}$, the final beam electron plasma frequency, is defined to be 
\begin{equation} \label{eq:omegabeam}
    \omega_{\rm pb} \equiv \sqrt{\frac{4\pi (n_{\rm b}/2) e^2}{m_{\rm e}}}\,,
\end{equation}
for $n_{\rm b}$ the total beam density at the end of pair injection. It is also useful to define the plasma frequency of the background protons,
\begin{equation} \label{eq:omegaproton}
    \omega_{\rm pi} \equiv \sqrt{\frac{4\pi (n_{\rm bg}/2) e^2}{m_{\rm i}}} = \omega_{\rm pb} \sqrt{\frac{m_{\rm e}}{m_{\rm i} \alpha}}\,.
\end{equation}
In our simulations, all quantities are normalized to the properties of the pair beam at the end of injection to ensure that all length-scales are well-resolved by our grid; since we only consider $\alpha \ge 1$, the skin depth of the protons ($c/\omega_{\rm pi}$) and the skin depth of the background electrons ($c/\omega_{\rm pe} \equiv \sqrt{m_{\rm e}/m_{\rm i}} \, c/\omega_{\rm pi}$) are always at least as large as the skin depth of the beam electrons. Throughout this paper, we cite results both in units of the beam electrons and in units of the background protons.

In our suite of simulations, we consider values of $\alpha$ equal to 1, 2, 4, 8, and 16; for each $\alpha$, we vary $t_{\rm f}$ among $t_{\rm f}\omega_{\rm pb} = 5 \times 10^3$, $10^4$, $2 \times 10^4$, and $4 \times 10^4$ (respectively corresponding to $t_{\rm f}\omega_{\rm pi} \sim 117/\sqrt{\alpha}$, $233/\sqrt{\alpha}$, $467/\sqrt{\alpha}$, and $934/\sqrt{\alpha}$). These values are small compared to those appropriate for real GRBs, since our realistic mass ratio prevents us from directly simulating larger $\alpha$ and $t_{\rm f}$. However, our range of $\alpha$ and $t_{\rm f}$ is broad enough to allow for reliable extrapolation to higher values.

For our production runs, we primarily use 36 final beam particles per cell per species; the number of background particles per cell per species is thus $36/\alpha$. For a few representative cases, we also use 72 final beam particles per cell per species (and $72/\alpha$ background particles per cell per species) to check for numerical convergence (see Appendix \ref{sec:convergence}). We resolve the final beam electron skin depth, $c/\omega_{\rm pb}$, using 10 cells. We employ square simulation boxes with side lengths of either 600 $c/\omega_{\rm pb}$ (6000 cells) or 1000 $c/\omega_{\rm pb}$ ($10^4$ cells); we use the larger boxes for our $t_{\rm f}\omega_{\rm pb} = 2 \times 10^4$ and $4 \times 10^4$ cases to ensure that the dominant scale of the magnetic field is not artificially limited by the box size. For our $\alpha=16$ simulations (i.e., the cases with the largest proton skin depth), $c/\omega_{\rm pi} \sim 200 \; c/\omega_{\rm pb}$, so our simulation boxes always fit at least three proton skin depths in each direction.

To check our 2D results, we run a single 3D simulation with our fiducial parameters, $\alpha=2$ and $t_{\rm f}\omega_{\rm pb}=10^4$; this simulation is analyzed in Appendix \ref{sec:3D}. For this 3D run, we use 18 final beam particles per cell per species, we resolve $c/\omega_{\rm pb}$ with 10 cells, and we employ a cubic simulation box with a 100 $c/\omega_{\rm pb}$ (1000 cell) side length.

In Appendix \ref{sec:convergence}, we show that our simulations converge numerically with respect to both particles per cell and spatial resolution, and we demonstrate that the growth of the magnetic field is not impeded by the size of our simulation box.

\section{Case study: $\lowercase{\alpha=2, \; t_{\rm f}\omega_{\rm pb} =10^4}$} \label{sec:case}

\begin{figure*}[h]
    \centering
    \includegraphics[scale=0.47]{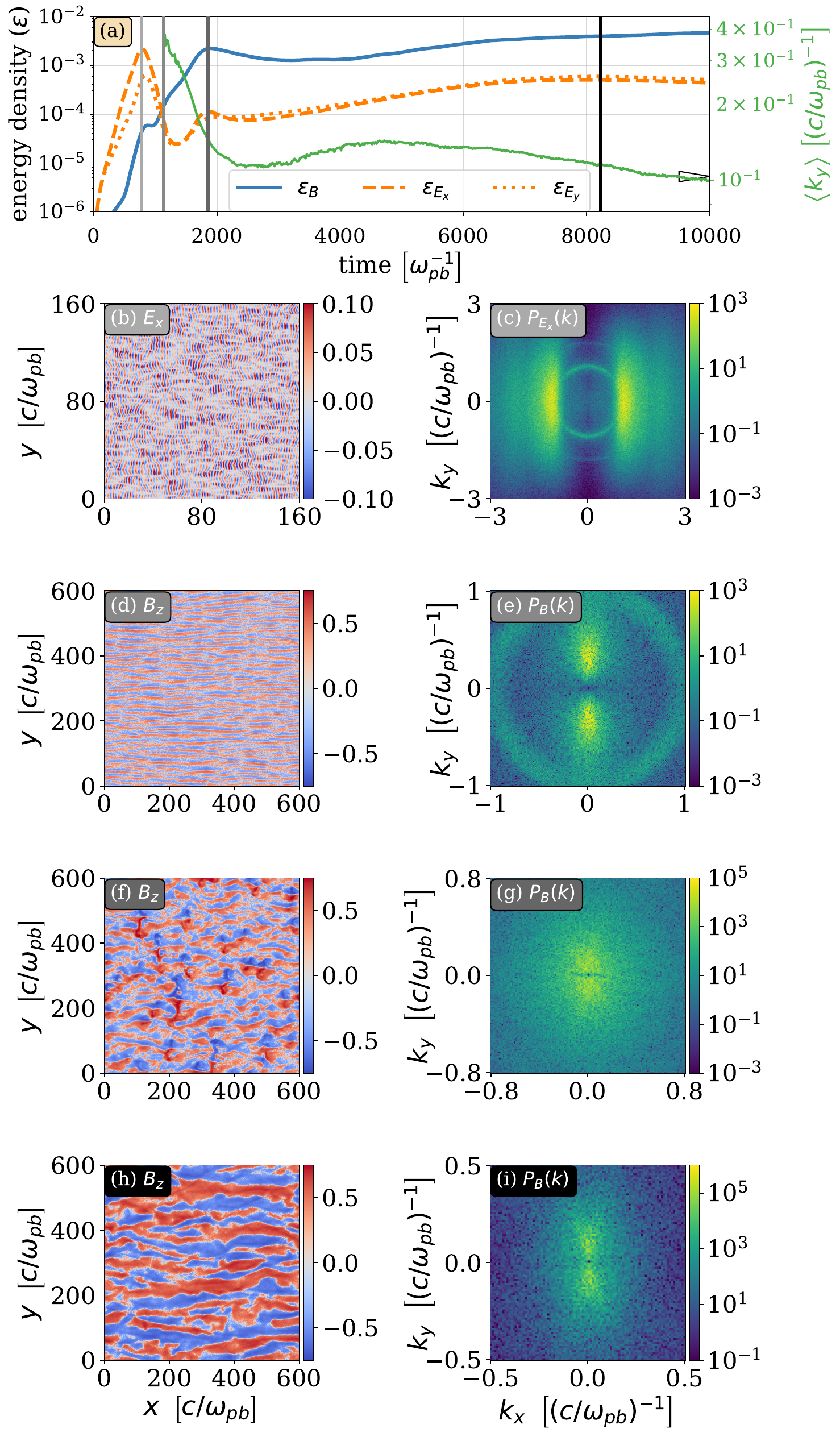}
    \caption{Time evolution and a few representative snapshots from our reference simulation with $\alpha=2$ and $t_{\rm f}\omega_{\rm pb} = 10^4$. Panel (a) shows the full time evolution of the box-averaged magnetic energy density ($\varepsilon_B$, solid blue curve) and electric energy densities ($\varepsilon_{E_x}$ and $\varepsilon_{E_y}$, dashed orange and dotted orange curves, respectively), as well as the average transverse magnetic wavenumber ($\left<k_y\right>c/\omega_{\rm pb}$, green curve, measured on the right vertical axis for times later than $t\omega_{\rm pb} \sim 1100$, when the magnetic energy overtakes the electric energy); the arrow at $\left<k_y\right> c/\omega_{\rm pb} \sim 10^{-1}$ on the right vertical axis indicates the characteristic wavenumber $2\pi/(c/\omega_{\rm pi})$. The four vertical grayscale lines mark the times corresponding to the snapshots below, with each line corresponding to a different row. Panels (b) and (c) show the 2D spatial profile and 2D power spectrum of $E_x$, respectively, at $t\omega_{\rm pb} \sim 770$, the peak of the electrostatic phase. Panels (d) and (e) show the 2D spatial profile and the 2D power spectrum of $B_z$, respectively, at $t\omega_{\rm pb} \sim 1100$, when the magnetic energy overtakes the electric energy. Panels (f) and (g) again show the spatial profile and power spectrum of $B_z$, but at $t\omega_{\rm pb} \sim 1850$, when the filamentation phase saturates. Finally, panels (h) and (i) show the spatial profile and power spectrum of $B_z$ at an arbitrarily late time, $t\omega_{\rm pb} \sim 8280$, deep into the secular phase. In panels (d), (f), and (h), the color bar is rescaled to $\left|B_z\right|^{0.2} \times {\rm sign}(B_z)$ to display a wider dynamic range.}
    \label{fig:field_evolution}
\end{figure*}

We begin by describing the salient features of a reference simulation with $\alpha=2$ and $t_{\rm f}\omega_{\rm pb} = 10^4$; this case demonstrates the various stages of magnetic field evolution that we see in each of our simulations. As we inject electron-positron pairs over the duration of the simulation, the magnetic field evolves through three general stages, as summarized in Figure \ref{fig:field_evolution}. Figure \ref{fig:field_evolution}a shows the complete time evolution of the energy densities of the magnetic and electric fields (blue and orange curves, respectively); we normalize these energy densities to the bulk kinetic energy of the injected pairs. We define the box-averaged magnetic energy density to be
\begin{equation} \label{eq:epsb}
    \varepsilon_B \equiv \frac{\left<B_z^2\right>/8\pi}{(\gamma_{\rm b}-1)m_{\rm e} n_{\rm b} c^2},
\end{equation}
where $B_z$ is the $z$-component of the magnetic field (the only non-zero component in our 2D unmagnetized plasma), $\gamma_{\rm b}$ is the bulk Lorentz factor of the injected pairs, and $n_{\rm b}$ is the total beam number density at the end of injection (in this case, $\gamma_{\rm b} = 1.5$ and $n_{\rm b}$ is twice the background number density). Similarly, we define the box-averaged electric energy densities to be
\begin{equation}
    \varepsilon_{E_i} \equiv \frac{\left<E_i^2\right>/8\pi}{(\gamma_{\rm b}-1)m_{\rm e} n_{\rm b} c^2},
\end{equation}
where $E_i$ can be either $E_x$ or $E_y$, the two non-zero components of the electric field; $\varepsilon_{E} \equiv \varepsilon_{E_x} + \varepsilon_{E_y}$.

Alongside the energy densities in Figure \ref{fig:field_evolution}a, we also show the time evolution of $\left<k_y\right>$ -- the average of the $y$-component of the magnetic field wavenumber (i.e., the magnetic wavenumber transverse to the beam) -- plotted in green and measured on the right vertical axis in units of $\left(c/\omega_{\rm pb}\right)^{-1}$; we start plotting $\left<k_y\right>$ at $t_{\rm f} \omega_{\rm pb} \sim 1100$, when the magnetic energy overtakes the electric energy. $\left<k_y\right>$ is derived from the 1D transverse power spectrum of the magnetic field, $P_B(k_y)$:
\begin{equation} \label{eq:avgk}
    \left<k_y\right> \equiv \frac{\int_0^{k_{\rm cut}} k_y \, P_B(k_y) \, dk_y}{\int_0^{k_{\rm cut}} P_B(k_y) \, dk_y}.
\end{equation}
We obtain $P_B(k_y)$ by averaging the 2D magnetic power spectrum over $k_x$. We place a finite upper bound, $k_{\rm cut}$, on the integrals in Equation \eqref{eq:avgk} to avoid contaminating our measurements with the high-$k_y$ end of the magnetic spectrum, which is affected by numerical shot noise; we choose $k_{\rm cut}$ to be the location where $k_y P_B(k_y)$ reaches a minimum between the peak of $k_y P_B(k_y)$ and $k_y c/\omega_{\rm pb}=1$.
We find that placing similar bounds on $k_x$ negligibly affects our results, so the quantities computed in this paper use the full range of $k_x$.\footnote{Alternatively, we could quantify the dominant magnetic wavenumber using the ``integral-scale'' definition (Equation \eqref{eq:kint}), which should be more robust against high-$k_y$ noise. In Appendix \ref{sec:kint_appendix}, we show that our results change negligibly if we use the integral-scale wavenumber instead of $\left<k_y\right>$.} 

The four vertical grayscale lines in Figure \ref{fig:field_evolution}a mark the times of a few illustrative snapshots, which are depicted in Figures \ref{fig:field_evolution}b through \ref{fig:field_evolution}i and characterize the various stages of field evolution; each vertical line corresponds to a different row in Figure \ref{fig:field_evolution}.

At early times (i.e., $t \omega_{\rm pb} \lesssim 1100$), the total energy of the system is dominated by the electric field, with $\varepsilon_{E_x} > \varepsilon_{E_y} \gg \varepsilon_B$. This early exponential growth of the electric energy is driven by the development of both the electrostatic two-stream instability and the quasi-electrostatic oblique instability \citep{Bret_2010a}, as evidenced by the spatial profile of $E_x$ (normalized by $\sqrt{8\pi\left(\gamma_{\rm b}-1\right)m_{\rm e} n_{\rm b} c^2}$) and its associated 2D power spectrum in Figures \ref{fig:field_evolution}b and \ref{fig:field_evolution}c, respectively. The oscillations with $x$-aligned (i.e., beam-aligned) wavevectors in Figure \ref{fig:field_evolution}b and the power at $(k_x c / \omega_{\rm pb} \sim \pm 1, \, k_y c / \omega_{\rm pb} = 0)$ in Figure \ref{fig:field_evolution}c are indicative of two-stream modes, while the oblique oscillations in Figure \ref{fig:field_evolution}b and the power at $(k_x c / \omega_{\rm pb} \sim \pm 1, \, k_y c / \omega_{\rm pb} \sim \pm 1)$ in Figure \ref{fig:field_evolution}c are indicative of oblique modes. The subdominant -- but still exponential -- growth of the magnetic energy is due to the electromagnetic component of the oblique mode. 

The rings around the center of the power spectrum in Figure \ref{fig:field_evolution}c seem to be physical: the rings remain at the same scale and amplitude despite changes to the parallelization of the domain, the numerical speed of light, the order of the FDTD scheme, the number of passes of the current filter, the number of particles per cell, or the spatial resolution. While these isotropic high-$k$ modes could be produced by early Langmuir wave collapse \citep{zakharov}, we do not see the small-scale magnetic cavities indicative of this process, even in our highest-resolution simulations. We thus leave the characterization of these rings to future work. 

\begin{figure}[h]
    \includegraphics[width=0.47\textwidth]{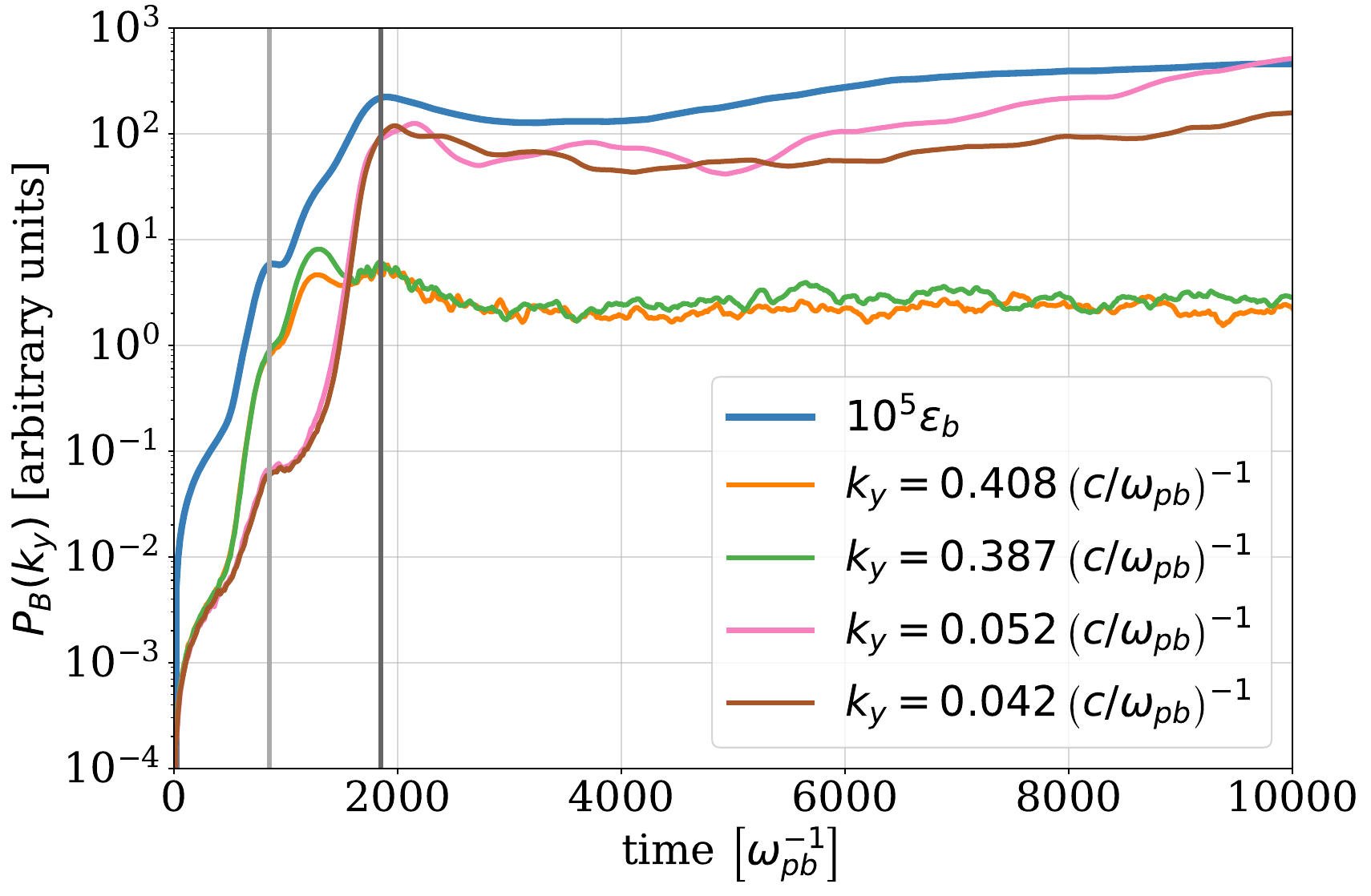}
    \caption{Time evolution of a few representative wavemodes plotted alongside $\varepsilon_B$ (blue curve; arbitrarily rescaled) for our $\alpha=2$, $t_{\rm f}\omega_{\rm pb}=10^4$ reference simulation. $k_y c/\omega_{\rm pb} \sim 0.408$ (orange curve) and $k_y c/\omega_{\rm pb} \sim 0.387$ (green curve) -- the two modes that carry the most magnetic power at the saturation of the electrostatic phase, $t_{\rm f}\omega_{\rm pb} \sim 850$ (vertical light gray line) -- are quickly overtaken by the much larger-scale modes $k_y c/\omega_{\rm pb} \sim 0.052$ (pink curve) and $k_y c/\omega_{\rm pb} \sim 0.042$ (brown curve) -- the two modes that carry the most power at the saturation of the filamentation phase, $t\omega_{\rm pb} \sim 1850$ (vertical dark gray line).}
    \label{fig:wavemodes}
\end{figure}

After the two-stream and oblique modes saturate around $t \omega_{\rm pb} \sim 850$, the magnetic energy enters a second stage of exponential growth from $1000 \lesssim t\omega_{\rm pb} \lesssim 1850$, with $\varepsilon_B$ quickly overtaking $\varepsilon_{E_x}$ and $\varepsilon_{E_y}$ by $t\omega_{\rm pb} \sim 1100$. Figures \ref{fig:field_evolution}d and \ref{fig:field_evolution}e show, respectively, the spatial profile of $B_z$ (normalized by $\sqrt{8\pi\left(\gamma_{\rm b}-1\right)m_{\rm e} n_{\rm b} c^2}$) and the 2D power spectrum of $B_z$ at the time when the magnetic energy surpasses the electric energy. The thin structures stretched along the $x$-direction in Figure \ref{fig:field_evolution}d and the dumbbell shape oriented along $k_x c/\omega_{\rm pb} = 0$ in Figure \ref{fig:field_evolution}e are characteristic of the filamentation instability. Figure \ref{fig:wavemodes} plots the power carried by a few individual transverse wavemodes over time, revealing that the envelope of $\varepsilon_B$ as plotted in Figure \ref{fig:field_evolution}a is shaped by progressively larger-scale modes as the growth proceeds. 

As the phase of filamentation growth proceeds from $1100 \lesssim t\omega_{\rm pb} \lesssim 1850$, the average transverse scale of the magnetic field increases by more than a factor of two, as shown by the drop in $\left<k_y\right>c/\omega_{\rm pb}$ in Figure \ref{fig:field_evolution}a. This increase in scale can also be seen in Figures \ref{fig:field_evolution}f and \ref{fig:field_evolution}g, which depict the spatial profile and 2D power spectrum of $B_z$ at $t\omega_{\rm pb} \sim 1850$, where the filamentation growth saturates; Figures \ref{fig:field_evolution}f and \ref{fig:field_evolution}g can be directly compared to the earlier snapshots in Figures \ref{fig:field_evolution}d and \ref{fig:field_evolution}e. Interestingly, the formerly ordered filamentary structures of Figure \ref{fig:field_evolution}d appear to be turbulently mixed as the filamentation growth saturates (Figure \ref{fig:field_evolution}f), but this order is restored at later times, albeit on much larger scales (Figure \ref{fig:field_evolution}h).  

\begin{figure}[h]
    \includegraphics[width=0.47\textwidth]{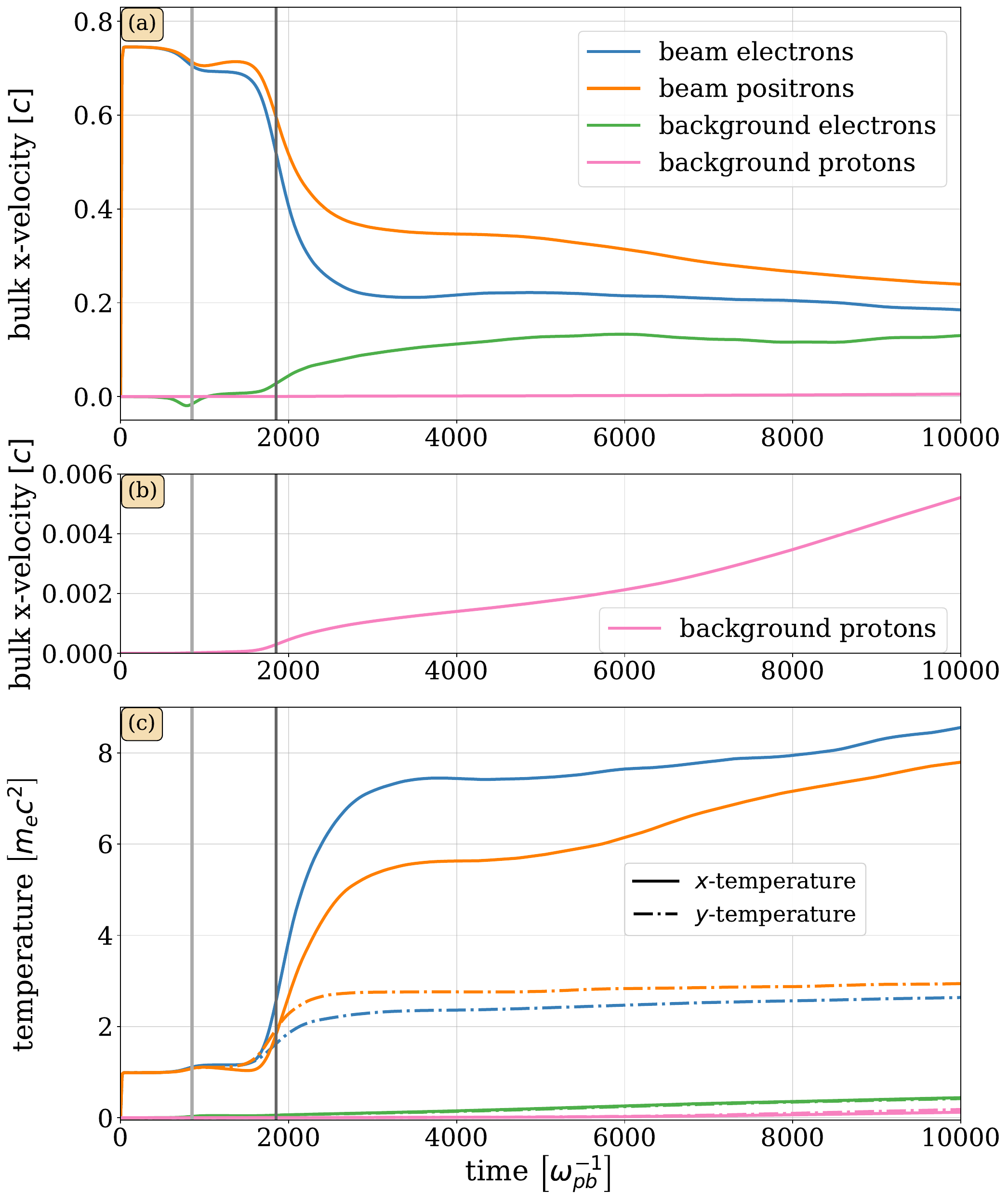}
    \caption{Time evolution of the box-averaged $x$-velocities ($v_x/c$, panel (a)) and temperatures ($k_{\rm B} T/m_{\rm e} c^2$, panel (c)) for the beam electrons (blue), beam positrons (orange), background electrons (green), and background protons (pink) in our $\alpha=2$, $t_{\rm f}\omega_{\rm pb}=10^4$ reference simulation. Panel (b) zooms in on the proton velocity curve from panel (a). In each panel, the light gray vertical line marks the time of saturation of the electrostatic phase and the dark gray vertical line marks the time of saturation of the filamentation phase.}
    \label{fig:velocities}
\end{figure}

After the saturation of the filamentation phase at $t\omega_{\rm pb} \sim 1850$, the magnetic energy dips slightly as magnetic filaments break up and form cavities. However, these large-scale filaments appear again at later times -- as illustrated by the late-time growth of the small-$k_y$ modes in Figure \ref{fig:wavemodes} -- leading to a final phase of slow, secular growth of the magnetic field from $t\omega_{\rm pb} \sim 4000$ until the end of the simulation at $t\omega_{\rm pb} = 10^4$.

As the filamentation phase saturates at $t\omega_{\rm pb} \sim 1850$, momentum is rapidly transferred from the beam to the background plasma, accelerating and heating the background electrons and protons; this is illustrated in Figure \ref{fig:velocities}, which shows the velocities and temperatures of each particle species over time in our reference simulation. The background electrons and protons -- which had largely been stationary at earlier times -- start picking up small bulk $x$-velocities at $t_{\rm f}\omega_{\rm pb} \sim 1600$, but decouple due to their mass difference. This background mass difference also causes the beam electrons to moderately decouple from the beam positrons; such symmetry breaking is known to drive the growth of low-density, high-magnetic-energy plasma cavities, which we see in Figure \ref{fig:field_evolution}f  \citep{peterson_21, peterson_22, groseljlong}.
Meanwhile, the relative drift between the protons and the leptons triggers a secondary filamentation instability that drives the observed late-time magnetic field growth, as evidenced by the formation of proton filaments that were absent at the peak of the first filamentation phase (see Section \ref{sec:scan}). This effect is more pronounced at large $\alpha$, where the properties of the background electrons become nearly the same as those of the beam species following saturation of the filamentation phase; as we show in Section \ref{sec:scan}, increasing $\alpha$ (at fixed $t_{\rm f}$) increases both the secular phase growth rate and the density contrast of the proton filaments.

As this secular phase proceeds, the transverse spatial scale of the magnetic field continues to increase, producing progressively thicker current filaments like those shown in Figure \ref{fig:field_evolution}h, at $t\omega_{\rm pb} \sim 8280$; by the end of the simulation, $\left<k_y\right> c/\omega_{\rm pb}$ is almost a factor of four smaller than it was in the early stages of the first filamentation phase. The final transverse magnetic field scale, $\left<\lambda_y\right>_{\rm f}  \equiv 2\pi/\left<k_y\right>_{\rm f}$ (where $\left<k_y\right>_{\rm f}$ is the final transverse magnetic wavenumber),
is $\sim 60 \; c/\omega_{\rm pb}$ for our reference simulation. This is almost twice the proton skin depth, which is $\sim 60 \; c/\omega_{\rm pb}$ for $\alpha=2$ (see Equation \eqref{eq:omegaproton}); in Figure \ref{fig:field_evolution}a, we add an arrow at $2\pi/(c/\omega_{\rm pi})$ on the right vertical axis for visual comparison of the scales. 

In summary, the evolution of the magnetic field proceeds through three phases: first, a primarily electrostatic phase dominated by two-stream and oblique modes ($t\omega_{\rm pb} \lesssim 1100$); second, an exponential electromagnetic phase driven by filamentation modes enhanced by continuous pair injection ($1100 \lesssim t\omega_{\rm pb} \lesssim 1850$), during which the scale of the magnetic field increases by a factor of more than two; and third, a phase of slow, secular growth driven by the relative drift between the leptons and the protons ($t\omega_{\rm pb} \gtrsim 4000$), at the end of which the scale of the magnetic field is roughly four times larger than at the start of the filamentation phase. In the next section, we show that these phases are generic across the range of $\alpha$ and $t_{\rm f}$ that we simulate.
%\vfill\eject
% \hfill

\section{Comparison across $\alpha$ and $\lowercase{t_{\rm f}}$} \label{sec:comparison}

\subsection{Parameter Scan} \label{sec:scan}
To draw general conclusions regarding the impact of continuous electron-positron pair injection on magnetic field evolution, we performed a parameter scan over $\alpha$ -- the final beam-to-background density ratio -- and $t_{\rm f}$ -- the pair injection duration -- with $\alpha$ varying between 1, 2, 4, 8, and 16 and $t_{\rm f}\omega_{\rm pb}$ varying between $5 \times 10^3$, $10^4$, $2 \times 10^4$, and $4 \times 10^4$. Here, we discuss our main findings from the parameter scan.

\begin{figure}[h]
    \includegraphics[width=0.47\textwidth]{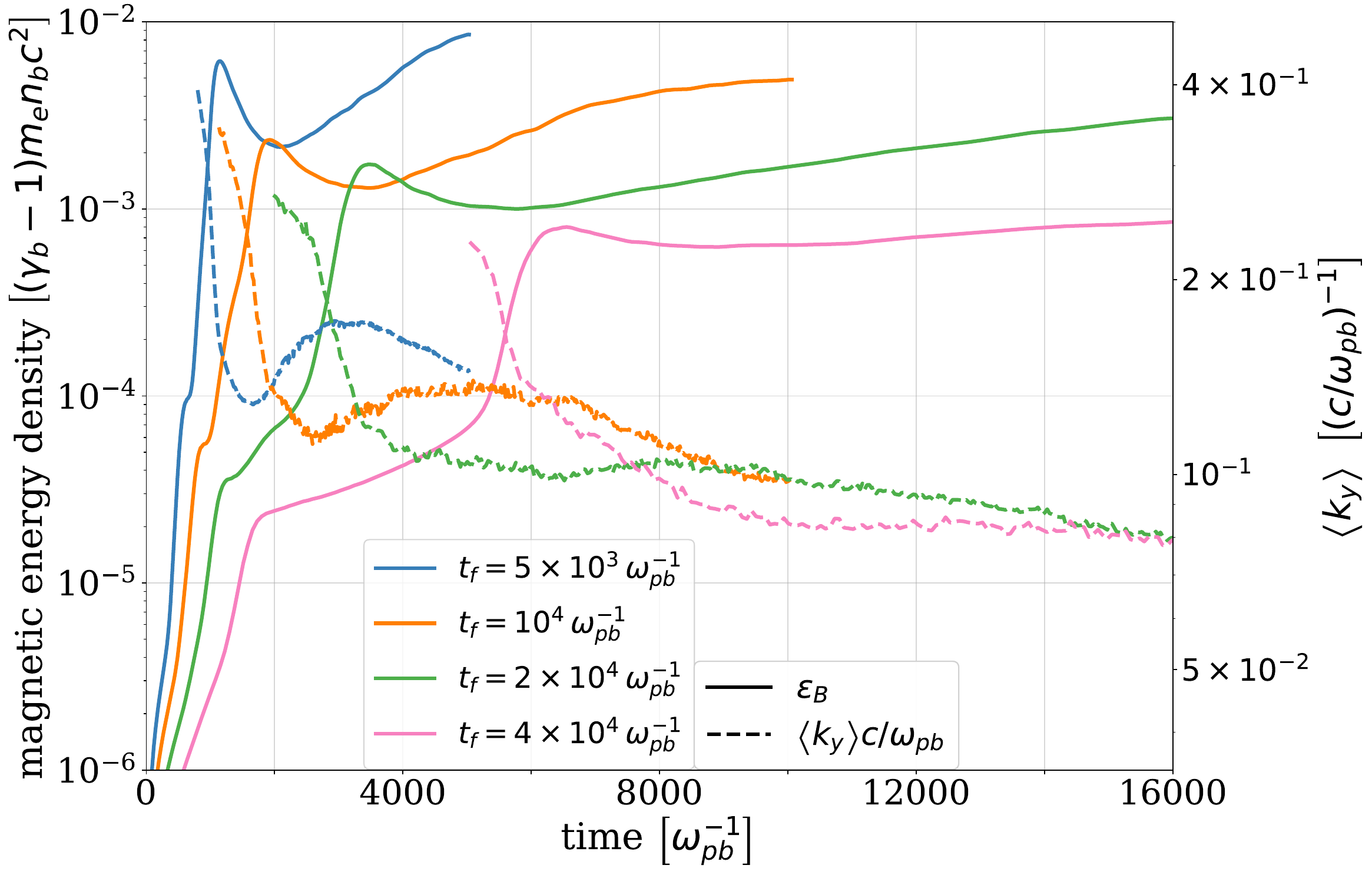}
    \caption{Time evolution of $\varepsilon_B$ (solid curves, measured on the left vertical axis) and $\left<k_y\right>c/\omega_{\rm pb}$ (dashed curves, measured on the right vertical axis) for four simulations with $\alpha=2$: $t_{\rm f}\omega_{\rm pb} = 5 \times 10^3$ (blue), $t_{\rm f}\omega_{\rm pb} = 10^4$ (orange), $t_{\rm f}\omega_{\rm pb} = 2 \times 10^4$ (green), and $t_{\rm f}\omega_{\rm pb} = 4 \times 10^4$ (pink). The green and pink curves extend well beyond the plotted time range.}
    \label{fig:alpha2_combined}
\end{figure}

Figure \ref{fig:alpha2_combined} compares the time evolution of the magnetic energy density (left vertical axis) and the average transverse wavenumber of the magnetic field (right vertical axis) for four simulations with $\alpha=2$: $t_{\rm f} \omega_{\rm pb} = 5 \times 10^3$, $10^4$, $2 \times 10^4$, and $4 \times 10^4$. Across each of these cases, the trajectories of $\varepsilon_B$ and of $\left<k_y\right>c/\omega_{\rm pb}$ over time are qualitatively the same: early growth of primarily electrostatic modes yields a weak magnetic field, which is then amplified in energy and scale by a filamentation instability enhanced by continuous pair injection, and is finally brought to even higher energies and larger scales by a relatively slow-growing secular phase. As $t_{\rm f}\omega_{\rm pb}$ increases, the timescales of each of these phases also increases: the electrostatic phase lasts longer, the filamentation phase starts later and develops more slowly, and the secular phase growth is more shallow. 
While the decoupling of the background electrons from the background protons and of the beam electrons from the beam positrons occurs at later times for higher $t_{\rm f}\omega_{\rm pb}$, the final velocities of the particles do not vary significantly between simulations at fixed $\alpha$.

\begin{figure}[h]
    \includegraphics[width=0.47\textwidth]{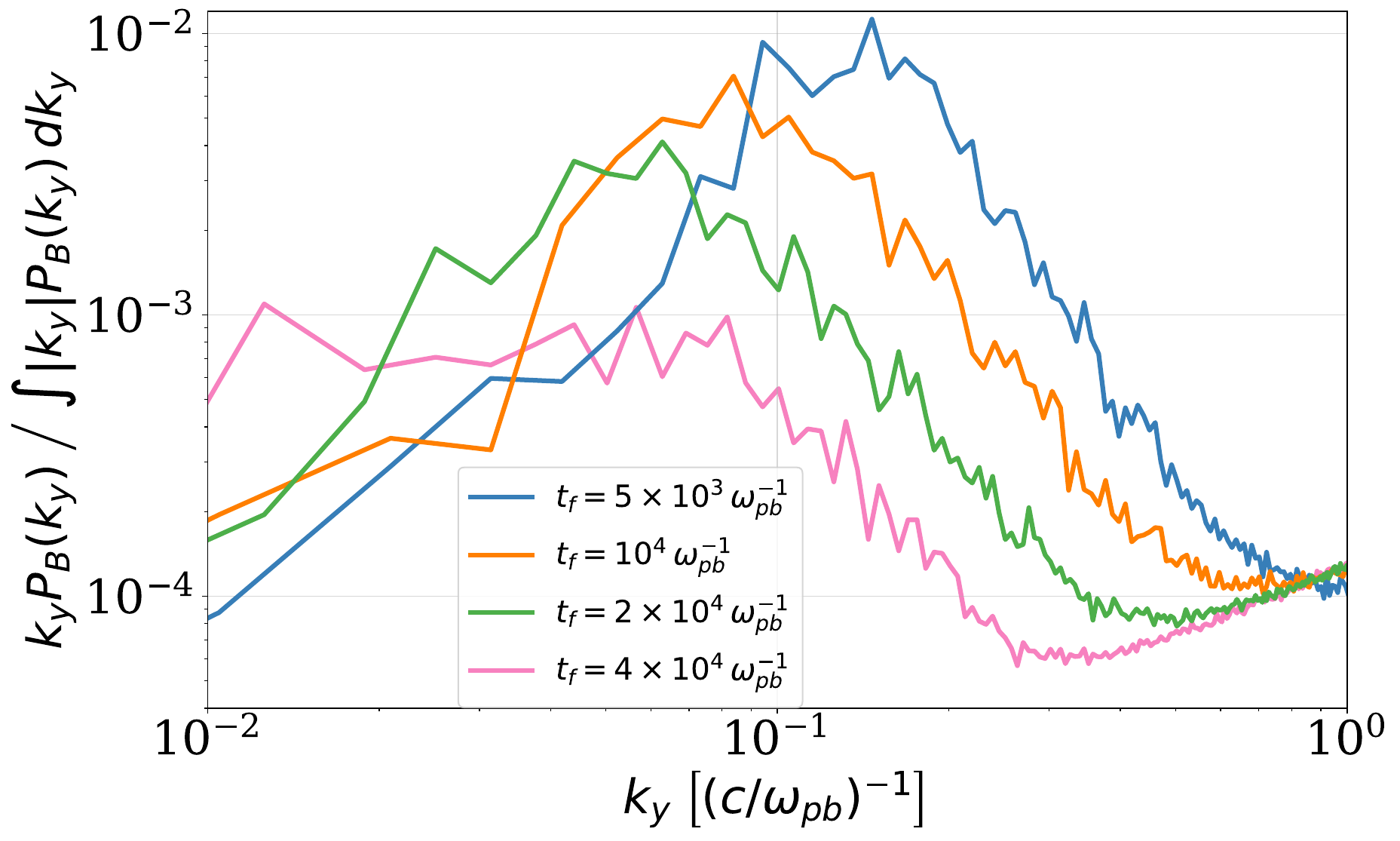}
    \caption{1D transverse magnetic power spectra at the end of pair injection for four simulations with $\alpha=2$: $t_{\rm f}\omega_{\rm pb} = 5 \times 10^3$ (blue), $t_{\rm f}\omega_{\rm pb} = 10^4$ (orange), $t_{\rm f}\omega_{\rm pb} = 2 \times 10^4$ (green), and $t_{\rm f}\omega_{\rm pb} = 4 \times 10^4$ (pink). Each power spectrum is normalized to the total area under the curve, $\int \left|k_y \right| P_B(k_y) \, dk_y$, to allow for fair comparison of the spectral peaks.}
    \label{fig:alpha2_pk}
\end{figure}

\begin{figure}[h]
    \includegraphics[width=0.47\textwidth]{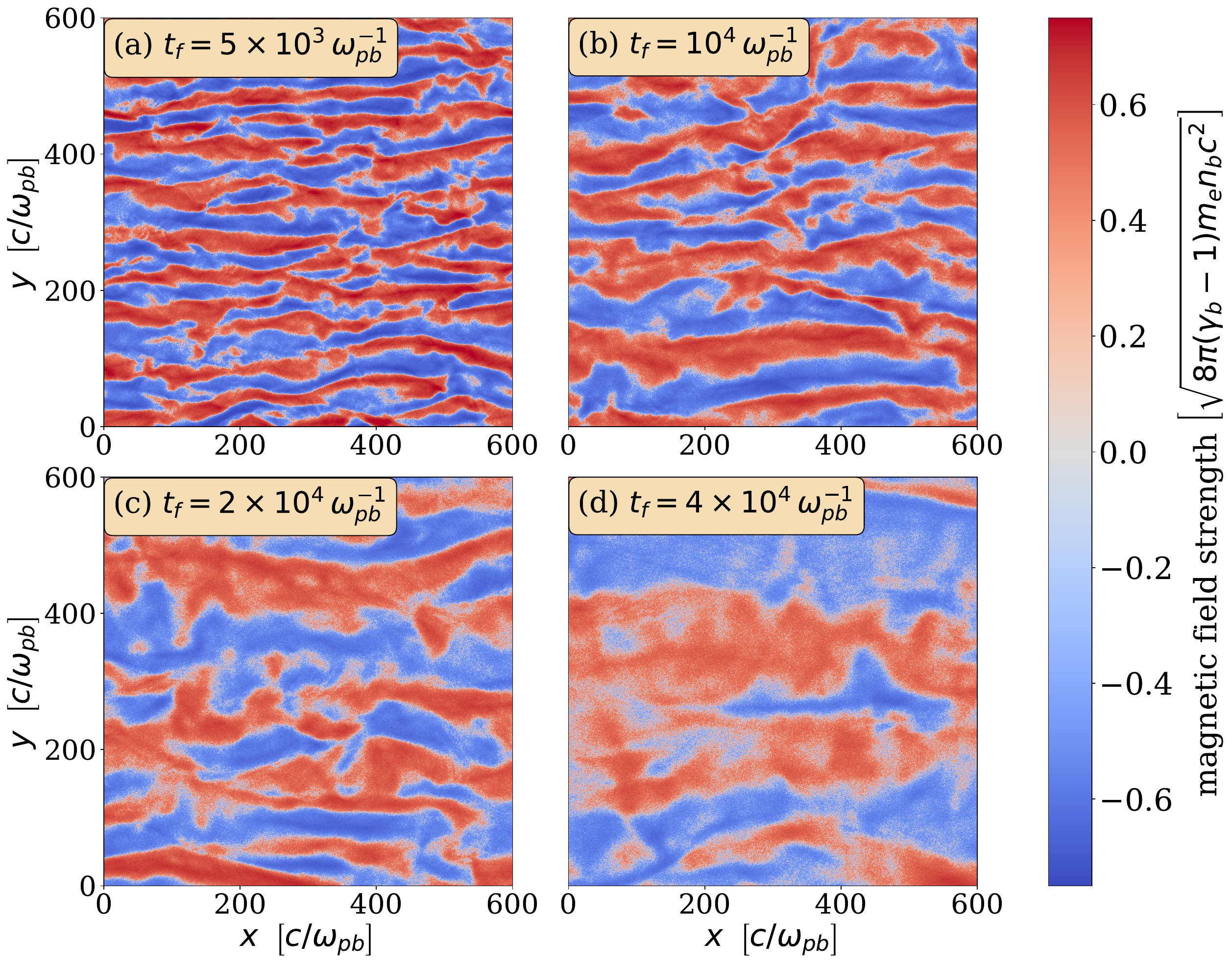}
    \caption{Spatial profiles of $B_z/\sqrt{8\pi\left(\gamma_{\rm b}-1\right)m_{\rm e} n_{\rm b} c^2}$ at the end of pair injection for four simulations with $\alpha=2$: (a) $t_{\rm f}\omega_{\rm pb} = 5 \times 10^3$, (b) $t_{\rm f}\omega_{\rm pb} = 10^4$, (c) $t_{\rm f}\omega_{\rm pb} = 2 \times 10^4$, and (d) $t_{\rm f}\omega_{\rm pb} = 4 \times 10^4$. The color bar is rescaled to $\left|B_z\right|^{0.2} \times {\rm sign}(B_z)$ to display a wider dynamic range.}
    \label{fig:alpha2_profiles}
\end{figure}

In agreement with \citet{garasev}, as we hold $\alpha$ fixed and increase $t_{\rm f}\omega_{\rm pb}$, both the final $\varepsilon_B$ and the final $\left<k_y\right>c/\omega_{\rm pb}$ {decrease}. Since smaller-scale magnetic fluctuations dissipate more quickly, these two trends are linked: a longer $t_{\rm f}$ allows more time for smaller-scale modes to die out -- thus reducing the final magnetic energy -- yet also allows more time for larger-scale modes to grow. This trend towards larger scales at higher $t_{\rm f}\omega_{\rm pb}$ is illustrated quantitatively in Figure \ref{fig:alpha2_pk}, which shows 1D transverse power spectra of $B_z$ at the end of pair injection for our four $\alpha=2$ cases; as $t_{\rm f}\omega_{\rm pb}$ increases, the peak of the spectrum shifts towards smaller $k_y c/\omega_{\rm pb}$, or towards larger scales. This same trend is illustrated in Figure \ref{fig:alpha2_profiles}, which shows spatial profiles of $B_z$ at the end of injection; as $t_{\rm f}\omega_{\rm pb}$ increases, so does the width of the filamentary magnetic structures.

\begin{figure}[h]
    \includegraphics[width=0.47\textwidth]{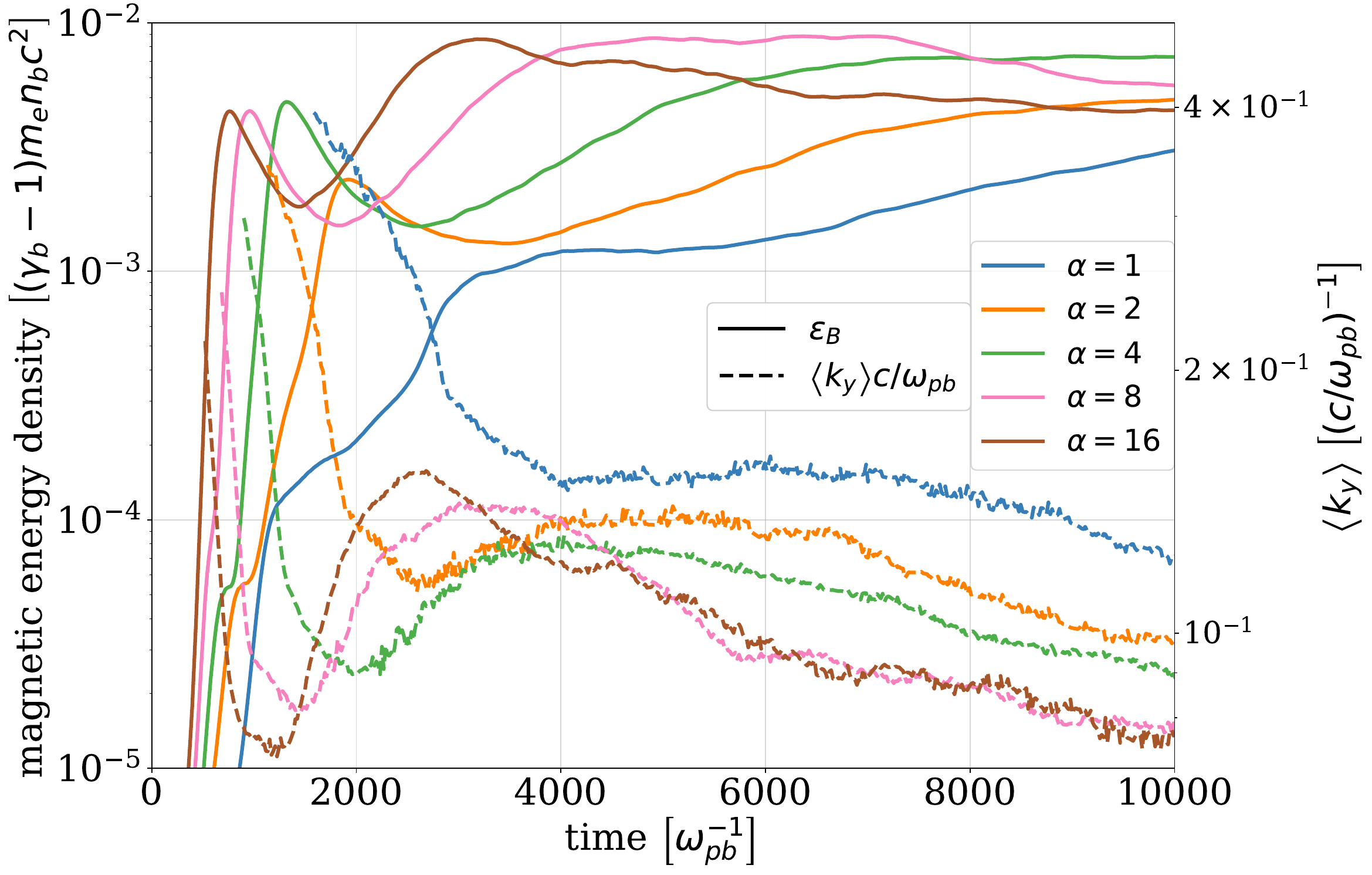}
    \caption{Time evolution of $\varepsilon_B$ (solid curves, measured on the left vertical axis) and $\left<k_y\right>c/\omega_{\rm pb}$ (dashed curves, measured on the right vertical axis) for five simulations with $t_{\rm f}\omega_{\rm pb} = 10^4$: $\alpha=1$ (blue), $\alpha=2$ (orange), $\alpha=4$ (green), $\alpha=8$ (pink), and $\alpha=16$ (brown).}
    \label{fig:tf10000_combined}
\end{figure}

\begin{figure}[h]
    \includegraphics[width=0.47\textwidth]{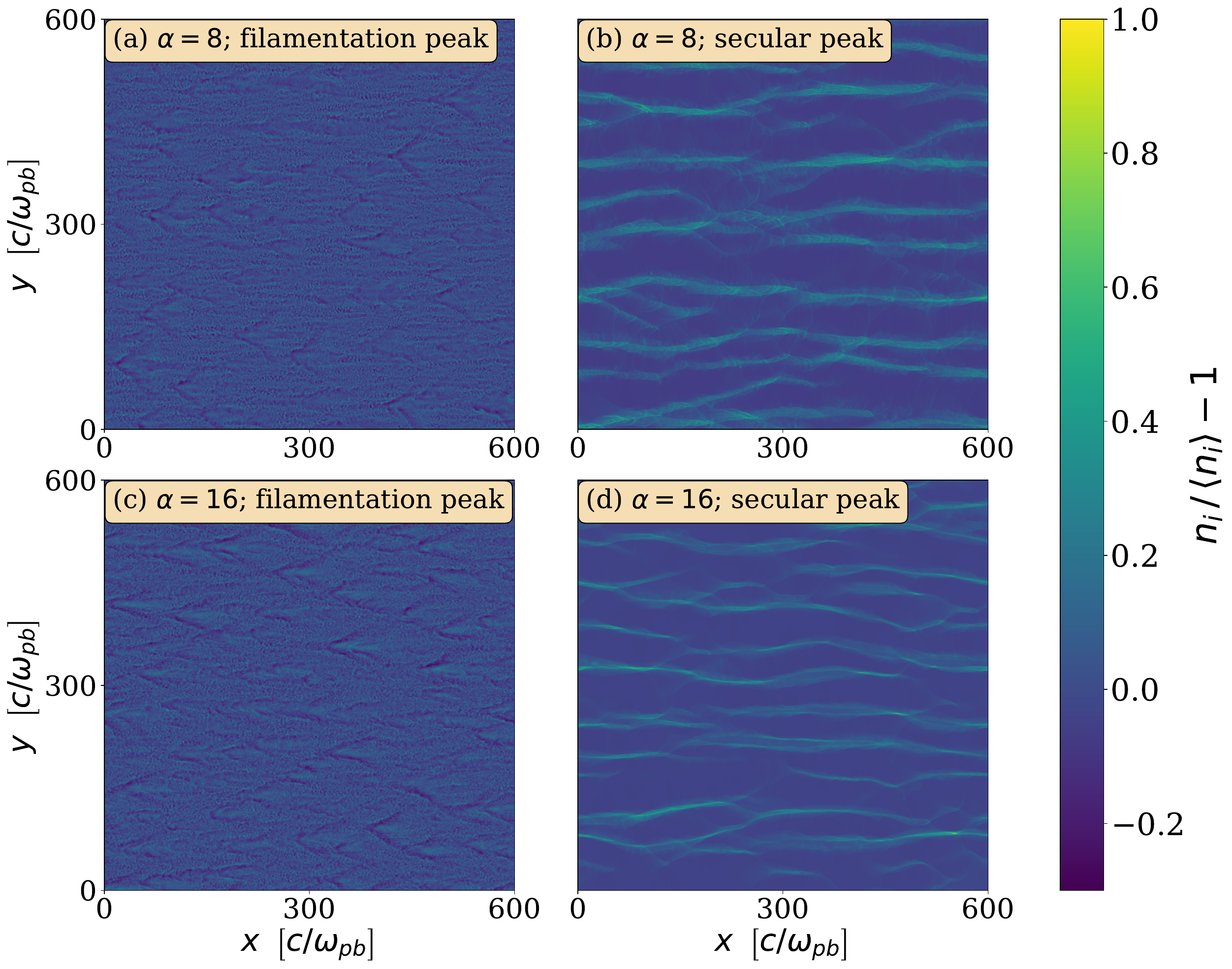}
    \caption{Comparison of the ion density contrast (i.e., $n_{\rm i}/\left<n_{\rm i}\right>-1$, normalized to the maximum value in each panel) at saturation of the first filamentation phase and at saturation of the secular phase for the $t_{\rm f}\omega_{\rm pb}=10^4$ simulations with $\alpha=8$ (panels (a) and (b)) and $\alpha=16$ (panels (c) and (d)). The $\alpha=8$ profiles are taken at $t\omega_{\rm pb} = 950$ (panel (a)) and $t\omega_{\rm pb} = 4100$ (panel (b)); the $\alpha=16$ profiles are taken at $t\omega_{\rm pb} = 750$ (panel (c)) and $t\omega_{\rm pb} = 3000$ (panel (d)).}
    \label{fig:protonfilaments}
\end{figure}

We now hold $t_{\rm f}\omega_{\rm pb}$ fixed and vary $\alpha$. Figure \ref{fig:tf10000_combined} shows the evolution of $\varepsilon_B$ and $\left<k_y\right>c/\omega_{\rm pb}$ for five simulations with $t_{\rm f}\omega_{\rm pb} = 10^4$ and with $\alpha$ varying between 1, 2, 4, 8, and 16. As $\alpha$ increases, the timescales of each of the phases decrease; for the $\alpha=8$ and $\alpha=16$ cases, the secular phase saturates early enough (at $t\omega_{\rm pb} \sim 4600$ for $\alpha=8$ and at $t\omega_{\rm pb} \sim 3200$ for $\alpha=16$) to allow for an extended period of magnetic field decay before the end of pair injection. As $\alpha$ increases, the decoupling of the background electrons from the protons becomes more dramatic; by the end of our $\alpha=16$ simulation, the background electrons have nearly formed a single population with the beam pairs in terms of velocity and temperature. The relative drift between this lepton beam and the protons triggers a slowly-growing filamentation instability that drives the secular phase; Figure \ref{fig:protonfilaments} shows the resulting proton filaments, which become more pronounced with higher $\alpha$.

\begin{figure}[h]
    \includegraphics[width=0.47\textwidth]{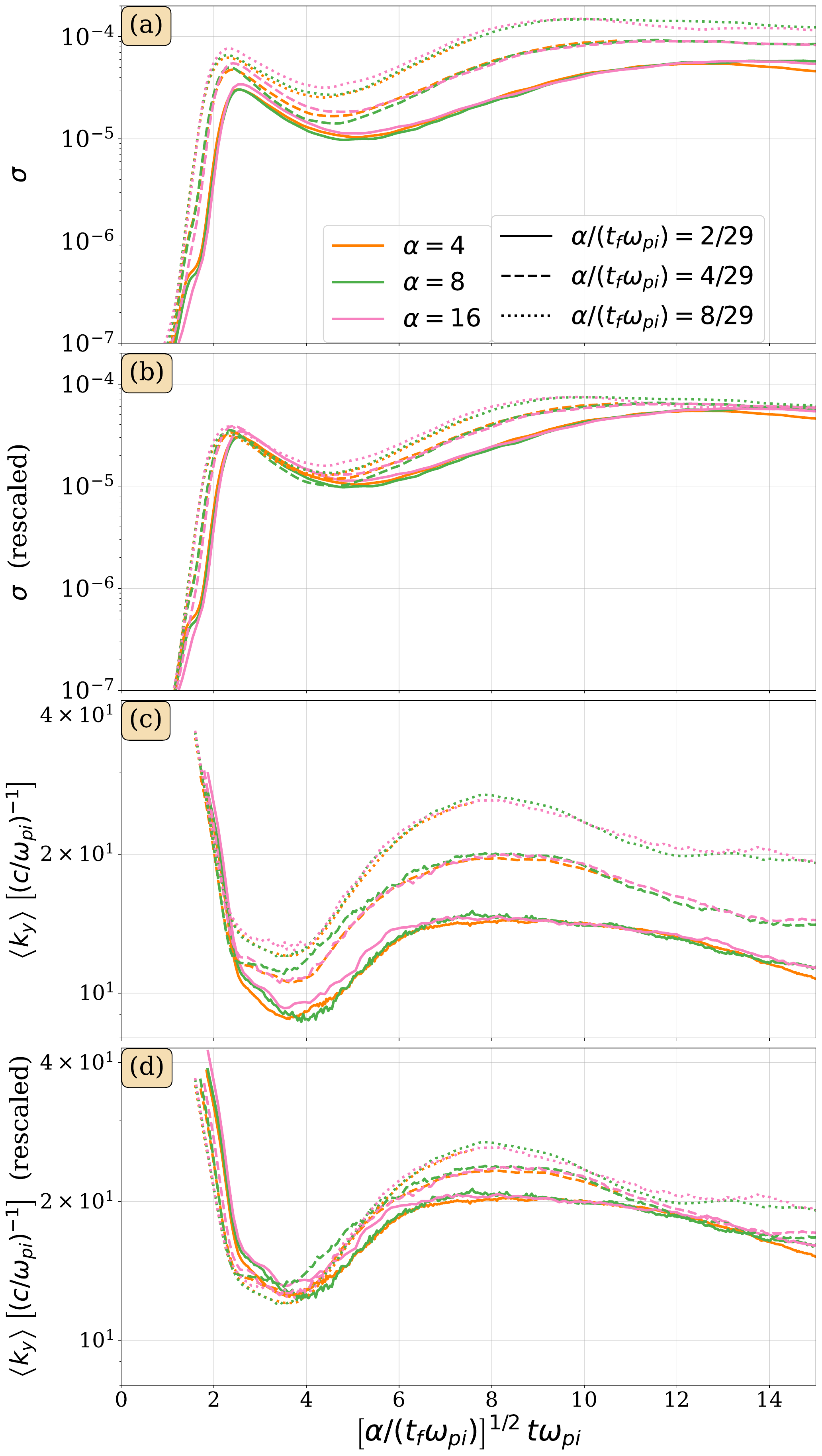}
    \caption{Time evolution of $\sigma$ (panels (a) and (b)) and $\left<k_y\right>c/\omega_{\rm pi}$ (panels (c) and (d)) for three simulations with $\alpha/(t_{\rm f}\omega_{\rm pi}) = 2/29$ (solid curves), three with $\alpha/(t_{\rm f}\omega_{\rm pi}) = 4/29$ (dashed curves), and three with $\alpha/(t_{\rm f}\omega_{\rm pi}) = 8/29$ (dotted curves); within each group of three, the simulation with $\alpha=4$ is shown in orange, $\alpha=8$ in green, and $\alpha=16$ in pink. The horizontal axis is rescaled to $\left[\alpha/(t_{\rm f}\omega_{\rm pi})\right]^{1/2} t\omega_{\rm pi}$ to capture the self-similar time evolution of $\sigma$ and $\left<k_y\right>c/\omega_{\rm pi}$. In panels (a) and (c), we show our simulation data as-is, with no vertical rescaling of $\sigma$ or $\left<k_y\right>c/\omega_{\rm pi}$; in panels (b) and (d), we rescale the normalization of $\sigma$ and $\left<k_y\right>c/\omega_{\rm pi}$ for each curve based on the pre-factors in Equations \eqref{eq:sigma_func} and \eqref{eq:ky_func}. This shows good agreement between our data and these analytic expressions.}
    \label{fig:rate_combined}
\end{figure}

Considering both $\alpha$ and $t_{\rm f}$ together, the quantity $\left[\alpha/(t_{\rm f} \omega_{\rm pi})\right]^{1/2}$ (for $\omega_{\rm pi}$ the plasma frequency of background protons) seems to play a key role in determining the evolution of the magnetic field.\footnote{In the following, we present scalings of various quantities with time measured as either $t \omega_{\rm pb}$ or as $t \omega_{\rm pi}$; the latter scales as $\sqrt{m_{\rm e}/m_{\rm i}} \, t \omega_{\rm pe}$.} In Figure \ref{fig:rate_combined}, we plot the time evolution of the upstream magnetization,
\begin{equation} \label{eq:sigma}
    \sigma \equiv \frac{\left<B_z^2\right>}{4\pi n_{\rm i} m_{\rm i} c^2} = 4(\gamma_{\rm b}-1) \alpha \frac{m_{\rm e}}{m_{\rm i}} \varepsilon_B,
\end{equation}
as well as the time evolution of $\left<k_y\right>c/\omega_{\rm pi}$ for a total of nine simulations with varying $\alpha$ and $t_{\rm f}\omega_{\rm pi}$: three simulations have $\alpha/(t_{\rm f} \omega_{\rm pi}) = 2/29$, three have $\alpha/(t_{\rm f} \omega_{\rm pi}) = 4/29$, and three have $\alpha/(t_{\rm f} \omega_{\rm pi}) = 8/29$. 
We rescale the units of time to $\left[\alpha/(t_{\rm f} \omega_{\rm pi})\right]^{1/2} \, t\omega_{\rm pi}$, resulting in an identical time evolution for all the curves: the onset time of the filamentation phase, the growth rate of the filamentation phase, and the growth rate of the secular phase are all constant across the nine cases plotted, suggesting that the evolution of the magnetic field under the action of continuous pair injection exhibits remarkable self-similarity. Cases with different $\alpha/(t_{\rm f} \omega_{\rm pi})$ differ by an overall normalization factor; our simulation data strongly imply that 
\begin{equation} \label{eq:sigma_func}
    \sigma = \left(\frac{\alpha}{t_{\rm f} \omega_{\rm pi}}\right)^{1/2} \, \mathcal{F}\left[t\omega_{\rm pi} \left(\frac{\alpha}{t_{\rm f} \omega_{\rm pi}}\right)^{1/2}\right]
\end{equation}
and
\begin{equation} \label{eq:ky_func}
    \left<k_y\right>c/\omega_{\rm pi} = \left(\frac{\alpha}{t_{\rm f} \omega_{\rm pi}}\right)^{1/4} \, \mathcal{G}\left[t\omega_{\rm pi} \left(\frac{\alpha}{t_{\rm f} \omega_{\rm pi}}\right)^{1/2}\right],
\end{equation}
where $\mathcal{F}$ and $\mathcal{G}$ are non-trivial functions of time that become exponential during the first filamentation phase. Normalized to the properties of the pair beam, Equations \eqref{eq:sigma_func} and \eqref{eq:ky_func} become
\begin{equation} \label{eq:epsb_func}
    \varepsilon_B = \frac{1}{2} \left(\frac{m_{\rm i}}{m_{\rm e}}\right)^{5/4} \frac{\alpha^{-1/4}}{\left(t_{\rm f} \omega_{\rm pb}\right)^{1/2}} \, \mathcal{F}\left[t\omega_{\rm pb} \frac{\alpha^{1/4}}{\left(t_{\rm f} \omega_{\rm pb}\right)^{1/2}} \left(\frac{m_{\rm i}}{m_{\rm e}}\right)^{-1/4} \right]
\end{equation}
and
\begin{align} \label{eq:ky_func2}
    \left<k_y\right>c/\omega_{\rm pb} &= \left(\frac{m_{\rm i}}{m_{\rm e}}\right)^{-3/8} \frac{\alpha^{-1/8}}{\left(t_{\rm f} \omega_{\rm pb}\right)^{1/4}} \nonumber \\ &\times \mathcal{G}\left[t\omega_{\rm pb} \frac{\alpha^{1/4}}{\left(t_{\rm f} \omega_{\rm pb}\right)^{1/2}} \left(\frac{m_{\rm i}}{m_{\rm e}}\right)^{-1/4} \right].
\end{align}
We note that, while our expressions for the magnetic wavenumber [Equations \eqref{eq:ky_func} and \eqref{eq:ky_func2}] nicely fit our simulations for all $\alpha \ge 2$, our expressions for the magnetic energy density [Equations \eqref{eq:sigma_func} and \eqref{eq:epsb_func}] fit better at larger $\alpha$ (i.e., $\alpha \ge 8$) than at smaller $\alpha$; for $2 \le \alpha < 8$, the normalization on $\sigma$ seems to scale with $\alpha$ rather than with $\alpha^{1/2}$, and, equivalently, the normalization on $\epsilon_B$ seems to scale with $\alpha^{1/2}$ rather than with $\alpha^{-1/4}$. This is likely because our cases with $\alpha \le 4$ are not yet in the asymptotic regime of large $\alpha$; for $\alpha > 16$, we expect Equations \eqref{eq:sigma_func}-\eqref{eq:ky_func2} to apply as written.

Equation \eqref{eq:sigma_func} allows us to make some general statements about the early growth of the filamentation mode under continuous pair injection.\footnote{In the following, we primarily focus on the exponential growth of the magnetic energy density; similar arguments hold for the exponential decrease of the mean wavenumber.} By reading off the argument of $\mathcal{F}$, we can write the filamentation growth rate as 
\begin{equation} \label{eq:growthrate}
    \frac{\Gamma}{\omega_{\rm pi}} \propto \left(\frac{\alpha}{t_{\rm f} \omega_{\rm pi}}\right)^{1/2}
\end{equation}
or, in units of $\omega_{\rm pb}$, as 
\begin{equation}
    \frac{\Gamma}{\omega_{\rm pb}} \propto \frac{\alpha^{1/4}}{\left(t_{\rm f} \omega_{\rm pb}\right)^{1/2}} \left(\frac{m_{\rm i}}{m_{\rm e}}\right)^{-1/4}.
\end{equation}
Additionally, if we define the onset time of the filamentation instability, $t_{\rm on}$, such that
\begin{equation}
    \sigma \propto \mathcal{F}\left(\frac{t}{t_{\rm on}}\right),
\end{equation}
then we can write the filamentation growth rate (Equation \eqref{eq:growthrate}) as  
\begin{equation}
    \frac{\Gamma}{\omega_{\rm pi}} \propto \frac{\alpha t_{\rm on}}{t_{\rm f}};
\end{equation}
this implies that the instantaneous beam-to-background density ratio at the onset time of the instability sets the growth rate. The linear dependence of the growth rate on the instantaneous density contrast, as well as the fact that the dominant wavelength grows exponentially at the same rate as the magnetic field, are evocative of the nonlinear streaming instability described by \citet{peterson_21,peterson_22}. This linear dependence is faster than the $\alpha^{1/2}$ scaling expected in the cold-beam limit of the filamentation instability, yet slower than the $\alpha^{3/2}$ scaling expected in the ultra-relativistic, hot-beam limit \citep{Bret_2010a}.

\begin{figure}[h]
    \centering
    \includegraphics[width=0.47\textwidth]{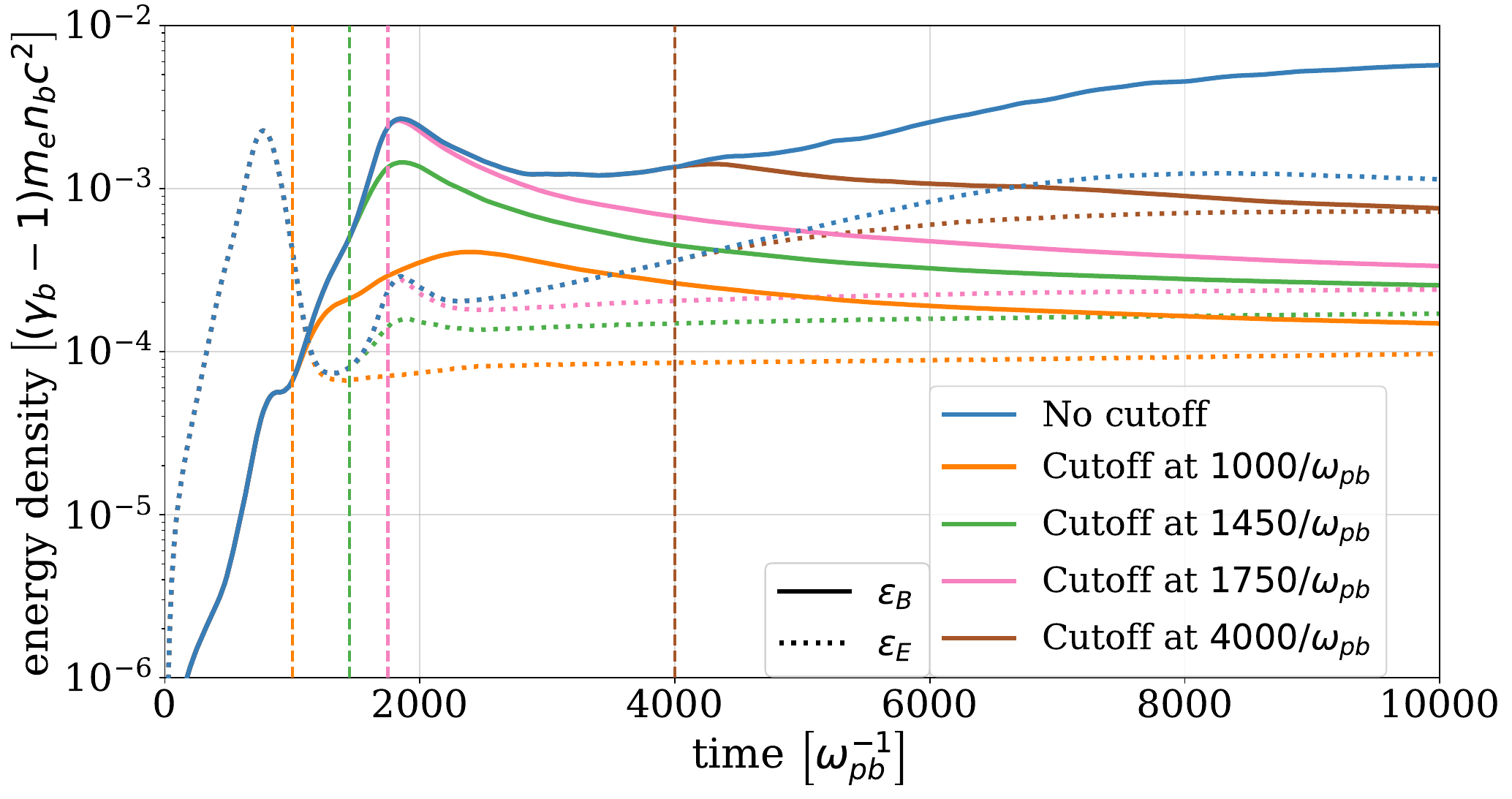}
    \caption{Comparison of simulations with $\alpha=2$ and $t_{\rm f}\omega_{\rm pb} = 10^4$, but with continuous pair injection prematurely cut off at different times. The blue curves represent a simulation with fully continuous injection (i.e., an injection cutoff at $t_{\rm cut}\omega_{\rm pb} = 10^4$), the orange curves represent a simulation with $t_{\rm cut}\omega_{\rm pb} = 1000$, the green curves a simulation with $t_{\rm cut}\omega_{\rm pb} = 1450$, the pink curves a simulation with $t_{\rm cut}\omega_{\rm pb} = 1750$, and the brown curves a simulation with $t_{\rm cut}\omega_{\rm pb} = 4000$. The solid curves show the evolution of $\varepsilon_B$, the dotted curves show the evolution of $\varepsilon_E$, and the dashed vertical lines show the times at which injection was cut off.}
    \label{fig:cutoffs}
\end{figure}

While the growth rate of the filamentation phase seems to depend only on the instantaneous beam-to-background density ratio at the onset time, continuous pair injection is still crucial for sustaining this growth over an extended period of time. In Figure \ref{fig:cutoffs}, we illustrate the importance of continuous pair injection by comparing multiple simulations (with $\alpha=2$ and $t_{\rm f}\omega_{\rm pb} = 10^4$) with pair injection cut off at different times before $t_{\rm f}$; we cut off injection at $t_{\rm cut}\omega_{\rm pb} = 1000$ (at the beginning of the filamentation phase), $t_{\rm cut}\omega_{\rm pb} = 1450$ (in the middle of the filamentation phase), $t_{\rm cut}\omega_{\rm pb} = 1750$ (immediately before the saturation of the filamentation phase), and $t_{\rm cut}\omega_{\rm pb} = 4000$ (near the beginning of the secular phase). In all cases, cutting off pair injection quenches subsequent growth of the magnetic and electric energies. Once injection is stopped, the energy briefly continues to increase (at a notably reduced growth rate) before saturating and decaying; evidently, any theoretical model of these systems must account for a growth rate with dynamical loading. This dependence on continuous injection applies to both the filamentation phase and the secular phase: the $t_{\rm cut}\omega_{\rm pb} = 1000$ and $t_{\rm cut}\omega_{\rm pb} = 1450$ cases prematurely halt the filamentation growth, while all the cut-off cases eliminate the secular growth, indicating that the secular phase is a unique element of our continuous-injection simulations.

\begin{figure}[h]
    \centering
    \includegraphics[width=0.47\textwidth]{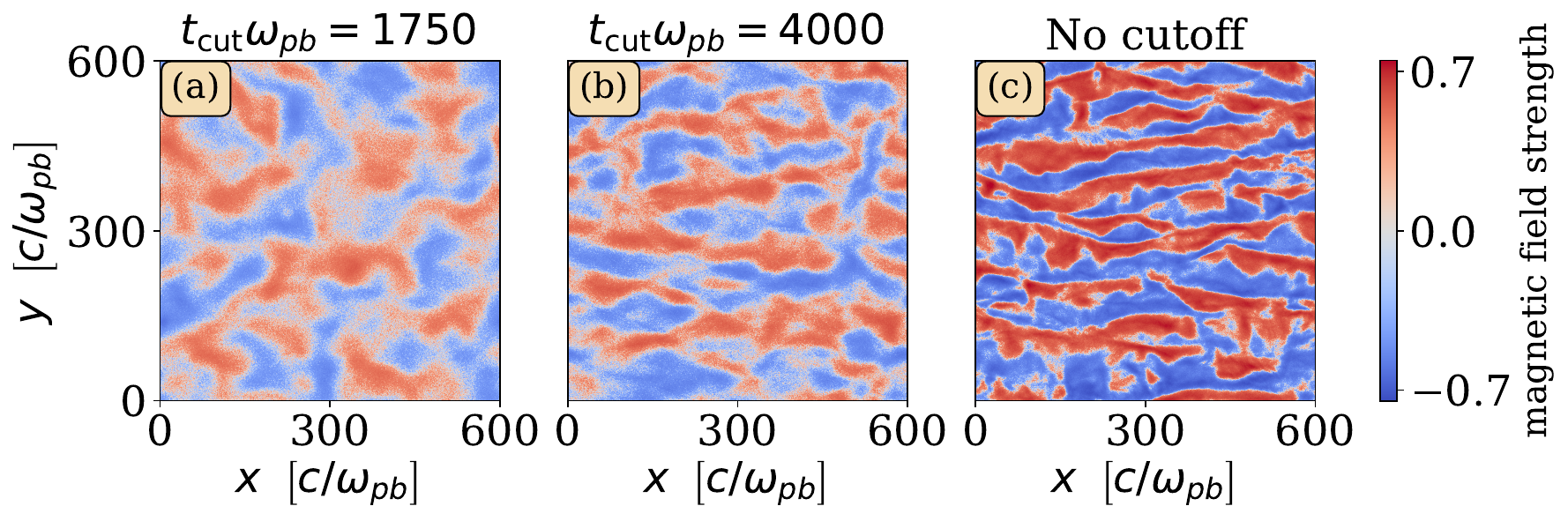}
    \caption{Spatial profiles of $B_z/\sqrt{8\pi\left(\gamma_{\rm b}-1\right)m_{\rm e} n_{\rm b} c^2}$ at $t\omega_{\rm pb} = 10^4$ for three simulations with $\alpha=2$ and $t_{\rm f} \omega_{\rm pb} = 10^4$ but with different pair injection cutoff times: (a) $t_{\rm cut} \omega_{\rm pb} = 1750$, (b) $t_{\rm cut} \omega_{\rm pb} = 4000$, and (c) $t_{\rm cut}\omega_{\rm pb} = 10^4$ (i.e., no cutoff). The color bar is rescaled to $\left|B_z\right|^{0.2} \times {\rm sign}(B_z)$ to display a wider dynamic range.}
    \label{fig:cutoffprofiles}
\end{figure}

Continuous injection is also vital for developing and maintaining the large-scale magnetic structures discussed above. Figure \ref{fig:cutoffprofiles} shows spatial profiles of the magnetic field at $t\omega_{\rm pb} = 10^4$ for the simulations with $t_{\rm cut}\omega_{\rm pb} = 1750$, with $t_{\rm cut}\omega_{\rm pb} = 4000$, and with no premature injection cutoff. These plots demonstrate that, in the absence of continuous injection, any large-scale magnetic structure fades away. 

\vfill\null

\subsection{Asymptotic Behavior} \label{sec:empirical}

While Equations \eqref{eq:sigma_func}-\eqref{eq:ky_func2} nicely capture the self-similarity of our system, we cannot use these equations to extrapolate our simulation data to arbitrary regimes due to the unknown functional forms of $\mathcal{F}$ and $\mathcal{G}$. That said, if these functions are weakly dependent on their arguments at late times (i.e., near $t=t_{\rm f}$), we would expect the magnetic energy density and the transverse magnetic wavenumber at the end of pair injection to scale with the pre-factors in Equations \eqref{eq:sigma_func}-\eqref{eq:ky_func2}: $\sigma_{\rm f}$ (i.e., $\sigma$ at the end of pair injection) would scale with $\alpha^{1/2} \, \left(t_{\rm f}\omega_{\rm pi}\right)^{-1/2}$ and $\left<k_y\right>_{\rm f} c/\omega_{\rm pi}$ would scale with $\alpha^{1/4} \, \left(t_{\rm f}\omega_{\rm pi}\right)^{-1/4}$. In this section, we 
measure
the asymptotic scalings for the \emph{final}   
magnetic energy and \emph{final} transverse wavenumber,
showing that the power-law dependencies on $t_{\rm f}$ and $\alpha$ are not 
vastly different 
from those 
given by the self-similar scalings~\eqref{eq:sigma_func}-\eqref{eq:ky_func2}. 
In Section \ref{sec:extrapolation}, we use these empirical scalings to extrapolate our simulation results to realistic $t_{\rm f}$ and $\alpha$.

\begin{figure}[h]
    \includegraphics[width=0.47\textwidth]{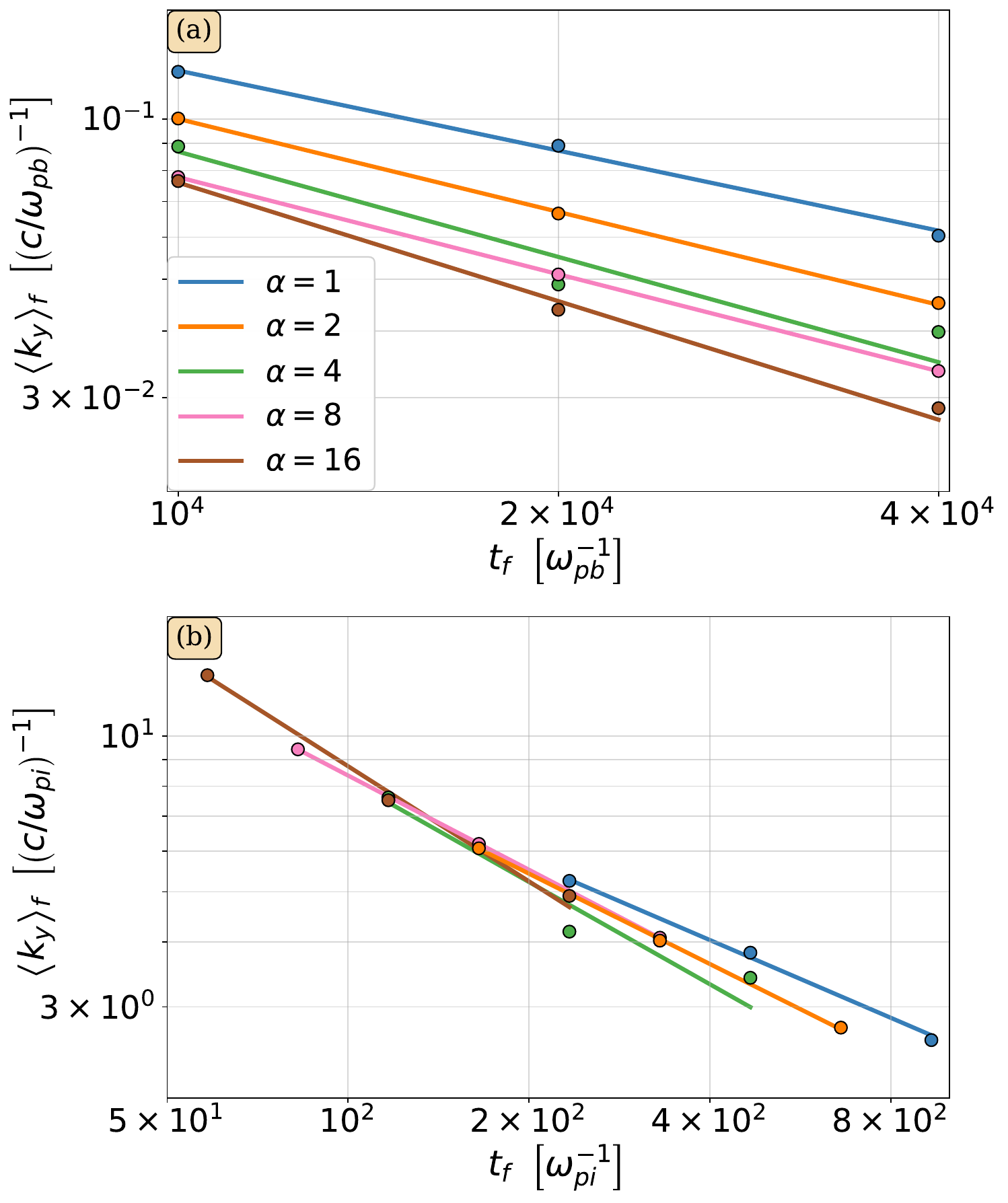}
    \caption{$\left<k_y \right>c/
    \omega_{\rm pb}$ vs. $t_{\rm f}\omega_{\rm pb}$ (panel (a)) and $\left<k_y\right>c/\omega_{\rm pi}$ vs. $t_{\rm f}\omega_{\rm pi}$ (panel (b)) at the end of pair injection for $\alpha = \{1, 2, 4, 8, 16\}$. Each dot represents one simulation, with blue dots for $\alpha=1$, orange for $\alpha=2$, green for $\alpha=4$, pink for $\alpha=8$, and brown for $\alpha=16$; each $\alpha$ has a dot for $t_{\rm f}\omega_{\rm pb} = 10^4, \, 2 \times 10^4,$ and $4 \times 10^4$ (respectively corresponding to $t_{\rm f}\omega_{\rm pi} \sim 233/\sqrt{\alpha}$, $467/\sqrt{\alpha}$, and $934/\sqrt{\alpha}$). The blue best-fit line (for $\alpha=1$) has a slope of $-0.50$, the orange best-fit line (for $\alpha=2$) has a slope of $-0.58$, the green best-fit line (for $\alpha=4$) has a slope of $-0.66$, the pink best-fit line (for $\alpha=8$) has a slope of $-0.60$, and the brown best-fit line (for $\alpha=16$) has a slope of $-0.74$.} 
    \label{fig:avgk_summary}
\end{figure}

Figure \ref{fig:avgk_summary} shows how $\left<k_y\right>_{\rm f} c/\omega_{\rm pb}$ scales with $t_{\rm f} \omega_{\rm pb}$ and how $\left<k_y\right>_{\rm f} c/\omega_{\rm pi}$ scales with $t_{\rm f} \omega_{\rm pi}$ for $\alpha = \{1, 2, 4, 8, 16\}$; both panels show the same simulations, but with the units appropriately converted. 
We do not include the $t_{\rm f}\omega_{\rm pb} = 5 \times 10^3$ (or $t_{\rm f}\omega_{\rm pi} \sim 117/\sqrt{\alpha}$) simulations in these plots, since we find that they generally do not follow the trends of the higher (and more realistic) $t_{\rm f}\omega_{\rm pb}$ cases (i.e., the $t_{\rm f}\omega_{\rm pb} = 5 \times 10^3$ cases are not yet in the asymptotic regime); for reference, we include the $t_{\rm f}\omega_{\rm pb} = 5 \times 10^3$ cases in Figure \ref{fig:avgk convergence} of Appendix \ref{sec:convergence}. With $t_{\rm f}\omega_{\rm pb} = 5 \times 10^3$ discarded, the $\left<k_y\right>_{\rm f} c/\omega_{\rm pb}$ dependence on $t_{\rm f}\omega_{\rm pb}$ at fixed $\alpha$ is well-fit by a power law $\propto \left(t_{\rm f} \omega_{\rm pb}\right)^{n}$, where $n$ ranges from $-0.50$ (for $\alpha=1$) to $-0.74$ (for $\alpha=16$); for $\alpha=2$, we obtain
%\vspace{-6mm}
\begin{equation} \label{eq:avgk_alpha2_beam}
\left<k_y\right>_{\rm f}(\alpha=2) \sim 0.10 \left(\frac{\omega_{\rm pb}}{c} \right) \left(\frac{t_{\rm f}\omega_{\rm pb}}{10^4} \right)^{-0.6}
\end{equation}
or, using units appropriate for the background protons,
\begin{equation} \label{eq:avgk_alpha2_ion}
    \left<k_y\right>_{\rm f}(\alpha=2) \sim 6.1 \left(\frac{\omega_{\rm pi}}{c} \right) \left(\frac{t_{\rm f}\omega_{\rm pi}}{165} \right)^{-0.6} \left(\frac{m_{\rm i}/m_{\rm e}}{1836} \right)^{0.2}.
\end{equation}

While the power-law slopes shown in Figure \ref{fig:avgk_summary} are steeper for higher $\alpha$, this trend is weak: within reasonable error, each of these power laws is consistent with a slope of $-2/3$. Because these slopes are largely insensitive to $\alpha$, we can also reliably infer the overall dependence of $\left<k_y\right>_{\rm f}$ on $\alpha$: for $t_{\rm f}\omega_{\rm pb} \ge 2 \times 10^4$, $\left<k_y\right>_{\rm f} c/\omega_{\rm pb}$ scales with $\alpha^{-1/4}$.  
We can thus modify Equations \eqref{eq:avgk_alpha2_beam} and \eqref{eq:avgk_alpha2_ion} to give
\begin{equation} \label{eq:avgk_beam}
    \left<k_y\right>_{\rm f} \sim 0.12 \left(\frac{\omega_{\rm pb}}{c} \right) \alpha^{-1/4} \left(\frac{t_{\rm f}\omega_{\rm pb}}{10^4} \right)^{-2/3}
\end{equation}
and
\begin{equation} \label{eq:avgk_ion}
    \left<k_y\right>_{\rm f} \sim 9.0 \left(\frac{\omega_{\rm pi}}{c} \right) \alpha^{-1/12} \left(\frac{t_{\rm f}\omega_{\rm pi}}{100} \right)^{-2/3} \left(\frac{m_{\rm i}/m_{\rm e}}{1836} \right)^{1/6}.
\end{equation}

\begin{figure}[h]
    \includegraphics[width=0.47\textwidth]{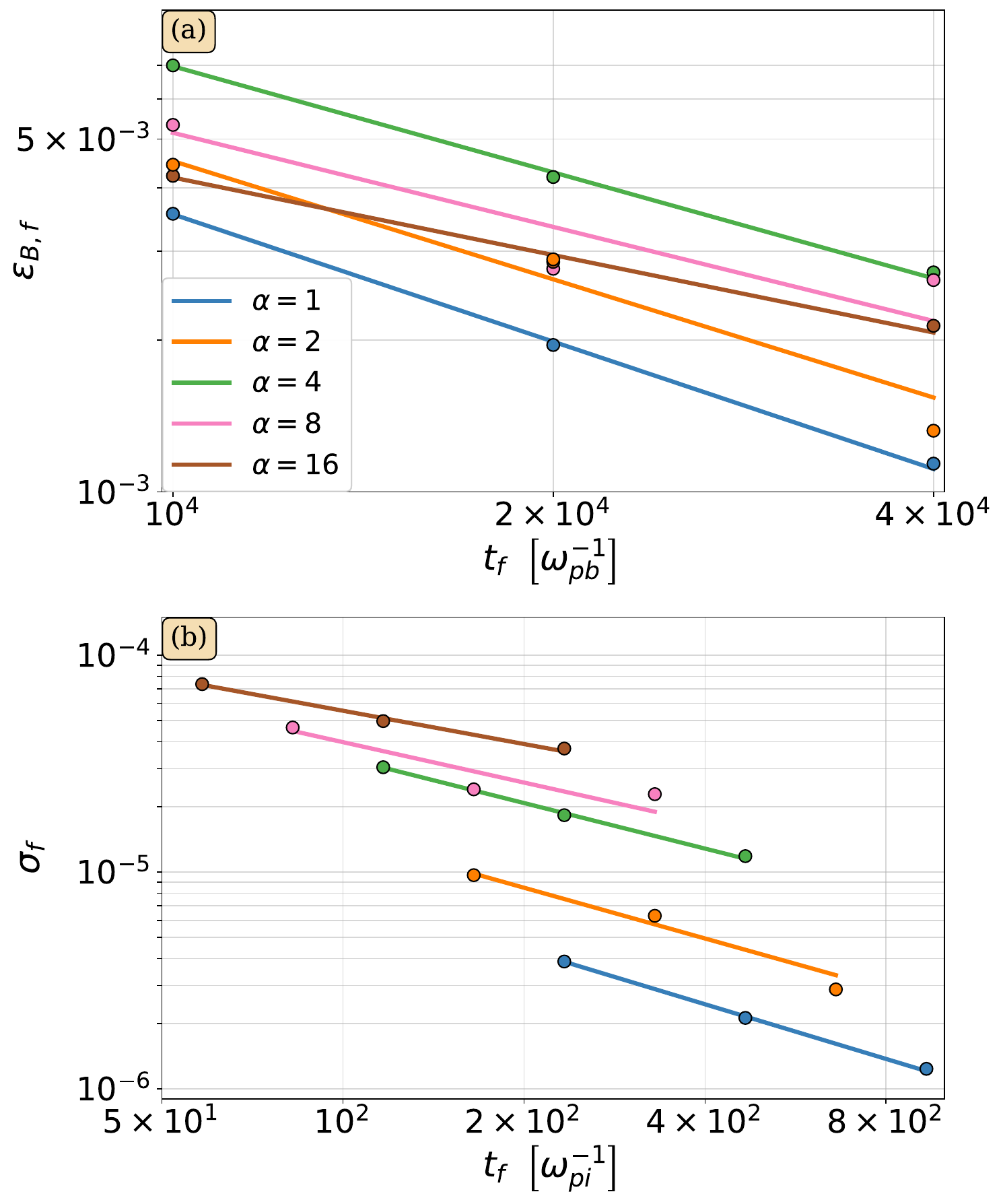}
    \caption{$\varepsilon_B$ vs. $t_{\rm f} \omega_{\rm pb}$ (panel (a)) and $\sigma$ vs. $t_{\rm f} \omega_{\rm pi}$ (panel (b)) at the end of pair injection for $\alpha = \{1, 2, 4, 8, 16\}$. Each dot represents one simulation, with blue dots for $\alpha=1$, orange for $\alpha=2$, green for $\alpha=4$, pink for $\alpha=8$, and brown for $\alpha=16$; each $\alpha$ has a dot for $t_{\rm f}\omega_{\rm pb} = 10^4, \, 2 \times 10^4,$ and $4 \times 10^4$ (respectively corresponding to $t_{\rm f}\omega_{\rm pi} \sim 233/\sqrt{\alpha}$, $467/\sqrt{\alpha}$, and $934/\sqrt{\alpha}$). The blue best-fit line (for $\alpha=1$) has a slope of $-0.84$, the orange best-fit line (for $\alpha=2$) has a slope of $-0.78$, the green best-fit line (for $\alpha=4$) has a slope of $-0.70$, the pink best-fit line (for $\alpha=8$) has a slope of $-0.62$, and the brown best-fit line (for $\alpha=16$) has a slope of $-0.51$.}
    \label{fig:epsb_summary}
\end{figure}

Similar to Figure \ref{fig:avgk_summary}, Figure \ref{fig:epsb_summary} plots the box-averaged magnetic energy density at the end of pair injection, with $\varepsilon_{B, \, \rm f}$ plotted against $t_{\rm f} \omega_{\rm pb}$ and $\sigma_{\rm f}$ plotted against $t_{\rm f} \omega_{\rm pi}$.\footnote{For precision, $\varepsilon_{B, \, \rm f}$ and $\sigma_{\rm f}$ are computed by integrating over the magnetic power spectrum, using the same $k_y$ integral bounds as those used in our computation of $\left<k_y\right>_{\rm f} c/\omega_{\rm pb}$ (Equation \eqref{eq:avgk}).} As with $\left<k_y\right>_{\rm f}$, we discard the $t_{\rm f} \omega_{\rm pb} = 5 \times 10^3$ simulations as they do not fit the trends of the higher $t_{\rm f} \omega_{\rm pb}$ cases (see Figure \ref{fig:epsb convergence} of Appendix \ref{sec:convergence} for the trends with $t_{\rm f} \omega_{\rm pb} = 5 \times 10^3$ included). The resulting dependence of $\varepsilon_{B, \, f}$ on $t_{\rm f} \omega_{\rm pb}$ at fixed $\alpha$ is well-fit by a power law $\propto \left(t_{\rm f} \omega_{\rm pb}\right)^{n}$, where $n$ ranges from $-0.84$ (for $\alpha=1$) to $-0.51$ (for $\alpha=16$); for $\alpha=2$, we obtain
\begin{equation} \label{eq:eps_alpha2}
    \varepsilon_{B, \, \rm f}(\alpha=2) \sim 4.4 \times 10^{-3} \left(\frac{t_{\rm f}\omega_{\rm pb}}{10^4} \right)^{-0.8}
\end{equation}
or, in units appropriate for the background protons,
\begin{equation} \label{eq:sigma_alpha2}
    \sigma_{\rm f}(\alpha=2) \sim 9.6 \times 10^{-6} \left(\frac{t_{\rm f}\omega_{\rm pi}}{165} \right)^{-0.8} \left(\frac{m_{\rm i}/m_{\rm e}}{1836} \right)^{-1.4}.
\end{equation}
Each of the power laws in Figure \ref{fig:epsb_summary} relating the final magnetic energy density to $t_{\rm f}$ is consistent with a slope of $-3/4$. Similar to what we found for $\left<k_y\right>_{\rm f}$, these power-law slopes are very weakly dependent on $\alpha$, marginally flattening out with increasing $\alpha$; extrapolating this trend suggests that $\varepsilon_{B, \, \rm f}$ and $\sigma_{\rm f}$ may become nearly independent of $t_{\rm f}\omega_{\rm pb}$ and $t_{\rm f}\omega_{\rm pi}$ (respectively) at high $\alpha$. In contrast to the dependence of $\left<k_y\right>_{\rm f}$ on $\alpha$, the dependence of $\varepsilon_{B, \, \rm f}$ on $\alpha$ is non-monotonic; however, the dependence of $\sigma_{\rm f}$ on $\alpha$ {is} monotonic. We find that 
\begin{equation} \label{eq:sigma_scaling}
    \sigma_{\rm f} \sim 9.2 \times 10^{-6} \, \alpha^{5/8} \left(\frac{t_{\rm f}\omega_{\rm pi}}{100} \right)^{-3/4} \left(\frac{m_{\rm i}/m_{\rm e}}{1836} \right)^{-11/8}.
\end{equation}

If $\mathcal{F}$ is weakly dependent on its argument at $t \sim t_{\rm f}$, then Equation \eqref{eq:sigma_func} implies that $\sigma_{\rm f}$ should scale with $\alpha^{1/2} \, \left(t_{\rm f}\omega_{\rm pi} \right)^{-1/2}$; if the argument of $\mathcal{F}$ is constant at $t=t_{\rm f}$ (i.e., if $\alpha t_{\rm f}\omega_{\rm pi}$ is constant), then $\sigma_{\rm f}$ should be proportional to $\alpha$ and to $(t_{\rm f}\omega_{\rm pi})^{-1}$. These scalings are not far off from the empirically derived Equation \eqref{eq:sigma_scaling}, which states that $\sigma_{\rm f} \propto \alpha^{5/8} \, \left(t_{\rm f}\omega_{\rm pi}\right)^{-3/4}$. Similarly, if $\mathcal{G}$ is weakly dependent on its argument at $t \sim t_{\rm f}$, then Equation \eqref{eq:ky_func} implies that $\left<k_y\right>_{\rm f} c/\omega_{\rm pi}$ should scale with $\alpha^{1/4} \, \left(t_{\rm f}\omega_{\rm pi} \right)^{-1/4}$; if the argument of $\mathcal{G}$ is constant at $t=t_{\rm f}$, then $\left<k_y\right>_{\rm f} c/\omega_{\rm pi}$ should be proportional to $\alpha^{1/2}$ and to $(t_{\rm f}\omega_{\rm pi})^{-1/2}$. These scalings differ from the empirically-derived Equation \eqref{eq:avgk_ion} -- which says that $\left<k_y\right>_{\rm f} c/\omega_{\rm pi} \propto \alpha^{-1/12} \, \left(t_{\rm f}\omega_{\rm pi} \right)^{-2/3}$ -- especially with regard to the $\alpha$-dependence. These discrepancies can be attributed to the fact that $\mathcal{F}$ and $\mathcal{G}$ are in fact not constant at late times, and therefore the scalings inferred from the pre-factors in Equations \eqref{eq:sigma_func} and \eqref{eq:ky_func} are just rough approximations.

In Figures \ref{fig:avgk convergence} and \ref{fig:epsb convergence} of Appendix \ref{sec:convergence}, we show that the trends in Figures \ref{fig:avgk_summary} and \ref{fig:epsb_summary} are robust to the number of particles per cell. We can thus use these scaling relations for the final transverse magnetic wavenumber and for the final magnetic energy density to reliably extrapolate our simulation results to values of $t_{\rm f}$ and $\alpha$ that are realistic for long GRBs, as we discuss in Section \ref{sec:extrapolation}.

\vfill\null
\section{Implications for GRB Afterglows} \label{sec:extrapolation}

We can now extrapolate the final transverse magnetic wavenumber and the final magnetic energy density to parameter regimes realistic for long GRBs. To compute a realistic $t_{\rm f}$, we use the time lag between the passage of the first prompt photons at radius $R$ and the passage of the GRB's external shock at radius $R$ (i.e., the time over which a given upstream fluid element will be enriched with electron-positron pairs before encountering the shock):
\begin{equation} \label{eq:tgev}
    t_{f, \, {\rm GRB}}(R) \approx \frac{R}{2 \Gamma^2 c},
\end{equation}
where $\Gamma$ is the Lorentz factor of the blast wave and $c$ is the speed of light \citep{beloborodovradiation}. 
$\alpha$ is a monotonically decreasing function of $R$ and is of order unity at the characteristic radius $R_{\rm load}$, where
\begin{equation}
    R_{\rm load} \approx 10^{17} \, {\rm cm} \, \left(\frac{E_{\rm GRB}}{10^{54} \, {\rm erg}}\right)^{1/2},
\end{equation}
for $E_{\rm GRB}$ the isotropic equivalent of the prompt GRB energy ahead of the external shock (see Equation (20) in \citet{beloborodovgev}); in the calculations that follow, we use $R \sim 10^{17} \, {\rm cm}$, the radius at which $\alpha \sim 1$ for a bright burst with $E_{\rm GRB} \sim 10^{54} \, {\rm erg}$. With this $R$ and a typical blast wave Lorentz factor of $\Gamma \sim 200$, Equation \eqref{eq:tgev} gives $t_{f, {\rm GRB}}(\alpha \sim 1) \sim 40 \, {\rm s}$.
The mass density of the circumburst medium (a Wolf-Rayet wind) also depends on $R$:
\begin{equation} \label{eq:rhowr}
    \rho_{\rm WR}(R) \approx \frac{3 \times 10^{11} \, {\rm g \, cm^{-1}}}{R^2},
\end{equation}
from Equation (34) in \citet{beloborodovgev}. For $\alpha \sim 1$ (or, equivalently, $R \sim 10^{17} \, {\rm cm}$), we then have a mass density of $\rho_{\rm WR}(\alpha \sim 1) \sim 3 \times 10^{-23} \, {\rm g \, cm^{-3}}$ and thus a proton number density of $n_{\rm i}(\alpha \sim 1) \sim 20 \, {\rm cm^{-3}}$ and a proton plasma frequency of $\omega_{\rm pi}(\alpha \sim 1) \sim 6 \times 10^{3} \, {\rm s^{-1}}$. Therefore, in units of $\omega_{\rm pi}^{-1}$, a realistic $t_{\rm f}$ for $\alpha \sim 1$ is
\begin{equation}
    \left[t_{f, \, {\rm GRB}} \, \omega_{\rm pi}\right](\alpha \sim 1) \sim 2 \times 10^5;
\end{equation}
this is two orders of magnitude larger than our largest simulated $t_{\rm f}\omega_{\rm pi}$ at $\alpha=2$, so the gap over which we must extrapolate our results is not unreasonably large.

By using $\alpha \sim 1$, $\omega_{\rm pi}(\alpha \sim 1) \sim 6 \times 10^3 \, {\rm s^{-1}}$ and $\left[t_{f, \, {\rm GRB}} \, \omega_{\rm pi}\right](\alpha \sim 1) \sim 2 \times 10^5$ in Equations \eqref{eq:avgk_ion} and \eqref{eq:sigma_scaling}, we can obtain realistic values for the scale and strength of the magnetic fields generated via the upstream pre-conditioning process studied in this paper. We can then compare these values to the expected scale and strength of the magnetic fields generated {by the shock itself}; this comparison will allow us to gauge the extent to which upstream pair loading modifies the physics of GRB afterglow shocks at early times.  

Equation \eqref{eq:avgk_ion} provides a realistic transverse magnetic wavenumber for $\alpha \sim 1$:
\begin{equation}
    \left<k_y\right>_{\rm f}(\alpha \sim 1) \sim 10^{-8} \, {\rm cm^{-1}},
\end{equation}
which implies a transverse spatial scale of  
\begin{equation} \label{eq:corrlength}
    \left<\lambda_y\right>_{\rm f}(\alpha \sim 1) \equiv \frac{2\pi}{\left<k_y\right>_{\rm f}(\alpha \sim 1)} \sim 6 \times 10^8 \, {\rm cm};
\end{equation} 
this exceeds the proton skin depth, $c/\omega_{\rm pi}(\alpha \sim 1) \sim 5 \times 10^6 \, {\rm cm}$, by two orders of magnitude. Given that the self-generated magnetic structures in our main suite of simulations never reach scales significantly larger than $c/\omega_{\rm pi}$, it may appear that our extrapolation of $\left<\lambda_y\right>_{\rm f}$ to super-ion scales is not fully justified. In Appendix \ref{sec:lowmass}, we demonstrate -- using simulations with reduced mass ratios -- that continuous pair injection can definitively produce magnetic structures larger than $c/\omega_{\rm pi}$, thus justifying our extrapolation.

\citet{groseljpair} found that a relativistic shock propagating in a pair-loaded medium would self-generate magnetic fields with transverse correlation length given by 
\begin{equation} \label{eq:lambdashock}
    \lambda_{\rm shock}(\alpha) \sim \frac{c}{\omega_{\rm pi}} \left(1+2\alpha\right)^{-2/3},
\end{equation}
which is $\sim 3 \times 10^6 \, {\rm cm}$ for $\alpha \sim 1$. For $\alpha \sim 1$, the magnetic field generated by pair enrichment ahead of the shock possesses a transverse correlation length that is more than two orders of magnitude larger than that of the magnetic field produced by the shock itself. We remark that \citet{groseljpair} did not explicitly model the pair injection process, but instead initialized their (weakly magnetized) shock simulations with a given number of electrons, positrons, and ions at rest in the upstream frame. For future work, one should repeat the simulations of \citet{groseljpair} but with {continuously injected} pairs carrying {net momentum} in the upstream. 

According to Equations \eqref{eq:avgk_ion} and \eqref{eq:lambdashock}, $\left<\lambda_y\right>_{\rm f} \, \left(c/\omega_{\rm pi}\right)^{-1}$ should {increase} weakly with $\alpha$ while $\lambda_{\rm shock} \, \left(c/\omega_{\rm pi}\right)^{-1}$ should {decrease} with $\alpha$ (as $\sim \alpha^{-2/3}$); as such, the gap between the two scales should {increase} with $\alpha$. 

To obtain a realistic value for $\sigma_{\rm f}$, we plug $\alpha \sim 1$ and $\left[t_{f, \, {\rm GRB}} \, \omega_{\rm pi}\right](\alpha \sim 1) \sim 2 \times 10^5$ into Equation \eqref{eq:sigma_scaling}, yielding
\begin{equation}
    \sigma_{\rm f}(\alpha \sim 1) \sim 3 \times 10^{-8};
\end{equation}
this magnetization is three orders of magnitude below the critical magnetization, $\sigma_{\rm cr} \sim 3 \times 10^{-5}$, above which particle acceleration is inhibited in relativistic electron-proton shocks \citep[e.g.,][]{sironimaxenergy,plotnikov_18}. Therefore, the magnetic field generated by pair loading at $R \sim 10^{17} \, {\rm cm}$ (corresponding to $\alpha \sim 1$) will not inhibit particle acceleration. According to Equation \eqref{eq:sigma_scaling}, $\sigma_{\rm f}$ should increase monotonically with $\alpha$; for $\alpha \sim 10$, we obtain $\sigma_{\rm f}(\alpha \sim 10) \sim 10^{-7}$, which is still well within the regime of efficient particle acceleration. 

Finally, we can assess whether shock-accelerated ions are magnetized in the field produced by upstream pair enrichment. In the downstream region, the Larmor radius of an ion gyrating in the field generated by upstream pair enrichment 
%in the upstream field 
is given by
\begin{equation} \label{eq:larmor}
    r_L = \frac{\gamma/\Gamma}{\sqrt{\sigma_{\rm f}}} \frac{c}{\omega_{\rm pi}},
\end{equation}
where $\gamma$ is the Lorentz factor of the ion. In the self-generated magnetic field of magnetization $\sigma_{\rm f}(\alpha \sim 1) \sim 3 \times 10^{-8}$, a thermal ion (with $\gamma \sim \Gamma$) will gyrate with Larmor radius 
\begin{equation}
    r_L(\alpha \sim 1) \sim 6 \times 10^3 \frac{c}{\omega_{\rm pi}};
\end{equation} 
this is roughly 50 times larger than $\left<\lambda_y\right>_{\rm f}(\alpha \sim 1)$. Consequently, upstream pair enrichment does not magnetize shock-accelerated ions, but it may affect their transport.

\section{Summary \& Conclusions} \label{sec:conclusions}

We employed periodic, 2D, fully kinetic particle-in-cell simulations to study the effects of the gradual enrichment of a cold electron-proton plasma with warm electron-positron pairs injected at mildly relativistic bulk speeds; these simulations approximately modeled the properties of a fluid element in the circumburst medium as it approached the expanding external shock front of a long gamma-ray burst. We carried out an extensive scan over two key parameters: $t_{\rm f}$ -- the duration of pair injection (or, equivalently, the time for the fluid element to encounter the shock front) -- and $\alpha$ -- the ratio of the final pair beam density to the background plasma density. We found that, across a wide range of $t_{\rm f}$ (from $t_{\rm f}\omega_{\rm pb} = 5 \times 10^3$ to $t_{\rm f}\omega_{\rm pb} = 4 \times 10^4$) and $\alpha$ (from $\alpha=1$ to $\alpha=16$), the evolution of the magnetic field proceeded in three common phases: first a primarily electrostatic phase dominated by two-stream and oblique modes, then an exponential electromagnetic phase driven by filamentation modes enhanced by continuous pair injection, and finally a stage of slow, secular growth driven by the relative drift between the leptons and the protons. Moreover, we found the magnetic field evolution to be self-similar in $\alpha$ and $t_{\rm f}$, with the temporal evolution of the system depending on the single parameter $\left[\alpha/(t_{\rm f} \omega_{\rm pi})\right]^{1/2} t\omega_{\rm pi}$; from this self-similarity, it follows that the growth rate of the magnetic energy density during the filamentation phase, $\Gamma/\omega_{\rm pi}$, scales with  $\left[\alpha/(t_{\rm f} \omega_{\rm pi})\right]^{1/2}$ and with $\alpha t_{\rm on}/t_{\rm f}$, where $t_{\rm on}$ is the onset time of the filamentation instability. While the growth rate of the filamentation phase depends only on the instantaneous beam-to-background density ratio at the onset time, continuous pair injection is still crucial for sustaining this growth rate over an extended period of time; simulations in which pair loading is prematurely interrupted at selected times yield significant differences as compared to the case of continuous injection.

Throughout the filamentation and secular phases, both the magnetic energy density and the transverse spatial scale of the magnetic field increase. The average transverse magnetic wavenumber at the end of pair injection, $\left<k_y\right>_{\rm f} c/\omega_{\rm pi}$, scales with $\alpha^{-1/12} \, \left(t_{\rm f}\omega_{\rm pi}\right)^{-2/3}$, according to Equation \eqref{eq:avgk_ion}. The magnetic energy density at the end of pair injection, normalized to the proton rest mass energy density, scales as 
$\alpha^{5/8} \, \left(t_{\rm f}\omega_{\rm pi}\right)^{-3/4}$, according to Equation \eqref{eq:sigma_scaling}. 
Using these trends to extrapolate to realistic regimes, we find that upstream pair enrichment generates weak magnetic fields on scales much larger than the proton skin depth;
for bright bursts, the extrapolated coherence scale at a shock radius of $R \sim 10^{17} \, {\rm cm}$ is $\left<\lambda_y\right> \sim 100 \; c/\omega_{\rm pi}$ and the corresponding magnetization is $\sigma \sim 10^{-8}$ for typical GRB and circumburst parameters. Since larger-scale fields decay slower, our results may help explain the persistence of magnetic fields at large distances behind GRB shocks.

During the preparation of this paper, another paper was published -- \citet{groseljlong} -- that explored an alternative mechanism for generating large-scale magnetic fields at GRB shocks. In that paper, the authors used unprecedentedly long-duration 2D PIC simulations to study the long-term evolution of a relativistic {pair} shock propagating into an initially {unmagnetized} medium. The continual acceleration of particles at the shock allowed high-energy particles to penetrate deep into the upstream, generating large-scale magnetic fields; by the end of their longest simulation, this process had generated magnetic structures of up to 100 plasma skin depths. For the later phases of the GRB afterglow -- when the upstream is no longer being enriched by electron-positron pairs from prompt photons -- the physical picture described in \citet{groseljlong} is the only viable model (provided that downstream synchrotron emission does not appreciably enrich the upstream with pairs, as envisioned in \citet{derishev}). For the early afterglow, however, our results imply that the work by \citet{groseljlong} should be expanded to include a non-zero, weak initial magnetic field on scales much larger than the proton skin depth.

We note, however, that while our simulations model an initially unmagnetized medium, a realistic circumburst medium would possess a non-zero initial mean magnetic field; realistic magnetizations for a Wolf-Rayet wind could be as high as $\sigma_0 \sim 10^{-5}$, for $\sigma_0$ defined using the magnetic field of the wind. Without directly simulating this ambient magnetic field, we can estimate its effect by comparing the gyration timescale of the beam electrons, $t_{gb}\omega_{\rm pi} \sim m_{\rm e}/(m_{\rm i} \sqrt{\sigma_0})$, to the filamentation growth time, $t_{\rm on} \omega_{\rm pi} \sim \left[\alpha/(t_{f, \, {\rm GRB}} \, \omega_{\rm pi})\right]^{-1/2}$ (from Equation \eqref{eq:growthrate}).
For even a weak ambient field of $\sigma_0 = 10^{-9}$, the beam electron gyration timescale is an order of magnitude shorter than the filamentation growth time for $\alpha \le 16$, implying that a weak background field would interfere with the development of the filamentation phase and any subsequent evolution of the magnetic field.

We also note that our scheme of gradually injecting pairs into a periodic domain is overly simplistic and likely not fully representative of reality. First, the periodicity in our simulations allows injected particles to remain in the volume for the entirety of the simulation, while in reality these particles should leave the volume after some time; a more realistic injection scheme should replace beam particles with ``fresh" pairs on some timescale $t_{\rm refresh}$. Additionally, as the simulated fluid element approaches the shock front, the statistical properties of the injected particle population should change; as such, the distribution from which beam particles are drawn should vary in time, as opposed to the temporally constant Maxwellian (with fixed injection rate) that we have used in this paper. Both of these changes to our injection scheme could be implemented in future work without needing to explicitly model the shock (in contrast to, e.g., \citealt{groseljpair}, \citealt{cerutti23}, or \citealt{groseljlong}).         

In addition to the possible inclusion of a non-zero initial background field and the aforementioned revisions to our particle injection scheme, our study leaves a number of other avenues for future work. For instance, the instabilities present in our simulations may evolve differently for different temporal profiles of pair injection; while we injected pairs linearly with time, it may be more realistic to inject pairs exponentially with time. One could also treat the process of pair injection more realistically by directly incorporating the prompt photons and self-consistent pair production, which is possible with PIC codes like Tristan-MP v2 \citep{tristan_v2}; the momentum continually imparted by the prompt photons (in addition to the momentum supplied by the produced electron-positron pairs) could further amplify the magnetic field and alter how the beam couples to the background plasma \citep{radiation}. Each of these avenues is also interesting from a pure plasma physics perspective, as the behavior of a beam-plasma system under the influence of a gradually-injected pair beam has barely been explored (see, however, \citet{compton,Faure}).

\begin{acknowledgments}
We thank A. Beloborodov for several useful discussions related to this work. R.G. was supported by NASA FINESST grant 80NSSC24K1477. L.S. was supported by NASA ATP grant 80NSSC20K0565 and NSF grant PHY2409223. This research was facilitated by Multimessenger Plasma Physics Center (MPPC) NSF grants PHY2206607 and PHY2206609. The work was supported by a grant from the Simons Foundation (MP-SCMPS-0000147, to L.S.). L.S. acknowledges support from DoE Early Career Award DE-SC0023015. 
D.G.~is supported by the Research Foundation---Flanders (FWO) 
Senior Postdoctoral Fellowship 12B1424N.  
Simulations were performed on NASA Pleiades (GID: s2356, s2610).
\end{acknowledgments}

\software{TRISTAN-MP \citep{tristan}}

\appendix

\section{Comparison with 3D simulation} \label{sec:3D}
While the 2D simulations we discuss in the main text do not capture the full 3D physics of our beam-plasma systems of interest, running 3D simulations with sufficiently large numbers of particles per cell, sufficiently high spatial resolutions, and sufficiently large box sizes is expensive. As such, we run only one 3D simulation -- with $\alpha=2$ and $t_{\rm f} \omega_{\rm pb} = 10^4$ -- to compare with our reference 2D simulation (Figure \ref{fig:field_evolution}); this simulation uses 18 final beam particles per cell per species, a resolution of 10 cells per $c/\omega_{\rm pb}$, and a box size of $100 \times 100 \times 100 \, \left(c/\omega_{\rm pb}\right)^3$, which fits slightly more than one magnetic wavelength in the $y$- and $z$-directions by the end of the simulation. We have checked that a $100 \times 100 \, \left(c/\omega_{\rm pb}\right)^2$ box (in 2D) is large enough to give results in reasonable agreement with the larger boxes analyzed in the main text.

\begin{figure}[h]
    \centering
    \includegraphics[width=0.47\textwidth]{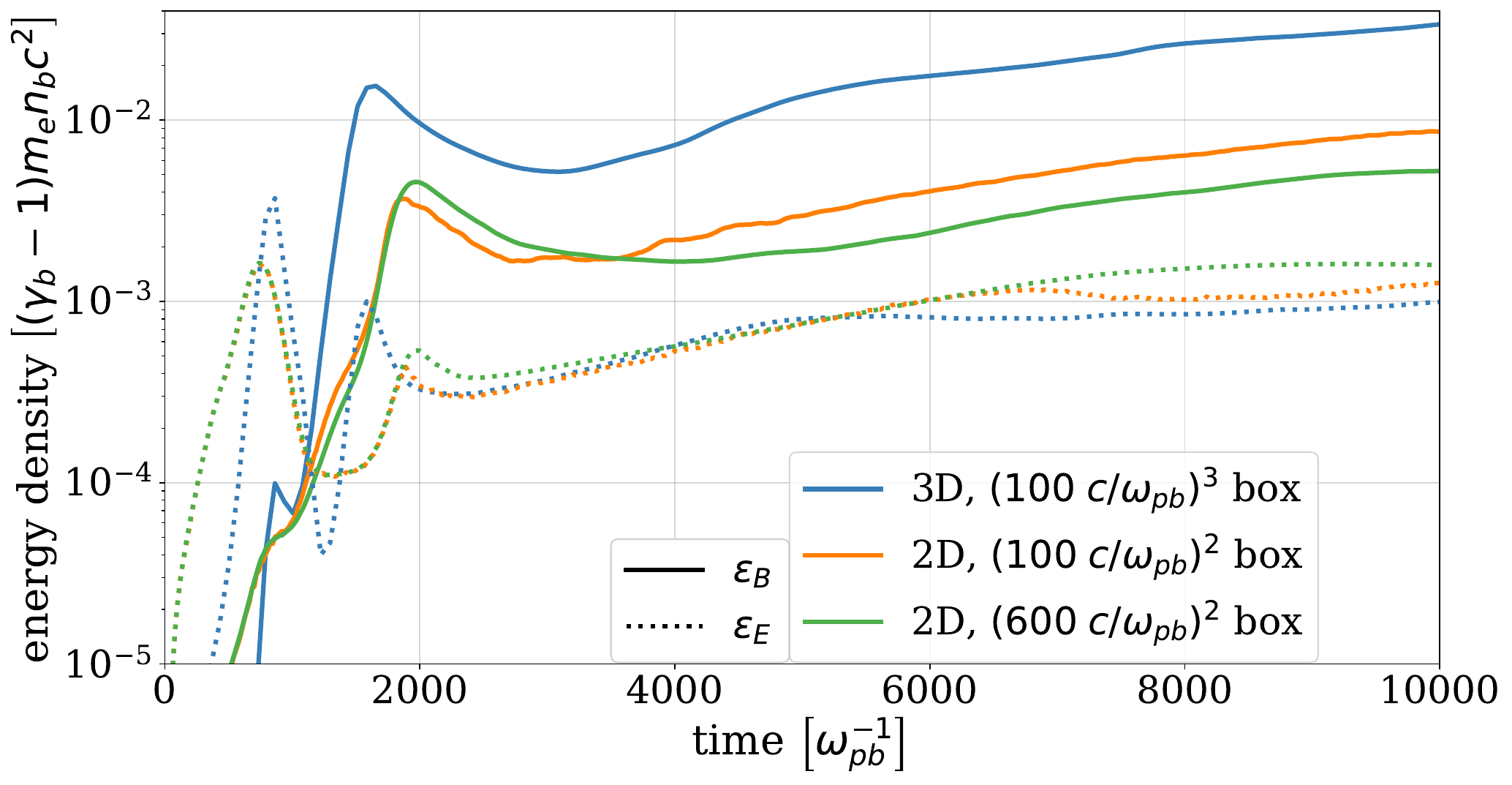}
    \caption{Comparison of our 3D simulation with $\alpha=2$ and $t_{\rm f}\omega_{\rm pb}=10^4$ (blue curves) to the equivalent 2D simulation with the same box side-length (orange curves) and to the equivalent 2D simulation with the fiducial box side-length (green curves). The solid curves show $\varepsilon_B$, which only includes the $z$-component of the magnetic field for the 2D simulations but includes all three components for the 3D simulation. Similarly, the dotted curves show $\varepsilon_E$, which includes the $x$- and $y$-components of the electric field for the 2D simulations but includes all three components for the 3D simulation.}
    \label{fig:3D_lines}
\end{figure}

\begin{figure}[h]
    \centering
    \includegraphics[width=0.47\textwidth]{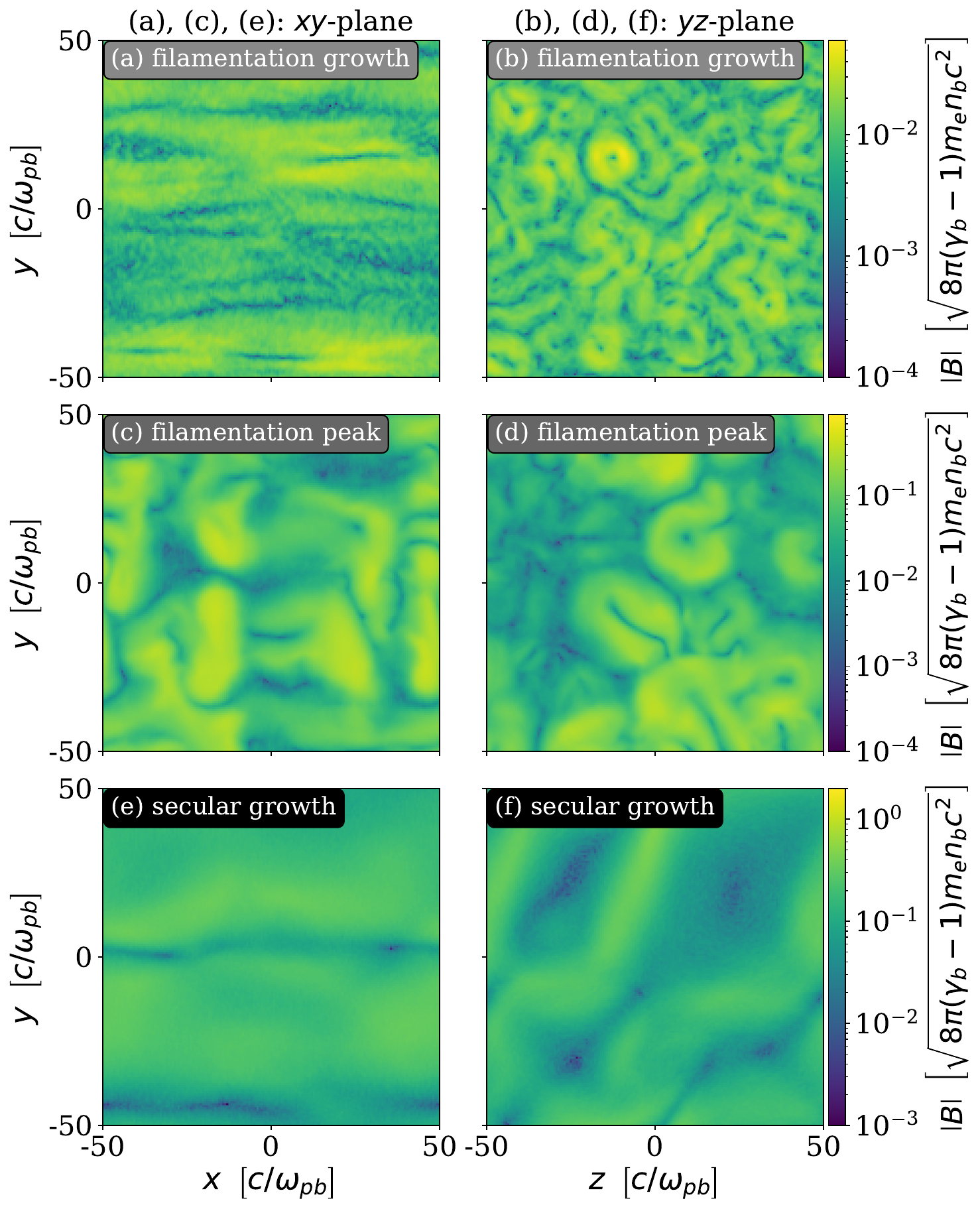}
    \caption{Spatial profiles of the magnitude of the magnetic field, $\left|B\right|/\sqrt{8\pi(\gamma_{\rm b}-1)m_{\rm e} n_{\rm b} c^2}$, in the $xy$-plane (left column) and in the $yz$-plane (right column) at three different times throughout our 3D simulation: panels (a) and (b) show snapshots at $t\omega_{\rm pb} \sim 1120$, during the exponential growth of the filamentation modes; panels (c) and (d) show snapshots at $t\omega_{\rm pb} \sim 1660$, at the saturation of the filamentation phase; and panels (e) and (f) show snapshots at $t\omega_{\rm pb} \sim 8000$, deep into the secular phase.}
    \label{fig:3D_profiles}
\end{figure}

In Figure \ref{fig:3D_lines} we compare our 3D simulation both to the equivalent 2D simulation in a $100 \times 100 \, \left(c/\omega_{\rm pb}\right)^2$ box and to the equivalent 2D simulation in a $600 \times 600 \, \left(c/\omega_{\rm pb}\right)^2$ box (all with the same spatial resolution and numbers of particles per cell). The 3D simulation is qualitatively similar to its 2D counterparts: the simulation starts with an electrostatic phase dominated by two-stream and oblique modes, transitions into an electromagnetic phase driven by filamentation modes, and then -- after the filamentation phase saturates -- ends with a prolonged phase of slower, secular growth. This is confirmed in Figure \ref{fig:3D_profiles}, which shows spatial profiles (in both the $xy$- and $yz$-planes, respectively in the left and right columns) of the magnitude of the magnetic field generated in the 3D simulation; the tube-like magnetic filaments that are visible in panels (a) and (b) grow in width as the exponential phase proceeds (panel (d)), becoming turbulent at the saturation of the filamentation phase (panel (c)) but later regenerating on even larger scales during the secular phase, filling the box with little more than one filament in panels (e) and (f).

Quantitatively, the 3D simulation differs from its 2D counterparts in predictable ways. In 3D, the filamentation instability can  
yield a magnetic field with non-zero $z$- {and} $y$-components, while in 2D only the $z$-component is allowed to grow. This explains why the box-averaged magnetic energy density in the 3D simulation is roughly two times larger than in the equivalent 2D simulation (see Figure \ref{fig:3D_lines}). According to \citet{peterson_21}, the differing geometry of the magnetic filaments (slab in 2D vs. cylindrical in 3D) is responsible for the small difference in the filamentation growth rate between 2D and 3D. 

\section{Reduced mass ratio simulations} \label{sec:lowmass}

In Section \ref{sec:extrapolation}, we used the empirical scaling law for the final transverse magnetic wavenumber ($\left<k_y\right>_{\rm f}$) derived in Section \ref{sec:empirical} to extrapolate our simulated values of $\left<k_y\right>_{\rm f}$ to realistic scales. This extrapolation predicted, for realistic GRB parameters, a self-generated magnetic field on scales much larger than the ion skin depth, $c/\omega_{\rm pi}$; however, given that the magnetic field scale never significantly exceeds $c/\omega_{\rm pi}$ in the simulation suite analyzed in the main text, it is unclear whether the extrapolation of our results to scales beyond $c/\omega_{\rm pi}$ is entirely justified. In this appendix, we show results from simulations with a reduced mass ratio of $m_{\rm i}/m_{\rm e} = 25$ to demonstrate that continuous pair injection can definitively produce magnetic fields at scales much larger than $c/\omega_{\rm pi}$, thus validating our extrapolation. Since $\omega_{\rm pb}^{-1} \propto (m_{\rm i}/m_{\rm e})^{-1/2} \, \omega_{\rm pi}^{-1}$, fixing the pair injection time in units of $\omega_{\rm pb}^{-1}$ and lowering the mass ratio yields a longer injection time in units of $\omega_{\rm pi}^{-1}$, allowing the self-generated magnetic structures to grow to larger scales.

\begin{figure}[h]
    \centering
    \includegraphics[width=0.47\textwidth]{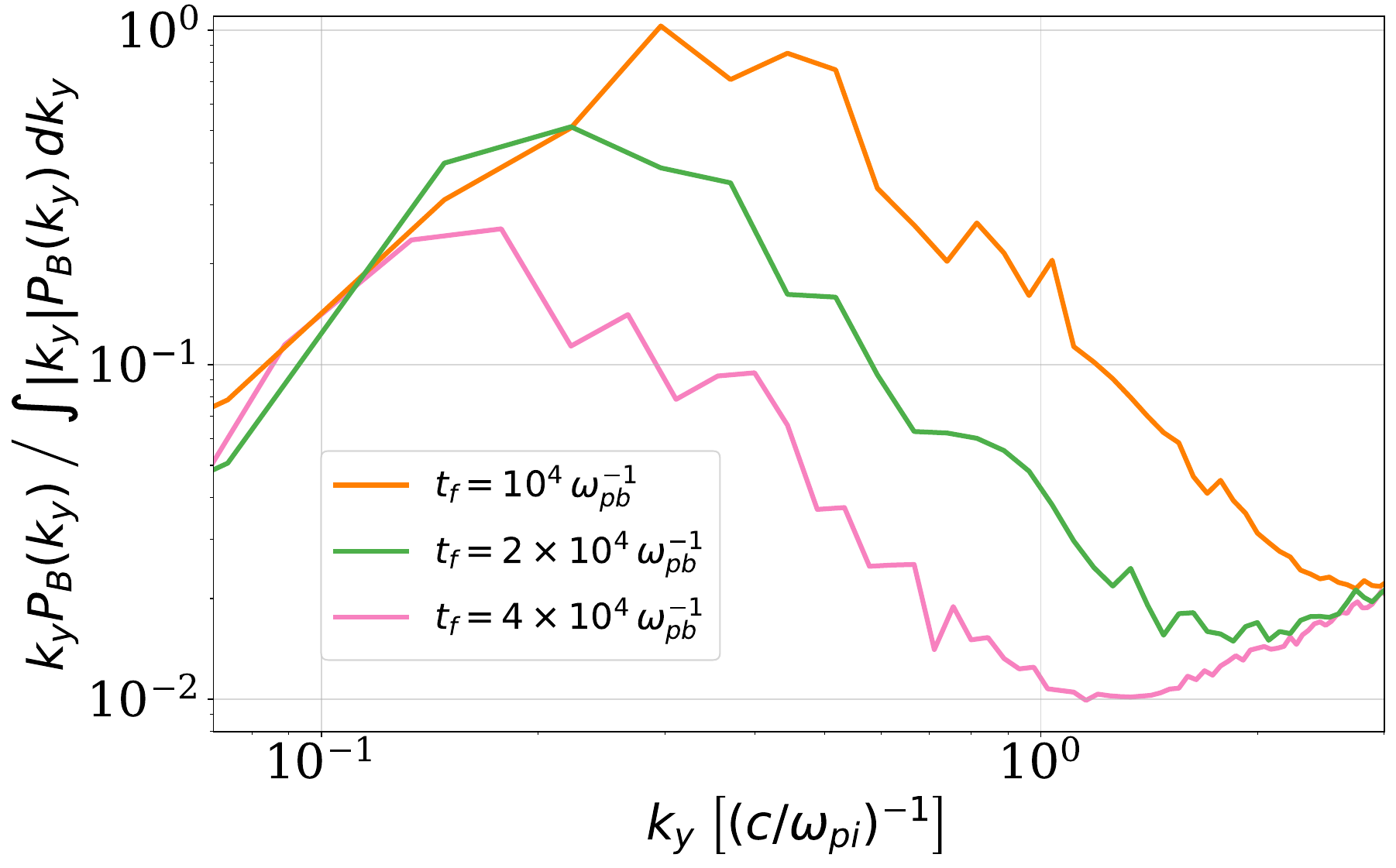}
    \caption{1D transverse magnetic power spectra at the end of pair injection for three simulations with $m_{\rm i}/m_{\rm e} = 25$ and $\alpha=2$: $t_{\rm f}\omega_{\rm pb} = 10^4$ (orange), $t_{\rm f}\omega_{\rm pb} = 2 \times 10^4$ (green), and $t_{\rm f}\omega_{\rm pb} = 4 \times 10^4$ (pink). Note that the horizontal axis is in units of inverse ion skin depths.}
    \label{fig:reducedpowspec}
\end{figure}

\begin{figure}[h]
    \centering
    \includegraphics[width=0.47\textwidth]{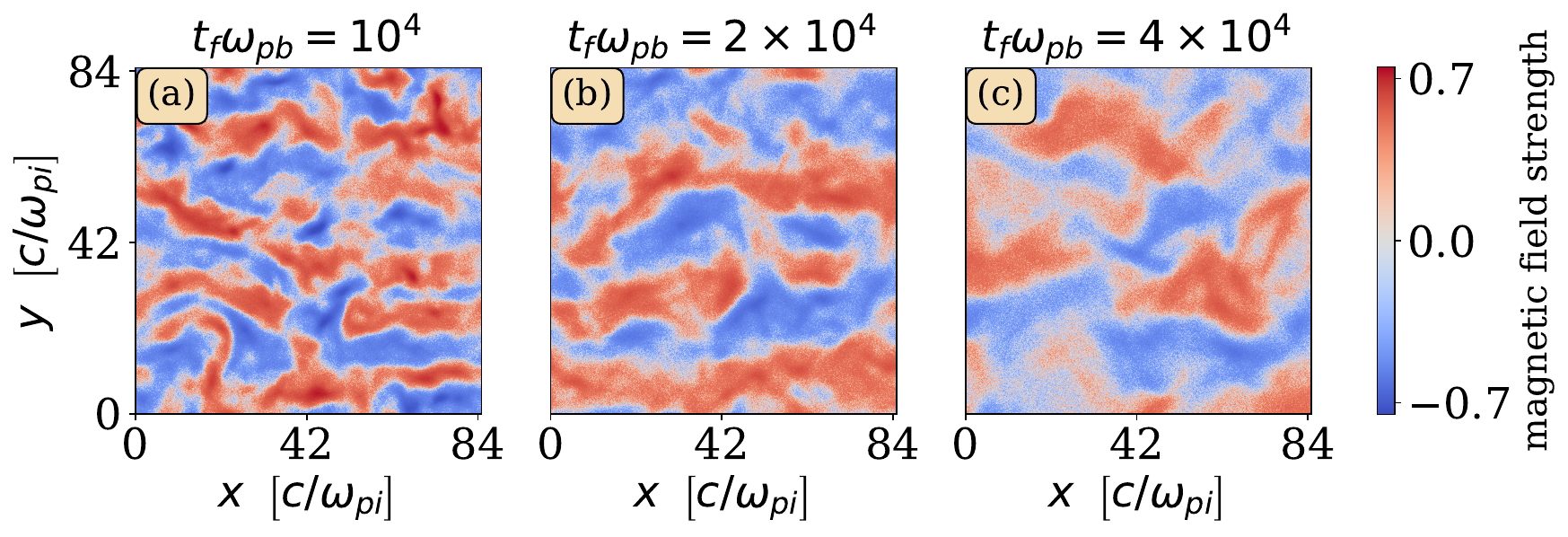}
    \caption{Spatial profiles of $B_z/\sqrt{8\pi\left(\gamma_{\rm b}-1\right)m_{\rm e} n_{\rm b} c^2}$ at the end of pair injection for three simulations with $m_{\rm i}/m_{\rm e}=25$, $\alpha=2$, and $t_{\rm f} \omega_{\rm pb}$ varying between $10^4$ (panel (a)), $2\times 10^4$ (panel (b)), and $4\times 10^4$ (panel (c)). Note that the units of length are ion skin depths, not beam electron skin depths.}
    \label{fig:reducedprofiles}
\end{figure}

In Figure \ref{fig:reducedpowspec}, we show the 1D transverse magnetic power spectra at the end of pair injection for three simulations with $m_{\rm i}/m_{\rm e} = 25$, $\alpha=2$, and $t_{\rm f}\omega_{\rm pb}$ varying between $10^4$, $2 \times 10^4$, and $4 \times 10^4$; these simulations are run in a 2D box of dimensions $\left(600 \, c/\omega_{\rm pb}\right)^2$ (or $\left(60\sqrt{2} \, c/\omega_{\rm pi}\right)^2$) with 36 particles per cell per species and all other parameters the same as in our main suite of simulations. In Figure \ref{fig:reducedprofiles}, we show the corresponding spatial profiles of the out-of-plane magnetic field ($B_z$). In both of these plots -- which have axes in units of the ion plasma scales -- it is evident that the scale of the self-generated magnetic structures, $2\pi/\left<k_y\right>_{\rm f}$, exceeds the ion skin depth by at least one order of magnitude. The fact that these $m_{\rm i}/m_{\rm e}=25$ simulations show that continuous pair injection can produce magnetic structures at super-ion scales gives us confidence that we can extrapolate our simulation results to realistic, macroscopic scales.

\section{Higher beam Lorentz factors}
\label{sec:gammab}

\begin{figure}[h]
    \centering
    \includegraphics[width=0.47\textwidth]{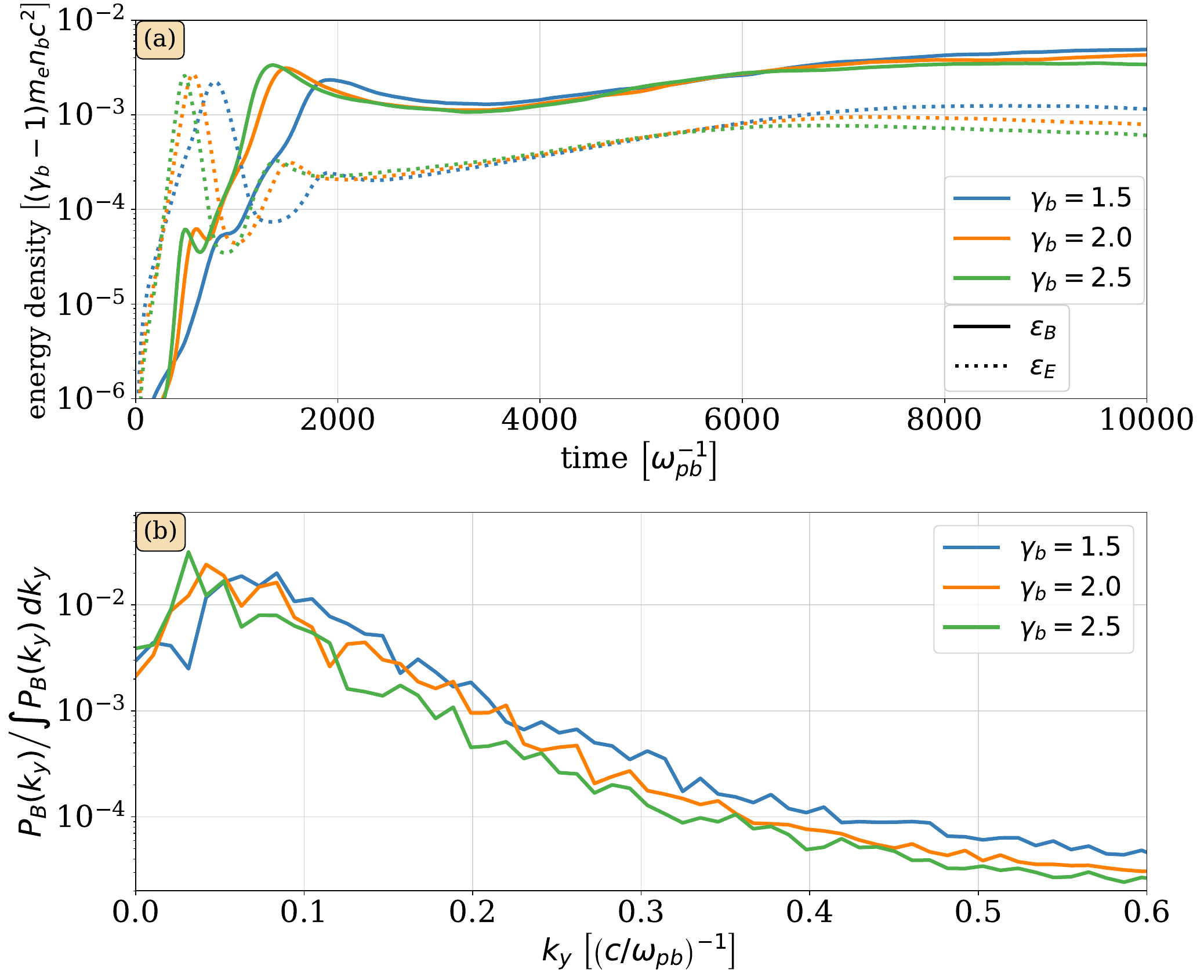}
    \caption{Comparison of our reference simulation ($\alpha=2$, $t_{\rm f}\omega_{\rm pb} = 10^4$) across multiple beam Lorentz factors, $\gamma_{\rm b}$. Panel (a) shows the time evolution of $\varepsilon_B$ (solid curves) and $\varepsilon_E$ (dotted curves) for simulations with $\gamma_{\rm b}=1.5$ (blue), $\gamma_{\rm b}=2.0$ (orange), and $\gamma_{\rm b}=2.5$ (green). Panel (b) shows the normalized 1D transverse magnetic power spectra at the end of pair injection for the same three simulations.}
    \label{fig:gammab compare}
\end{figure}

In the simulations presented in the main text of this paper, we assume that the electron-positron pair beam drifts with a Lorentz factor of $\gamma_{\rm b} = 1.5$ in the upstream frame of a GRB shock (i.e., our simulation frame). Here, we check how our results vary for slightly larger $\gamma_{\rm b}$. Figure \ref{fig:gammab compare} shows the magnetic and electric energy densities (in the top panel) and the final transverse magnetic power spectra (in the bottom panel) for three simulations with $\gamma_{\rm b} =$ 1.5, 2.0, and 2.5. Our results vary only slightly among the three simulations. As expected, as we increase $\gamma_{\rm b}$, increasing the free energy of the beam, the growth rates of the electrostatic and filamentation phases increase; however, the growth rate of the secular phase is largely unchanged. $\varepsilon_B$ and $\left<k_y\right> c/\omega_{\rm pb}$ at the end of pair injection are both monotonic in $\gamma_{\rm b}$: as $\gamma_{\rm b}$ increases, both $\varepsilon_{B, \, \rm f}$ and $\left<k_y\right>_{\rm f} c/\omega_{\rm pb}$ decrease by small factors. 

\section{Integral-scale wavenumber vs. $\lowercase{\left<k_y\right>}$} \label{sec:kint_appendix}

Throughout this paper, we have used $\left<k_y\right>$ -- as defined in Equation \eqref{eq:avgk} -- to quantify the dominant transverse wavenumber of the self-generated magnetic field. Because $\left<k_y\right>$ is susceptible to contamination by numerical shot noise at the high-$k_y$ end of the magnetic spectrum, we have applied a finite upper bound on the integrals in Equation \eqref{eq:avgk}. Alternatively, we could have used the ``integral-scale'' definition of the dominant wavenumber:
\begin{equation} \label{eq:kint}
    k_{\rm int} \equiv \frac{\int_0^\infty P_B(k_y) \, dk_y}{\int_0^\infty  k_y^{-1}P_B(k_y) \, dk_y}.
\end{equation}
Due to the factor of $k_y^{-1}$ in the denominator, $k_{\rm int}$ should be more resilient to high-$k_y$ noise, even with an infinite integral upper bound. We show that our main results change negligibly if we use $k_{\rm int}$ instead of $\left<k_y\right>$.

\begin{figure}[h]
    \centering
    \includegraphics[width=0.47\textwidth]{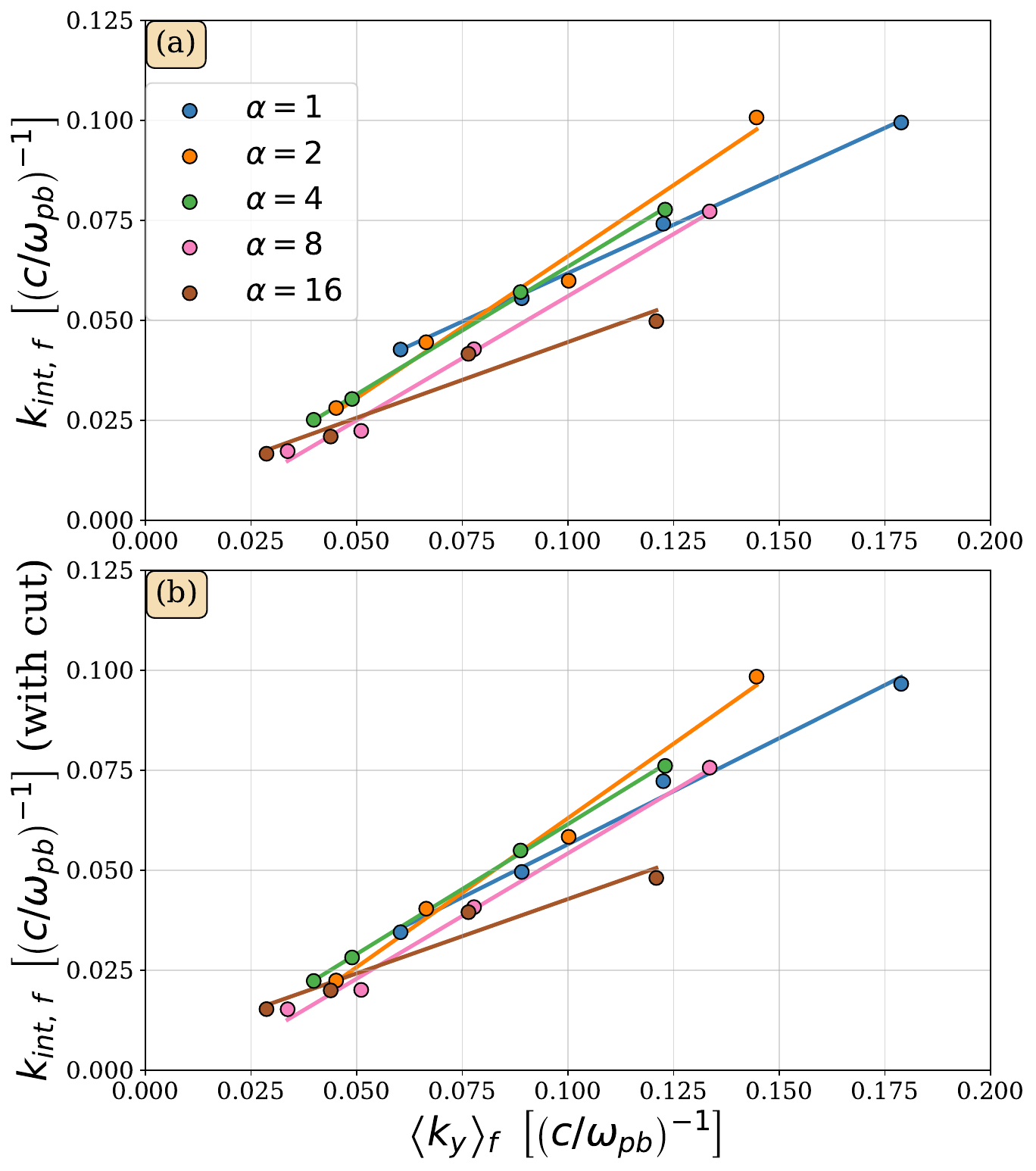}
    \caption{Comparison of the integral-scale wavenumber, $k_{\rm int, \, f}$ (as defined in \eqref{eq:kint}), and the average transverse wavenumber, $\left<k_y\right>_{\rm f}$ (as defined in \eqref{eq:avgk}), at the end of pair injection for each of the simulations in our main suite; each point corresponds to a different simulation -- with blue points for $\alpha=1$, orange for $\alpha=2$, green for $\alpha=4$, pink for $\alpha=8$, and brown for $\alpha=16$ -- and the solid lines indicate the lines of best fit for each $\alpha$. Panel (a) uses the values of $k_{\rm int, \, f}$ computed with no upper bounds on the integrals, while panel (b) uses the values of $k_{\rm int, \, f}$ computed with the same high-$k_y$ integral cutoffs as used for $\left<k_y\right>_{\rm f}$.}
    \label{fig:kint_vs_avgk}
\end{figure}

In Figure \ref{fig:kint_vs_avgk}, we compare the integral-scale wavenumber, $k_{\rm int, \, f}$, to the average transverse wavenumber, $\left<k_y\right>_{\rm f}$, at the end of pair injection for each of the simulations in our main suite; in panel (a) we compute $k_{\rm int, \, f}$ with no upper bounds on the integrals in Equation \eqref{eq:kint}, while in panel (b) we compute $k_{\rm int, \, f}$ using the same high-$k_y$ integral cutoffs as we have used for $\left<k_y\right>$ throughout this paper. By making scatter plots of $k_{\rm int, \, f}$ vs. $\left<k_y\right>_{\rm f}$, we notice that switching from $\left<k_y\right>_{\rm f}$ to $k_{\rm int, \, f}$ simply decreases the dominant magnetic wavenumber by a factor of roughly two; this factor varies weakly with $\alpha$, so the dependence of the dominant wavenumber on both $t_{\rm f}$ and $\alpha$ should change negligibly with the choice of $k_{\rm int, \, f}$ vs. $\left<k_y\right>_{\rm f}$. We also note that there is negligible difference between $k_{\rm int, \, f}$ computed with and without a finite integral upper bound. We therefore conclude that the main results of this paper are robust to the choice of definition for the dominant magnetic wavenumber.

\section{Numerical Convergence} \label{sec:convergence}
The fidelity of our simulations relies primarily on three numerical parameters: the number of particles per cell, the spatial resolution (i.e., the number of cells that fit into one final beam electron skin depth, $c/\omega_{\rm pb}$), and the size of the simulation box. Here, we show scans over each of these three parameters to demonstrate that the simulations analyzed in the main text are converged numerically.

\begin{figure}[htbp]
    \centering
    \includegraphics[width=0.47\textwidth]{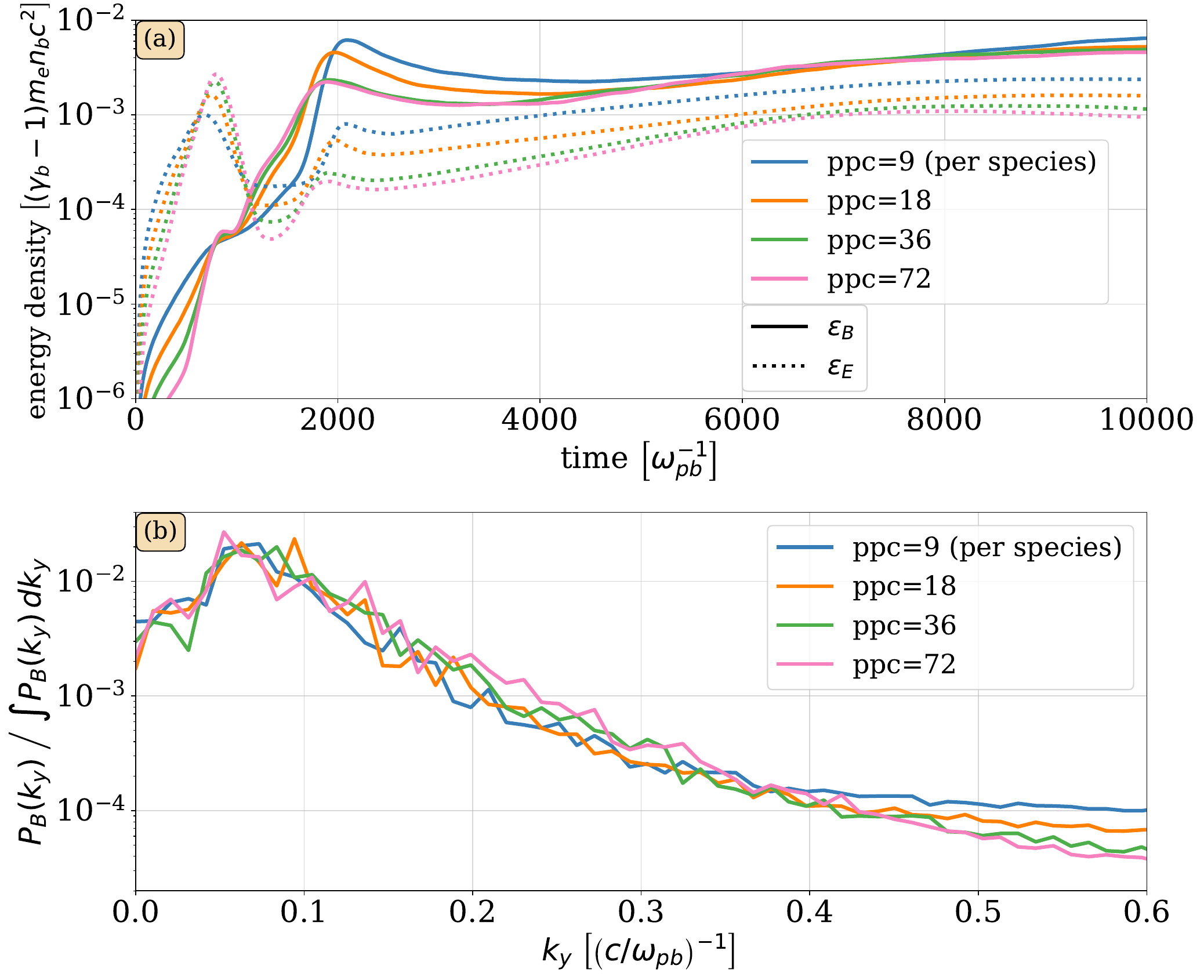}
    \caption{Tests to determine whether our $\alpha=2$, $t_{\rm f}\omega_{\rm pb} = 10^4$ reference simulation is numerically converged with respect to particles per cell per species. Panel (a) shows the time evolution of $\varepsilon_B$ (solid curves) and $\varepsilon_E$ (dotted curves) for simulations with 9 (blue), 18 (orange), 36 (green), and 72 (pink) final beam particles per cell per species. Panel (b) shows the normalized 1D transverse magnetic power spectra at the end of pair injection for the same four simulations.}
    \label{fig:ppc convergence}
\end{figure}

Figure \ref{fig:ppc convergence} shows our reference simulation, $\alpha=2$ and $t_{\rm f}\omega_{\rm pb} = 10^4$ (with $c/\omega_{\rm pb}$ resolved by 10 cells and with a box size of $600 \times 600 \, \left(c/\omega_{\rm pb}\right)^2$), with the final number of beam particles per cell {per species} varying between 9, 18, 36, and 72. For 36 beam particles per cell per species and above, the difference between the simulations is negligible: between 36 beam particles and 72 beam particles per species, the respective curves for $\varepsilon_B$ are practically overlapping, the curves for $\varepsilon_E$ differ by a very small factor, and the final $P_B(k_y)$ start to diverge due to numerical noise only above $k_y c/\omega_{\rm pb} \sim 0.5$. In the plots in the main text, we only show simulations with final beam particles per cell per species of at least 36. 

\begin{figure}[htbp]
    \centering
    \includegraphics[width=0.47\textwidth]{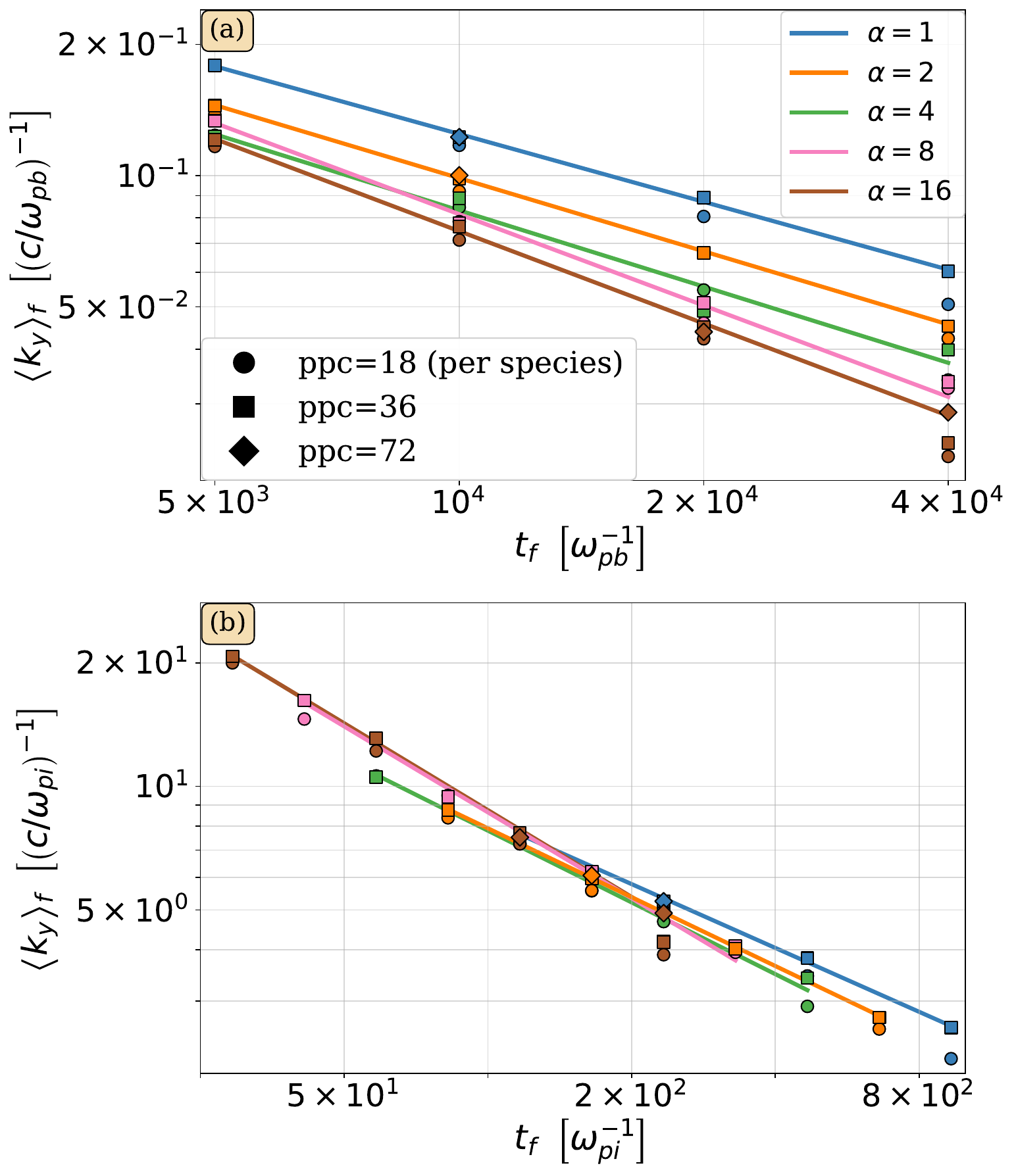}
    \caption{$\left<k_y\right>c/\omega_{\rm pb}$ vs. $t_{\rm f}\omega_{\rm pb}$ (top panel) and $\left<k_y\right>c/\omega_{\rm pi}$ vs. $t_{\rm f}\omega_{\rm pi}$ (bottom panel) at the end of pair injection for all $\alpha$ and for varying numbers of particles per cell. Each dot represents one simulation, with blue dots for $\alpha=1$, orange for $\alpha=2$, green for $\alpha=4$, pink for $\alpha=8$, and brown for $\alpha=16$. Simulations with 18 final beam particles per cell per species are represented by circles, those with 36 final beam particles per cell per species are represented by squares, and those with 72 final beam particles per cell per species are represented by diamonds. The lines are fit to the cases with highest ppc: the blue best-fit line (for $\alpha=1$) has a slope of $-0.52$, the orange best-fit line (for $\alpha=2$) has a slope of $-0.56$, the green best-fit line (for $\alpha=4$) has a slope of $-0.58$, the pink best-fit line (for $\alpha=8$) has a slope of $-0.70$, and the brown best-fit line (for $\alpha=16$) has a slope of $-0.70$. In contrast to Figure \ref{fig:avgk_summary}, where only the simulations with $t_{\rm f}\omega_{\rm pb} \ge 10^4$ are considered, here the $t_{\rm f}\omega_{\rm pb} = 5 \times 10^3$ simulations are both included in the plots and used in computing the best-fits.}
    \label{fig:avgk convergence}
\end{figure}

\begin{figure}[htbp]
    \centering
    \includegraphics[width=0.47\textwidth]{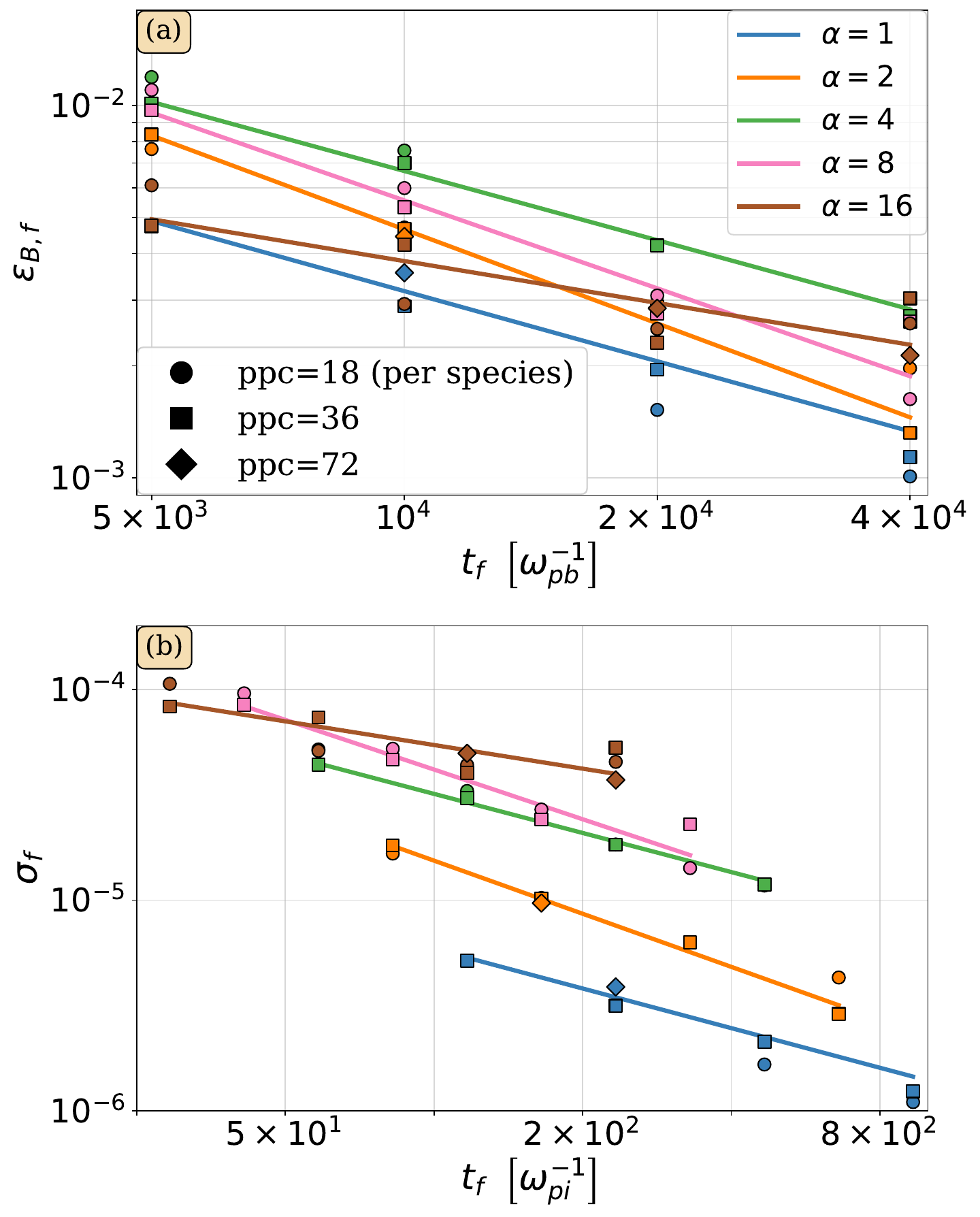}
    \caption{$\varepsilon_B$ vs. $t_{\rm f}\omega_{\rm pb}$ (top panel) and $\sigma$ vs. $t_{\rm f}\omega_{\rm pi}$ (bottom panel) at the end of pair injection for all $\alpha$ and for varying numbers of particles per cell. Each dot represents one simulation, with blue dots for $\alpha=1$, orange for $\alpha=2$, green for $\alpha=4$, pink for $\alpha=8$, and brown for $\alpha=16$. Simulations with 18 final beam particles per cell per species are represented by circles, those with 36 final beam particles per cell per species are represented by squares, and those with 72 final beam particles per cell per species are represented by diamonds. The lines are fit to the cases with highest ppc: the blue best-fit line (for $\alpha=1$) has a slope of $-0.62$, the orange best-fit line (for $\alpha=2$) has a slope of $-0.84$, the green best-fit line (for $\alpha=4$) has a slope of $-0.62$, the pink best-fit line (for $\alpha=8$) has a slope of $-0.78$, and the brown best-fit line (for $\alpha=16$) has a slope of $-0.37$. In contrast to Figure \ref{fig:epsb_summary}, where only the simulations with $t_{\rm f}\omega_{\rm pb} \ge 10^4$ are considered, here the $t_{\rm f}\omega_{\rm pb} = 5 \times 10^3$ simulations are both included in the plots and used in computing the best-fits.}
    \label{fig:epsb convergence}
\end{figure}

The transverse magnetic wavenumber and magnetic energy density at the end of pair injection are both sensitive to the number of particles per cell. To gauge the uncertainty on our scaling relations for $\left<k_y\right>_{\rm f}$, $\varepsilon_{B, \, \rm f}$, and $\sigma_{\rm f}$ derived from Figures \ref{fig:avgk_summary} and \ref{fig:epsb_summary} (which show our highest-ppc cases), we plot these quantities for simulations with a wide range of particles-per-cell in Figures \ref{fig:avgk convergence} (for $\left<k_y\right>_{\rm f} c/\omega_{\rm pb}$ and $\left<k_y\right>_{\rm f} c/\omega_{\rm pi}$) and \ref{fig:epsb convergence} (for $\varepsilon_{B, \, \rm f}$ and $\sigma_{\rm f}$). Given the relatively small differences between our cases with 18, 36, and 72 final beam particles per cell per species, we can conclude that the trends we report in the main paper are robust.   

\begin{figure}[htbp]
    \centering    \includegraphics[width=0.47\textwidth]{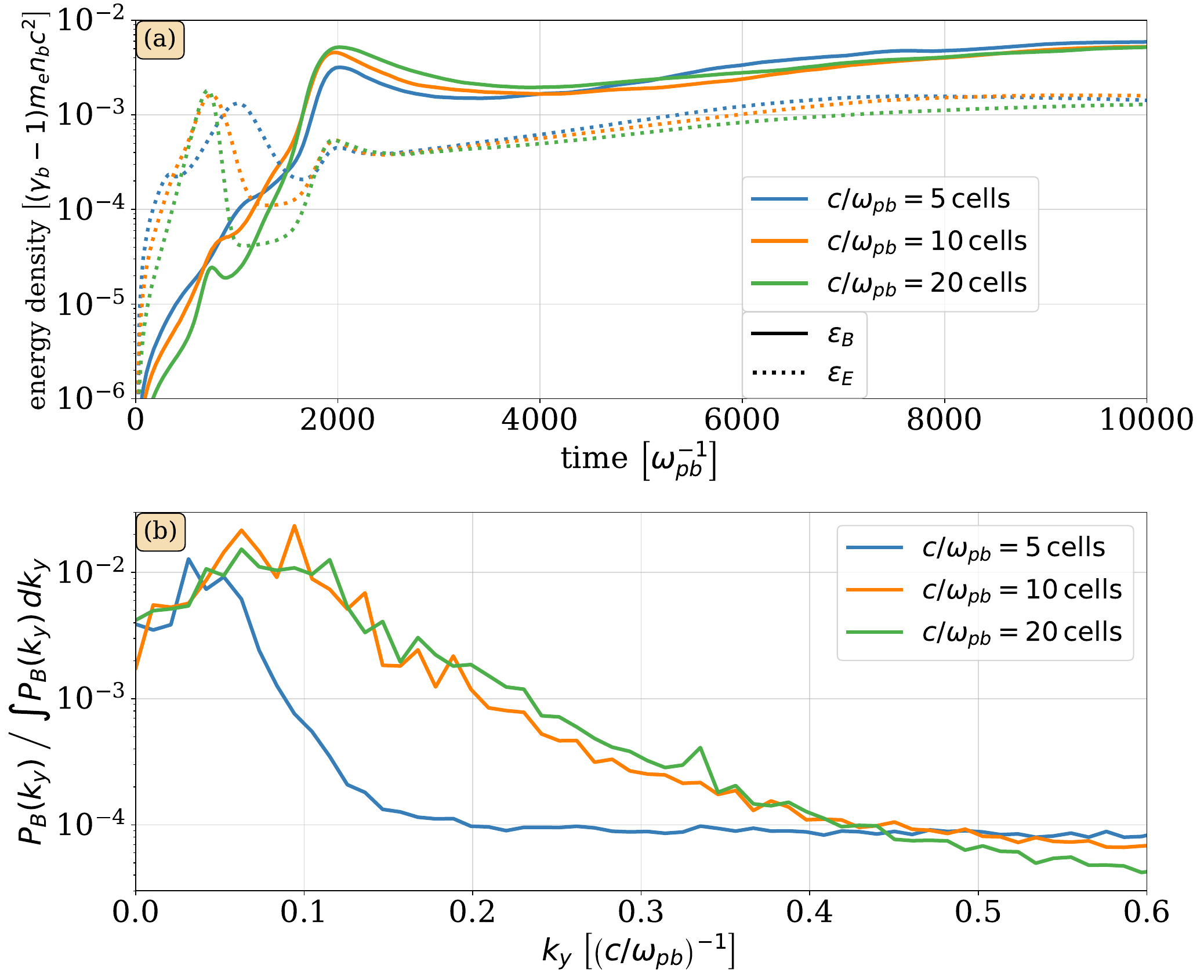}
    \caption{Tests to determine whether our $\alpha=2$, $t_{\rm f}\omega_{\rm pb} = 10^4$ reference simulation is numerically converged with respect to spatial resolution. Panel (a) shows the time evolution of $\varepsilon_B$ (solid curves) and $\varepsilon_E$ (dotted curves) for simulations that resolve the final beam electron skin depth, $c/\omega_{\rm pb}$, with 5 (blue), 10 (orange), and 20 (green) cells. Panel (b) shows the normalized 1D transverse magnetic power spectra at the end of pair injection for the same three simulations.}
    \label{fig:comp convergence}
\end{figure}

Figure \ref{fig:comp convergence} shows the $\alpha=2$, $t_{\rm f}\omega_{\rm pb} = 10^4$ case (with 18 final beam particles per cell per species and with a box size of $600 \times 600 \, \left(c/\omega_{\rm pb}\right)^2$) for resolutions varying between 5, 10, and 20 cells per $c/\omega_{\rm pb}$. While the simulation with 5 cells per $c/\omega_{\rm pb}$ is substantially noisier than the other two simulations, these other two cases appear well-converged; while the power spectrum for the $c/\omega_{\rm pb} = 5 \, {\rm cells}$ case becomes noise-limited above $k_y c/\omega_{\rm pb} \sim 0.15$, the power spectra for the $c/\omega_{\rm pb} = 10 \, {\rm cells}$ and $c/\omega_{\rm pb} = 20 \, {\rm cells}$ cases do not become noise-limited until at least $k_y c/\omega_{\rm pb} \sim 0.6$. In each of the plots in the main text, we only show simulations that resolve $c/\omega_{\rm pb}$ with 10 cells.

\begin{figure}[htbp]
    \centering
    \includegraphics[width=0.47\textwidth]{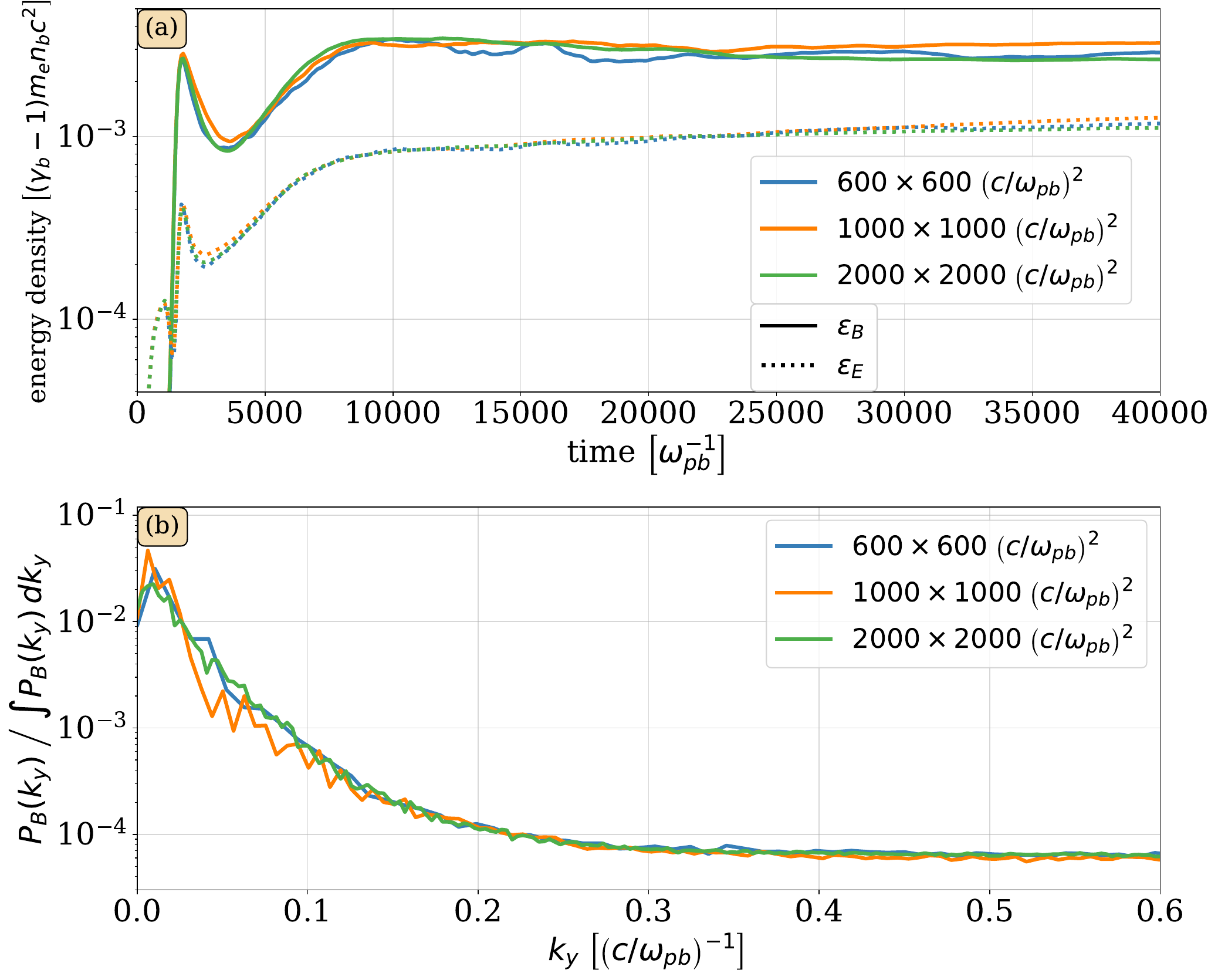}
    \caption{Tests to determine whether our $\alpha=16$, $t_{\rm f}\omega_{\rm pb} = 4 \times 10^4$ simulation is numerically converged with respect to simulation box size. Panel (a) shows the time evolution of $\varepsilon_B$ (solid lines) and $\varepsilon_E$ (dotted lines) for simulations using box sizes of $600 \times 600 \, \left(c/\omega_{\rm pb}\right)^2$ (blue), $1000 \times 1000 \, \left(c/\omega_{\rm pb}\right)^2$ (orange), and $2000 \times 2000 \, \left(c/\omega_{\rm pb}\right)^2$ (green). Panel (b) shows the normalized 1D transverse magnetic power spectra at the end of pair injection for the same three simulations.}
    \label{fig:boxsize convergence}
\end{figure}

For our simulations with high $\alpha$ and high $t_{\rm f}$, there is a concern that the large-scale magnetic structures produced by the end of our simulations will not fit within the simulation box; we therefore also check that the size of our box does not stifle the growth of the magnetic field's spatial scale.  Figure \ref{fig:boxsize convergence} shows our $\alpha=16, \, t_{\rm f}\omega_{\rm pb}=4 \times 10^4$ case (with 18 final beam particles per cell per species and with $c/\omega_{\rm pb}$ resolved by 10 cells) for box sizes varying between $600 \times 600 \, \left(c/\omega_{\rm pb}\right)^2$, $1000 \times 1000 \, \left(c/\omega_{\rm pb}\right)^2$, and $2000 \times 2000 \, \left(c/\omega_{\rm pb}\right)^2$. While $\varepsilon_B$ for the smallest box shows some deviations -- possibly indicating that the magnetic structures are hitting the box scale -- $\varepsilon_B$, $\varepsilon_E$, and $P_B(k_y)$ appear well-behaved for the two larger boxes. Therefore, for our simulations that produce the largest magnetic structures we employ a $1000 \times 1000 \, \left(c/\omega_{\rm pb}\right)^2$ box; for the plots in the main text, we use a $1000 \times 1000 \, \left(c/\omega_{\rm pb}\right)^2$ box for each of the $t_{\rm f}\omega_{\rm pb}=2 \times 10^4$ and $t_{\rm f}\omega_{\rm pb}=4 \times 10^4$ cases, and we use a $600 \times 600 \, \left(c/\omega_{\rm pb}\right)^2$ box for everything else.

\bibliography{main.bib}{}

\begin{thebibliography}{}
\expandafter\ifx\csname natexlab\endcsname\relax\def\natexlab#1{#1}\fi
\providecommand{\url}[1]{\href{#1}{#1}}
\providecommand{\dodoi}[1]{doi:~\href{http://doi.org/#1}{\nolinkurl{#1}}}
\providecommand{\doeprint}[1]{\href{http://ascl.net/#1}{\nolinkurl{http://ascl.net/#1}}}
\providecommand{\doarXiv}[1]{\href{https://arxiv.org/abs/#1}{\nolinkurl{https://arxiv.org/abs/#1}}}

\bibitem[{{Achterberg} \& {Wiersma}(2007)}]{achterberg1}
{Achterberg}, A., \& {Wiersma}, J. 2007, \aap, 475, 1, \dodoi{10.1051/0004-6361:20065365}

\bibitem[{{Achterberg} {et~al.}(2007){Achterberg}, {Wiersma}, \& {Norman}}]{achterberg2}
{Achterberg}, A., {Wiersma}, J., \& {Norman}, C.~A. 2007, \aap, 475, 19, \dodoi{10.1051/0004-6361:20065366}

\bibitem[{{Beloborodov}(2002)}]{beloborodovradiation}
{Beloborodov}, A.~M. 2002, \apj, 565, 808, \dodoi{10.1086/324195}

\bibitem[{{Beloborodov} {et~al.}(2014){Beloborodov}, {Hasco{\"e}t}, \& {Vurm}}]{beloborodovgev}
{Beloborodov}, A.~M., {Hasco{\"e}t}, R., \& {Vurm}, I. 2014, \apj, 788, 36, \dodoi{10.1088/0004-637X/788/1/36}

\bibitem[{{Birdsall} \& {Langdon}(1991)}]{birdsall}
{Birdsall}, C.~K., \& {Langdon}, A.~B. 1991, {Plasma Physics via Computer Simulation}

\bibitem[{{Blandford} \& {Eichler}(1987)}]{blandford}
{Blandford}, R., \& {Eichler}, D. 1987, \physrep, 154, 1, \dodoi{10.1016/0370-1573(87)90134-7}

\bibitem[{{Boris}(1970)}]{boris}
{Boris}, J.~P. 1970, Proc. 4th Conf. Num. Sim. Plasmas, 3.
\newblock \url{https://apps.dtic.mil/sti/citations/ADA023511}

\bibitem[{{Brainerd}(2000)}]{brainerd}
{Brainerd}, J.~J. 2000, \apj, 538, 628, \dodoi{10.1086/309136}

\bibitem[{{Bret} {et~al.}(2010){Bret}, {Gremillet}, \& {Dieckman}}]{Bret_2010a}
{Bret}, A., {Gremillet}, L., \& {Dieckman}, M.~E. 2010, Phys. Plasmas, 17, 120501, \dodoi{10.1063/1.3514586}

\bibitem[{{Bret} {et~al.}(2013){Bret}, {Stockem}, {Fiuza}, {Ruyer}, {Gremillet}, {Narayan}, \& {Silva}}]{Bret_2013}
{Bret}, A., {Stockem}, A., {Fiuza}, D., {et~al.} 2013, Phys. Plasmas, 20, 042102, \dodoi{10.1063/1.4798541}

\bibitem[{{Bret} {et~al.}(2014){Bret}, {Stockem}, {Narayan}, \& {Silva}}]{Bret_2014}
{Bret}, A., {Stockem}, A., {Narayan}, R., \& {Silva}, L.~O. 2014, Physics of Plasmas, 21, 072301, \dodoi{10.1063/1.4886121}

\bibitem[{{Cerutti} \& {Giacinti}(2023)}]{cerutti23}
{Cerutti}, B., \& {Giacinti}, G. 2023, \aap, 676, A23, \dodoi{10.1051/0004-6361/202346481}

\bibitem[{{Chang} {et~al.}(2008){Chang}, {Spitkovsky}, \& {Arons}}]{chang}
{Chang}, P., {Spitkovsky}, A., \& {Arons}, J. 2008, \apj, 674, 378, \dodoi{10.1086/524764}

\bibitem[{{Derishev} \& {Piran}(2016)}]{derishev}
{Derishev}, E.~V., \& {Piran}, T. 2016, \mnras, 460, 2036, \dodoi{10.1093/mnras/stw1175}

\bibitem[{{Faure} {et~al.}(2024){Faure}, {Tordeux}, {Gremillet}, \& {Lemoine}}]{Faure}
{Faure}, J.~C., {Tordeux}, D., {Gremillet}, L., \& {Lemoine}, M. 2024, \pre, 109, 015203, \dodoi{10.1103/PhysRevE.109.015203}

\bibitem[{{Fried}(1959)}]{fried}
{Fried}, B.~D. 1959, Physics of Fluids, 2, 337, \dodoi{10.1063/1.1705933}

\bibitem[{Garasev \& Derishev(2016)}]{garasev}
Garasev, M., \& Derishev, E. 2016, Mon. Not. Roy. Astron. Soc., 461, 641, \dodoi{10.1093/mnras/stw1345}

\bibitem[{{Gro{\v{s}}elj} {et~al.}(2022){Gro{\v{s}}elj}, {Sironi}, \& {Beloborodov}}]{groseljpair}
{Gro{\v{s}}elj}, D., {Sironi}, L., \& {Beloborodov}, A.~M. 2022, \apj, 933, 74, \dodoi{10.3847/1538-4357/ac713e}

\bibitem[{{Gro{\v{s}}elj} {et~al.}(2024){Gro{\v{s}}elj}, {Sironi}, \& {Spitkovsky}}]{groseljlong}
{Gro{\v{s}}elj}, D., {Sironi}, L., \& {Spitkovsky}, A. 2024, \apjl, 963, L44, \dodoi{10.3847/2041-8213/ad2c8c}

\bibitem[{Gruzinov(2001)}]{gruzinovdecay}
Gruzinov, A. 2001, The Astrophysical Journal, 563, L15, \dodoi{10.1086/324223}

\bibitem[{Gruzinov \& Waxman(1999)}]{gruzinov1999}
Gruzinov, A., \& Waxman, E. 1999, The Astrophysical Journal, 511, 852, \dodoi{10.1086/306720}

\bibitem[{Hakobyan {et~al.}(2023)Hakobyan, Spitkovsky, Chernoglazov, Philippov, Groselj, \& Mahlmann}]{tristan_v2}
Hakobyan, H., Spitkovsky, A., Chernoglazov, A., {et~al.} 2023, PrincetonUniversity/tristan-mp-v2: v2.6, v2.6,  Zenodo, \dodoi{10.5281/zenodo.7566725}

\bibitem[{{Huang} {et~al.}(2022){Huang}, {Kirk}, {Giacinti}, \& {Reville}}]{Huang2022}
{Huang}, Z.-Q., {Kirk}, J.~G., {Giacinti}, G., \& {Reville}, B. 2022, \apj, 925, 182, \dodoi{10.3847/1538-4357/ac3f38}

\bibitem[{{Kato}(2007)}]{kato}
{Kato}, T.~N. 2007, \apj, 668, 974, \dodoi{10.1086/521297}

\bibitem[{{Keshet} {et~al.}(2009){Keshet}, {Katz}, {Spitkovsky}, \& {Waxman}}]{keshet2009}
{Keshet}, U., {Katz}, B., {Spitkovsky}, A., \& {Waxman}, E. 2009, \apjl, 693, L127, \dodoi{10.1088/0004-637X/693/2/L127}

\bibitem[{{Keshet} \& {Waxman}(2005)}]{keshet2005}
{Keshet}, U., \& {Waxman}, E. 2005, \prl, 94, 111102, \dodoi{10.1103/PhysRevLett.94.111102}

\bibitem[{{Kirk} {et~al.}(2000){Kirk}, {Guthmann}, {Gallant}, \& {Achterberg}}]{kirk}
{Kirk}, J.~G., {Guthmann}, A.~W., {Gallant}, Y.~A., \& {Achterberg}, A. 2000, \apj, 542, 235, \dodoi{10.1086/309533}

\bibitem[{{Kumar} \& {Panaitescu}(2004)}]{kumarwind}
{Kumar}, P., \& {Panaitescu}, A. 2004, \mnras, 354, 252, \dodoi{10.1111/j.1365-2966.2004.08185.x}

\bibitem[{{Kumar} \& {Zhang}(2015)}]{kumar}
{Kumar}, P., \& {Zhang}, B. 2015, \physrep, 561, 1, \dodoi{10.1016/j.physrep.2014.09.008}

\bibitem[{{Lemoine}(2015)}]{lemoinedecay}
{Lemoine}, M. 2015, Journal of Plasma Physics, 81, 455810101, \dodoi{10.1017/S0022377814000920}

\bibitem[{{Lemoine} {et~al.}(2019{\natexlab{a}}){Lemoine}, {Gremillet}, {Pelletier}, \& {Vanthieghem}}]{Lemoine_2019_PRL}
{Lemoine}, M., {Gremillet}, L., {Pelletier}, G., \& {Vanthieghem}, A. 2019{\natexlab{a}}, Phys. Rev. Lett., 123, 035101, \dodoi{10.1103/PhysRevLett.123.035101}

\bibitem[{{Lemoine} \& {Pelletier}(2010)}]{lemoine1}
{Lemoine}, M., \& {Pelletier}, G. 2010, \mnras, 402, 321, \dodoi{10.1111/j.1365-2966.2009.15869.x}

\bibitem[{{Lemoine} \& {Pelletier}(2011)}]{lemoine2}
---. 2011, \mnras, 417, 1148, \dodoi{10.1111/j.1365-2966.2011.19331.x}

\bibitem[{{Lemoine} {et~al.}(2019{\natexlab{b}}){Lemoine}, {Pelletier}, {Vanthieghem}, \& {Gremillet}}]{Lemoine_2019_III}
{Lemoine}, M., {Pelletier}, G., {Vanthieghem}, A., \& {Gremillet}, L. 2019{\natexlab{b}}, Phys. Rev. E, 100, 033210, \dodoi{10.1103/PhysRevE.100.033210}

\bibitem[{{Lemoine} {et~al.}(2019{\natexlab{c}}){Lemoine}, {Vanthieghem}, {Pelletier}, \& {Gremillet}}]{Lemoine_2019_II}
{Lemoine}, M., {Vanthieghem}, A., {Pelletier}, G., \& {Gremillet}, L. 2019{\natexlab{c}}, Phys. Rev. E, 100, 033209, \dodoi{10.1103/PhysRevE.100.033209}

\bibitem[{{Lyubarsky} \& {Eichler}(2006)}]{lyubarsky}
{Lyubarsky}, Y., \& {Eichler}, D. 2006, \apj, 647, 1250, \dodoi{10.1086/505523}

\bibitem[{{Martinez} {et~al.}(2021){Martinez}, {Grismayer}, \& {Silva}}]{compton}
{Martinez}, B., {Grismayer}, T., \& {Silva}, L.~O. 2021, Journal of Plasma Physics, 87, 905870313, \dodoi{10.1017/S0022377821000660}

\bibitem[{{Matsumoto} \& {Omura}(1993)}]{buneman}
{Matsumoto}, H., \& {Omura}, Y. 1993.
\newblock \url{https://books.google.com/books?id=BRLePAAACAAJ}

\bibitem[{{Medvedev} \& {Loeb}(1999)}]{medvedev}
{Medvedev}, M.~V., \& {Loeb}, A. 1999, \apj, 526, 697, \dodoi{10.1086/308038}

\bibitem[{{M{\'e}sz{\'a}ros}(2002)}]{meszaros2002}
{M{\'e}sz{\'a}ros}, P. 2002, \araa, 40, 137, \dodoi{10.1146/annurev.astro.40.060401.093821}

\bibitem[{{M{\'e}sz{\'a}ros}(2006)}]{meszaros2006}
---. 2006, Reports on Progress in Physics, 69, 2259, \dodoi{10.1088/0034-4885/69/8/R01}

\bibitem[{{Moiseev} \& {Sagdeev}(1963)}]{Moiseev_1963}
{Moiseev}, S.~S., \& {Sagdeev}, R.~Z. 1963, Journal of Nuclear Energy. Part C, Plasma Physics, Accelerators, Thermonuclear Research, 5, 43

\bibitem[{{Nakar} {et~al.}(2011){Nakar}, {Bret}, \& {Milosavljevi{\'c}}}]{nakar}
{Nakar}, E., {Bret}, A., \& {Milosavljevi{\'c}}, M. 2011, \apj, 738, 93, \dodoi{10.1088/0004-637X/738/1/93}

\bibitem[{{Pelletier} {et~al.}(2019){Pelletier}, {Gremillet}, {Vanthieghem}, \& {Lemoine}}]{Pelletier_2019}
{Pelletier}, G., {Gremillet}, L., {Vanthieghem}, A., \& {Lemoine}, M. 2019, Phys. Rev. E, 100, 013205, \dodoi{10.1103/PhysRevE.100.013205}

\bibitem[{{Peterson} {et~al.}(2021){Peterson}, {Glenzer}, \& {Fiuza}}]{peterson_21}
{Peterson}, J.~R., {Glenzer}, S., \& {Fiuza}, F. 2021, \prl, 126, 215101, \dodoi{10.1103/PhysRevLett.126.215101}

\bibitem[{{Peterson} {et~al.}(2022){Peterson}, {Glenzer}, \& {Fiuza}}]{peterson_22}
---. 2022, \apjl, 924, L12, \dodoi{10.3847/2041-8213/ac44a2}

\bibitem[{{Piran}(1999)}]{fireball}
{Piran}, T. 1999, \physrep, 314, 575, \dodoi{10.1016/S0370-1573(98)00127-6}

\bibitem[{Piran(2005)}]{piran}
Piran, T. 2005, Rev. Mod. Phys., 76, 1143, \dodoi{10.1103/RevModPhys.76.1143}

\bibitem[{{Plotnikov} {et~al.}(2018){Plotnikov}, {Grassi}, \& {Grech}}]{plotnikov_18}
{Plotnikov}, I., {Grassi}, A., \& {Grech}, M. 2018, \mnras, 477, 5238, \dodoi{10.1093/mnras/sty979}

\bibitem[{{Rabinak} {et~al.}(2011){Rabinak}, {Katz}, \& {Waxman}}]{rabinak}
{Rabinak}, I., {Katz}, B., \& {Waxman}, E. 2011, \apj, 736, 157, \dodoi{10.1088/0004-637X/736/2/157}

\bibitem[{{Ramirez-Ruiz} {et~al.}(2007){Ramirez-Ruiz}, {Nishikawa}, \& {Hededal}}]{ramirezruiz}
{Ramirez-Ruiz}, E., {Nishikawa}, K.-I., \& {Hededal}, C.~B. 2007, \apj, 671, 1877, \dodoi{10.1086/522072}

\bibitem[{{Reville} \& {Bell}(2014)}]{Reville2014}
{Reville}, B., \& {Bell}, A.~R. 2014, \mnras, 439, 2050, \dodoi{10.1093/mnras/stu088}

\bibitem[{{Shaisultanov} {et~al.}(2012){Shaisultanov}, {Lyubarsky}, \& {Eichler}}]{shaisultanov}
{Shaisultanov}, R., {Lyubarsky}, Y., \& {Eichler}, D. 2012, \apj, 744, 182, \dodoi{10.1088/0004-637X/744/2/182}

\bibitem[{{Sironi} {et~al.}(2015){Sironi}, {Keshet}, \& {Lemoine}}]{sironishocks}
{Sironi}, L., {Keshet}, U., \& {Lemoine}, M. 2015, \ssr, 191, 519, \dodoi{10.1007/s11214-015-0181-8}

\bibitem[{{Sironi} {et~al.}(2013){Sironi}, {Spitkovsky}, \& {Arons}}]{sironimaxenergy}
{Sironi}, L., {Spitkovsky}, A., \& {Arons}, J. 2013, \apj, 771, 54, \dodoi{10.1088/0004-637X/771/1/54}

\bibitem[{{Spitkovsky}(2005)}]{spitkovsky}
{Spitkovsky}, A. 2005, in American Institute of Physics Conference Series, Vol. 801, Astrophysical Sources of High Energy Particles and Radiation, ed. T.~{Bulik}, B.~{Rudak}, \& G.~{Madejski}, 345--350, \dodoi{10.1063/1.2141897}

\bibitem[{{Spitkovsky} {et~al.}(2019){Spitkovsky}, {Gargate}, {Park}, \& {Sironi}}]{tristan}
{Spitkovsky}, A., {Gargate}, L., {Park}, J., \& {Sironi}, L. 2019, {TRISTAN-MP: TRIdimensional STANford - Massively Parallel code}, Astrophysics Source Code Library, record ascl:1908.008.
\newblock \doeprint{1908.008}

\bibitem[{{Thompson} \& {Madau}(2000)}]{thompson}
{Thompson}, C., \& {Madau}, P. 2000, \apj, 538, 105, \dodoi{10.1086/309100}

\bibitem[{{Umeda} {et~al.}(2003){Umeda}, {Omura}, {Tominaga}, \& {Matsumoto}}]{zigzag}
{Umeda}, T., {Omura}, Y., {Tominaga}, T., \& {Matsumoto}, H. 2003, Computer Physics Communications, 156, 73, \dodoi{10.1016/S0010-4655(03)00437-5}

\bibitem[{{Vanthieghem} {et~al.}(2020){Vanthieghem}, {Lemoine}, {Plotnikov}, {Grassi}, {Grech}, {Gremillet}, \& {Pelletier}}]{shock_rev}
{Vanthieghem}, A., {Lemoine}, M., {Plotnikov}, I., {et~al.} 2020, Galaxies, 8, 33, \dodoi{10.3390/galaxies8020033}

\bibitem[{{Vanthieghem} {et~al.}(2022){Vanthieghem}, {Mahlmann}, {Levinson}, {Philippov}, {Nakar}, \& {Fiuza}}]{radiation}
{Vanthieghem}, A., {Mahlmann}, J.~F., {Levinson}, A., {et~al.} 2022, \mnras, 511, 3034, \dodoi{10.1093/mnras/stac162}

\bibitem[{{Waxman}(2006)}]{waxman}
{Waxman}, E. 2006, Plasma Physics and Controlled Fusion, 48, B137, \dodoi{10.1088/0741-3335/48/12B/S14}

\bibitem[{Weibel(1959)}]{weibel}
Weibel, E.~S. 1959, Phys. Rev. Lett., 2, 83, \dodoi{10.1103/PhysRevLett.2.83}

\bibitem[{{Wiersma} \& {Achterberg}(2004)}]{wiersma}
{Wiersma}, J., \& {Achterberg}, A. 2004, \aap, 428, 365, \dodoi{10.1051/0004-6361:20041882}

\bibitem[{{Yee}(1966)}]{yee}
{Yee}, K. 1966, IEEE Transactions on Antennas and Propagation, 14, 302, \dodoi{10.1109/TAP.1966.1138693}

\bibitem[{{Zakharov}(1972)}]{zakharov}
{Zakharov}, V.~E. 1972, Soviet Journal of Experimental and Theoretical Physics, 35, 908

\end{thebibliography}
\bibliographystyle{aasjournal}

\end{document}